\documentclass[useAMS,usenatbib]{mn2e}
\newcommand \url{}
\newcommand{\ts}{\textsuperscript}
\newcommand{\ion}[1]{~\textsc{#1}}

\usepackage{longtable}
\usepackage{graphicx}
\usepackage{lscape}
\usepackage{rotating}
\usepackage{amssymb}

\def \nodata{. . .}

\def \HI{H\ion{i}}

\def \Lbol{L$_{\rm AGN}$}
\def \Mstar{M$_{\star}$}
\def \logMstar{log(M$_{\star}$/M$_{\odot}$)}
\def \logRimp{log($\rho_{\rm imp}$/kpc)}
\def \logsSFR{log(sSFR / yr$^{-1}$)}
\def \logLbol{log(L$_{\rm AGN}$/erg s$^{-1}$)}
\def \kms{km~s$^{-1}$}

\def \Rimp{$\rho_{\rm imp}$}

\def \deltaEW{${\rm \Delta log(EW/m\AA)}$}
\def \RadRat{$f_{\rm AGN}/f_{\rm HM01}$}

\begin{document}

\title[Probing the CGM around AGN hosts]{The COS-AGN survey: Revealing the nature of circum-galactic gas around hosts of active galactic nuclei}
\author[Berg et al.] {
\parbox[t]{\textwidth}{
Trystyn A. M. Berg$^1$, Sara L. Ellison$^1$,  Jason Tumlinson$^2$, Benjamin D. Oppenheimer$^{3}$,  Ryan Horton$^{3}$, Rongmon Bordoloi$^{4,6}$, Joop Schaye$^5$}\\\\
$^1$ Department of Physics and Astronomy, University of Victoria, Victoria, British Columbia, V8P 1A1, Canada.\\
$^2$ Space Telescope Science Institute, 3700 San Martin Drive, Baltimore, MD 21218, USA.\\
$^3$ CASA, Department of Astrophysical and Planetary Sciences, University of Colorado, Boulder, Colorado, 80309, USA. \\
$^4$ MIT-Kavli Center for Astrophysics and Space Research, 77 Massachusetts Avenue, Cambridge, MA, 02139, USA\\
$^5$ Leiden Observatory, Leiden University, PO Box 9513, NL-2300 RA Leiden, the Netherlands\\
$^6$ Hubble Fellow\\
}

\maketitle
\begin{abstract}
Active galactic nuclei (AGN) are thought to play a critical role in shaping galaxies, but their effect on the circumgalactic medium (CGM) is not well studied. We present results from the COS-AGN survey: 19 quasar sightlines that probe the CGM of 20 optically-selected AGN host galaxies with impact parameters $80<$\Rimp{}$< 300$~kpc.  Absorption lines from a variety of species are measured and compared to a stellar mass and impact parameter matched sample of sightlines through non-AGN galaxies. Amongst the observed species in the COS-AGN sample (Ly$\alpha$, C\ion{ii}, Si\ion{ii}, Si\ion{iii}, C\ion{iv}, Si\ion{iv}, N\ion{v}), only Ly$\alpha$ shows a high covering fraction ($94^{+6}_{-23}$\% for rest-frame equivalent widths EW$\geq124$~m\AA{}) whilst many of the metal ions are not detected in individual sightlines. A sightline-by-sightline comparison between COS-AGN and the control sample yields no significant difference in EW distribution. However, stacked spectra of the COS-AGN and control samples show significant ($>3 \sigma$) enhancements in the EW of both Ly$\alpha$ and Si\ion{iii}  at impact parameters $>164$~kpc by a factor of $+0.45\pm0.05$~dex and $>+0.75$~dex respectively. The lack of detections of both high-ionization species near the AGN and strong kinematic offsets between the absorption systemic galaxy redshifts indicates that neither the AGN's ionization nor its outflows are the origin of these differences. Instead, we suggest the observed differences could result from either AGN hosts residing in haloes with intrinsically distinct gas properties, or that their CGM has been affected by a previous event, such as a starburst, which may also have fuelled the nuclear activity.
\end{abstract}


\section{Introduction}

The circum-galactic medium (CGM) is the interface between cold flows from the intergalactic medium onto a galaxy, and hosts hot halo gas and material ejected from galaxies \citep[for reviews, see][]{Putman12,Tumlinson17}. With various processes in galaxy evolution consuming (e.g. star formation) and removing (e.g. winds) gas, the CGM is shaped by the processes internal to the galaxy. Early progress in the study of the CGM came from connecting absorption lines in quasar (QSO) spectra with galaxies imaged in the foreground, tracing the extent and properties of the CGM gas as a function of the host galaxy's properties \citep[e.g.][]{Bergeron86,Bowen95,Lanzetta95,Adelberger05,Chen10,Steidel10,Bordoloi11,Prochaska11,Turner14}. Building on these foundations, our understanding of the CGM has been significantly improved through surveys with the Hubble Space Telescope (HST) Cosmic Origins Spectrograph \citep[COS;][]{Green12}. The first of several surveys of the CGM surrounding low redshift galaxies was the COS-Halos survey \citep{Tumlinson13} which targetted the CGM around 44 $\sim$L$^{\star}$ galaxies, demonstrating that the properties of the CGM differ depending on whether the central galaxy is passive or star-forming \citep[defined using a specific star formation rate cut of sSFR$\rm{=10^{-11} yr^{-1}}$;][]{Tumlinson11,Werk13,Borthakur16}. The COS-Halos team found a distinct lack of O\ion{vi} around passive galaxies, while \HI{} was found at the same strength around all galaxies \citep{Tumlinson11,Thom12}. Additionally, connections have been made between the CGM and properties of the host galaxy, including: increased \HI{} content of the CGM with larger interstellar medium (ISM) gas masses \citep[COS-GASS;][]{Borthakur15}, the presence of extended gas reservoirs around galaxies of all stellar mass \citep[COS-Dwarfs;][]{Bordoloi14}, and enhanced metal content around starbursting hosts \citep[COS-Burst;][]{Borthakur13,Heckman17}.

An important stage in the evolution of galaxies is when their central supermassive black holes are actively accreting material. This active galactic nucleus (AGN) phase may be responsible for the removal of gas from star forming reservoirs within galaxies via winds and outflows \citep{Veilleux05,Tremonti07,Sturm11,Woo17}, and has been associated with the evolution of galaxies off the star-forming main sequence to passive galaxies \citep{Springel05,Schawinski07,Fabian12,Bluck14,Bluck16}. In addition, radio-mode feedback and radiation from the AGN keeps the CGM hot, buoyant, and consistently ionized \citep[][]{McNamara07,Bower17,Hani17}, as well as preventing gas from returning to the host galaxy. Such processes have been proposed to be responsible for O\ion{vi} bimodality seen in the CGM by COS-Halos without an active AGN \citep{Oppenheimer17}.

Observationally linking the environmental and feedback effects of AGN hosts with their CGM has primarily been done through the use of projected QSO-QSO pairs at higher redshifts. This technique has the added benefit of looking at the role of a stronger and weaker QSO radiation fields located in the respective transverse (background QSO) and line-of-sight (foreground QSO; i.e.~along the outflow) CGM \citep{Bowen06,Farina13,Johnson15}. Cool gas traced by \HI{} and Mg\ion{ii} is anisotropically distributed about the QSO, with larger column densities of \HI{} preferentially found along the transverse direction \citep{QPQ1,Farina13}; suggesting that radiation from the QSO does not affect the transverse medium \citep[][]{QPQ2,QPQ7,Farina14}. An excess of cool gas (relative to the intergalactic medium) has been found all the way out to one Mpc, with a stronger enhancement at smaller impact parameters \citep{QPQ6}. When the QSO-QSO pairs are split by the bolometric luminosity of the QSO host, the Mg\ion{ii} covering fraction is larger for high-luminosity QSOs (covering fraction of $\approx 60$\% for luminosities of ${\rm L_{Bol}\geq45.5}$ erg s$^{-1}$) compared to low luminosity QSOs \citep[$\approx20$\%, ${\rm L_{Bol}\leq45.5}$ erg s$^{-1}$;][]{Johnson15}. All of these observations of excess cool gas around luminous QSOs is suggestive of either a viewing angle effect with the ionizing radiation exciting cool gas along the line of sight to the QSO, or an environmental effect of haloes hosting massive QSOs such as debris from galaxy interactions fuelling QSO activity \citep[][]{QPQ6,Farina14,Johnson15}.

Most of the work described above has focused on high luminosity quasars. However, there has been little focus on how the less luminous but more common Seyfert-like AGN shape their surrounding CGM. In the only observational study of the CGM surrounding Seyfert galaxies, \cite{Kacprzak15} found a low (10\%) Mg\ion{ii} $\lambda$ 2796 \AA{} covering fraction around 14 AGN (between 100 and 200 kpc) in the transverse direction relative to field and QSO host galaxies, but a reservoir of cool gas still exists along the line of sight to the AGN. They suggest that AGN-driven outflows are destroying the cool gas in the transverse direction (i.e.~along the outflow), suggesting that the difference between their observations of the CGM of AGN-dominated galaxies with previous observations of QSOs \citep[e.g.][]{QPQ6} is caused by the viewing angle of the AGN.

Predictions from zoom-in simulations of galaxies taken from the EAGLE cosmological simulation \citep{Schaye15} suggest that radiative feedback from the AGN should ionize the gas out to a distance of two virial radii \citep{Oppenheimer13,Segers17}. After implementing non-equilibrium ionization into their models, \cite{Oppenheimer13} have predicted that  AGN proximity fossil zones exist around galaxies that host (or have hosted) bright AGN, with the metals remaining in an over ionized state for several megayears \citep[depending on the luminosity duty cycle and lifetime of the AGN;][]{Segers17,Oppenheimer18}. However, the detailed CGM properties in simulations can be quite sensitive to the size of the CGM clouds, implementation of feedback, and different recipes between codes \citep{Stinson12,Gutcke17,Nelson17}.

In this paper, we investigate the observational properties of the CGM around galaxies hosting Type II  Seyfert AGN (which we will henceforth simply refer to as AGN). We measure the rest-frame equivalent widths (EWs) of a range of ionization species present in the CGM material probed by QSO sightlines near 20 AGN-host galaxies. We provide a systematic comparison to non-AGN galaxies observed in the literature to quantify whether the CGM around AGN host galaxies is different from their counterparts. Throughout the paper, we assume a flat $\Lambda$CDM Universe with $H_{0}=67.8~{\rm km~s^{-1}~Mpc^{-1}}$ and $\Omega_{M}=0.308$ \citep{Planck15}.

\section{Data}
\subsection{Sample selection and properties}
\label{sec:SampProps}
\begin{figure}[h]
\begin{center}
\includegraphics[width=0.5\textwidth]{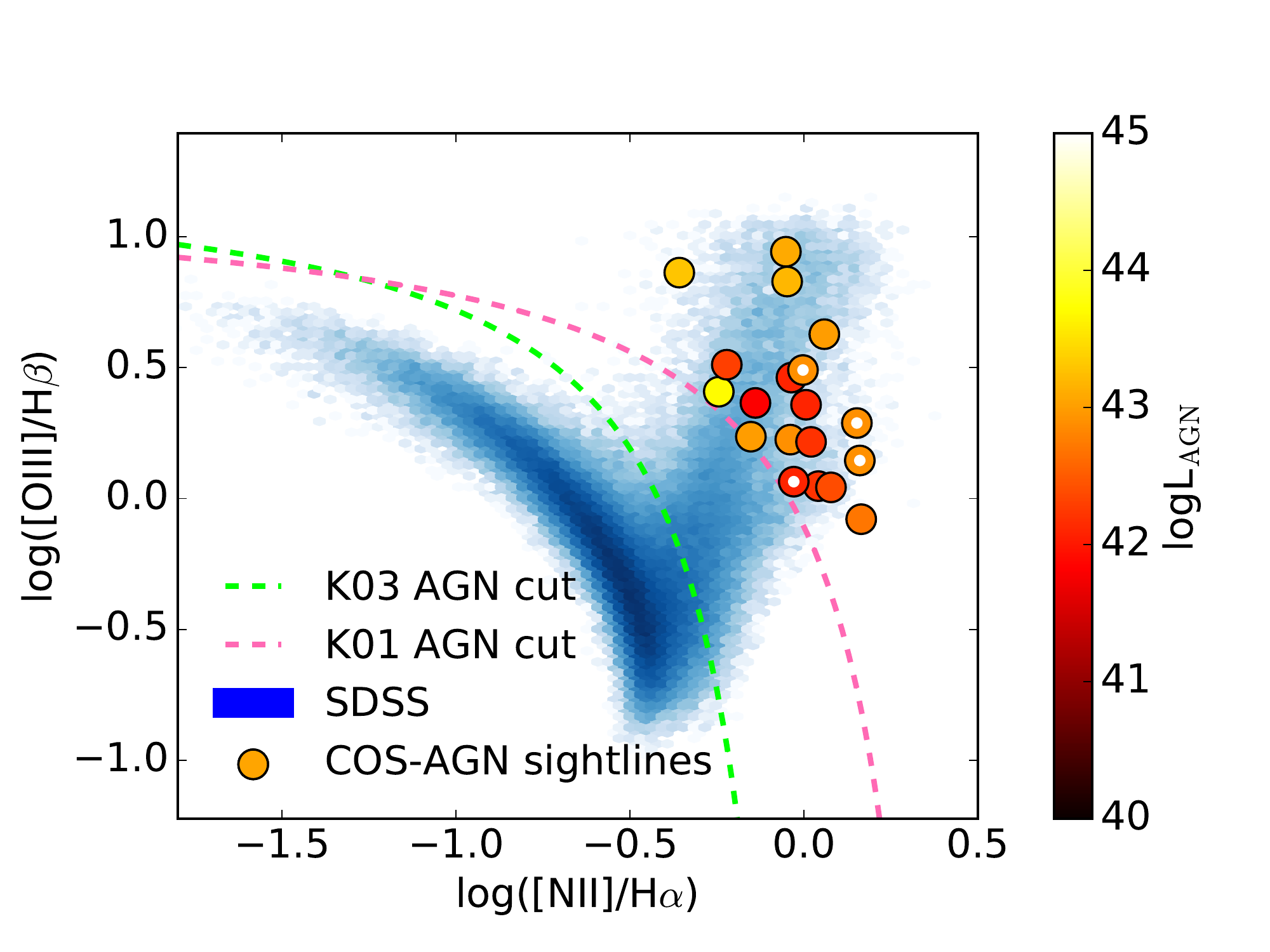}
\caption{The BPT diagram of all SDSS galaxies with spectroscopic observations (blue shaded region; only showing SDSS galaxies with $>5\sigma$ detections of diagnostic emission lines). The solid circles show the COS-AGN galaxies, and are coloured based on their AGN luminosity. The dashed pink and green lines denote the~\citet[K01]{Kewley01} and~\citet[K03]{Kauffmann03} cuts typically used to select AGN and composite galaxies.  LINERS classified using the \citet{Kewley06} emission line metrics (see Table \ref{tab:SightProps}) are denoted with a white dot on top of the datapoint.}
\label{fig:BPT}
\end{center}
\end{figure}

\begin{figure*}
\begin{center}
\includegraphics[width=\textwidth]{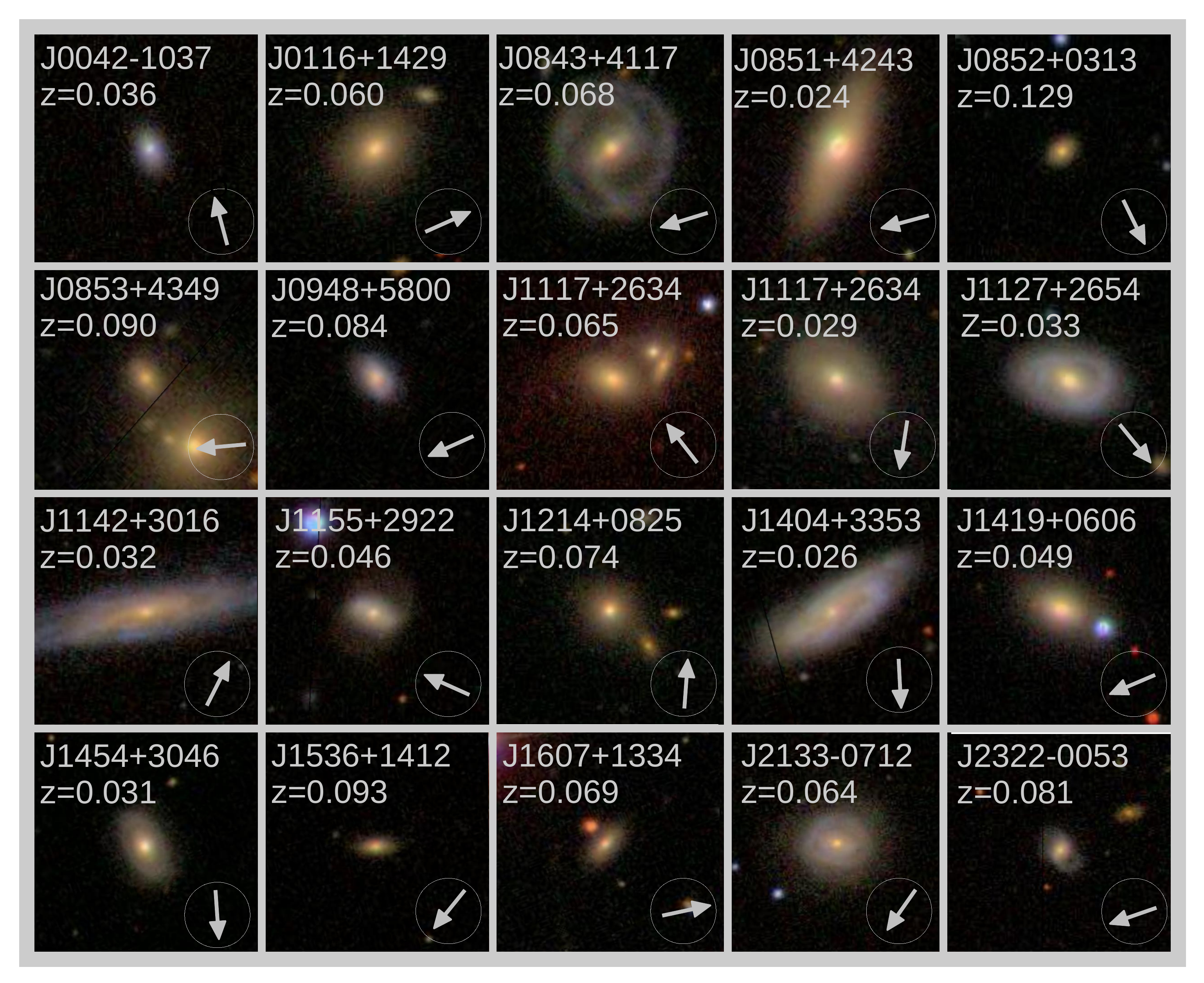}
\caption{SDSS postage stamp images of the entire COS-AGN sample. Each image is 50 arcsec $\times$ 50 arcsec. The QSO name and AGN host galaxy redshift are provided in each panel. For reference, the arrow in the bottom right of each panel gives the relative direction of the background QSO from the centre of the galaxy.}
\label{fig:postage}
\end{center}
\end{figure*}

The QSO sightlines through the CGM of AGN galaxies were selected by cross-matching coordinates of Sloan Digital Sky Survey \citep[SDSS;][]{Abazajian09} galaxies hosting AGN with the locations of UV-bright QSOs ($17<m_{FUV}<18.9$) identified in the Galaxy Evolution Explorer (GALEX) catalogue\footnote{http://galex.stsci.edu/} \citep{Martin05}.  AGN were selected using the emission line ratios [N\ion{ii}/H$\alpha$] and [O\ion{iii}/H$\beta$] measured in the SDSS\footnote{Emission line fluxes were taken from http://www.mpa-garching.mpg.de/SDSS/.} following the line ratio diagnostics provided in \cite{Kewley01}. We required that all four emission lines  be detected at $\geq5\sigma$, and that the background QSO sightline must probe within 300 kpc of the AGN host galaxy. Using these criteria, we identified ten AGN-QSO pairs (along nine QSO sightlines) that had been previously observed with HST whose data is located in the Hubble Spectroscopic Legacy Archive (HSLA; data release 1\footnote{https://archive.stsci.edu/hst/spectral\_legacy/}). We further selected ten more QSO sightlines to probe the inner 175 kpc of the CGM surrounding AGN, which we observed with HST/COS in Cycle-22. These combined 19 sightlines probing 20 AGN host galaxies make up our COS-AGN sample. Descriptions of the observations are presented in Section \ref{sec:Obs}. Figure \ref{fig:BPT} shows the so-called BPT diagram \citep{Baldwin81} of the COS-AGN host galaxies (coloured circles) compared to SDSS galaxies (blue shaded region), whilst Figure \ref{fig:postage} shows the SDSS thumbnail images of each AGN host. Of the 20 COS-AGN host galaxies, four are classified as LINERS based on the diagnostics presented in \citet[equations 1--15]{Kewley06} that encompass both the classic BPT diagram (Figure \ref{fig:BPT}) and line ratios that include [O\ion{ii}] $\lambda\lambda$ 3726\AA{} \& 3729\AA{},  and [S\ion{ii}] $\lambda \lambda$ 6717\AA{} \& 6731\AA{}. Although the LINER category was originally envisaged to identify low luminosity AGN \citep{Heckman80,Ho97}, alternative ionization mechanisms can also produce extended LINER-like emission \citep{Yan12,Belfiore16}. We therefore keep these sightlines in our sample as they may be bona fide AGN, but are flagged through-out the analysis to assess the effects of including or removing LINERs from the sample.

The properties of the COS-AGN galaxies are given in Table \ref{tab:SightProps}. We adopted the SDSS spectroscopic redshifts as the systemic redshifts of the galaxy ($z_{\rm gal}$). Stellar masses were taken from \cite{Mendel14}, which are based on spectral energy distribution fits to the SDSS $g$ and $r$-band photometry of a single S\'ersic fit to the galaxy, and assume a \cite{Chabrier03} initial mass function. The most commonly adopted SFRs for SDSS galaxies are those provided in the MPA/JHU catalogues \citep{Brinchmann04,Salim07}.  These SFRs are nominally based on fits to emission lines for star forming galaxies.  However, it is well known that the standard SFR conversions \citep[e.g.][]{Kennicutt98} will be incorrect when AGN contribute to emission line fluxes.  Therefore, for galaxies with an AGN, the MPA/JHU catalog presents a SFR based on the correlation between the 4000 \AA{} break and sSFR in star forming galaxies. However, SFRs derived from the 4000 \AA{} break have large uncertainties, and it has been recently shown that the far-IR can be used to obtain more accurate values \citep{Rosario16}. We therefore use the infrared luminosity (${\rm L_{IR}}$) of the SDSS galaxies predicted by an artificial neural network \citep{Ellison16} to calculate log(sSFR) using the conversion log(sSFR/yr$^{-1}$)$={\rm log L_{IR}} - 43.951 - {\rm log(M_{\star}/M_{\odot})}$.  All of the COS-AGN galaxies have a predicted ${\rm L_{IR}}$ confidence of $\approx0.1$ dex, passing the adopted quality control cut in \cite{Ellison16}. The  bolometric luminosity of the AGN (\Lbol{}) is calculated based on the AGN's [O\ion{iii}] line luminosity following \cite{Kauffmann09}, where the bolometric luminosity of the AGN is $600\times$ the [O\ion{iii}] line luminosity. The projected proper impact parameter of the QSO sightline (\Rimp{}) is calculated at the systemic redshift of the AGN host galaxy.  Lastly, the AGN type from the \cite{Kewley06} classification is also tabulated.

\begin{table*}
\tiny
\begin{center}
\caption{COS-AGN sightline properties}
\label{tab:SightProps}
\begin{tabular}{lccccccccccc}
\hline
QSO name& $z_{gal}$& Galaxy SDSS objid&  Gal. R.A.& Gal. Dec.& log(M$_\star$)& log(sSFR)& log(L$_{\rm AGN}$)& $\rho_{\rm imp}$& Program & Companion? & AGN type$^{\star}$\\
& &  & [$^\circ$]& [$^\circ$]& [log(M$_\odot$)]& [log(yr$^{-1}$)]& [log(erg s$^{-1}$)]&  [kpc]& &\\
\hline
J0042$-$1037& 0.036& 587727178454007913&  10.562& $-$10.738& 9.5& $-$10.2& 42.8&  299& COS$-$Halos & N & ---\\
J0116+1429& 0.060& 587724232641544435&  19.126& 14.482& 11.1& $-$11.6& 42.9&  136& COS$-$AGN & N & LINER\\
J0843+4117& 0.068& 587732048403824840&  130.898& 41.308& 11.0& $-$10.9& 42.9&  223& COS$-$Dwarfs & N & LINER\\
J0851+4243& 0.024& 587732048404676666&  132.757& 42.736& 10.7& $-$11.7& 42.2&  82& COS$-$AGN & Y & ---\\
J0852+0313& 0.129& 587728880331194569&  133.255& 3.240& 11.0& $-$10.7& 42.7&  170& COS$-$AGN & N & ---\\
J0853+4349& 0.090& 587731886277197945&  133.357& 43.820& 11.0& $-$11.0& 42.2&  164& HSLA & Y & ---\\
J0948+5800& 0.084& 587725468527231140&  147.118& 58.022& 10.7& $-$10.2& 43.7&  165& COS$-$AGN & N & ---\\
J1117+2634& 0.065& 587741708880773130&  169.437& 26.526& 11.2& $-$11.5& 42.4&  268& COS$-$Dwarfs & Y & ---\\
J1117+2634& 0.029& 587741602026029237&  169.461& 26.657& 10.4& $-$11.0& 41.8&  185& COS$-$Dwarfs & Y & ---\\
J1127+2654& 0.033& 587741602027012125&  171.943& 26.960& 10.5& $-$11.2& 42.1&  145& COS$-$AGN & N & ---\\
J1142+3016& 0.032& 587741490911576221&  175.575& 30.230& 10.4& $-$10.6& 42.3&  108& COS$-$AGN & N & Seyfert\\
J1155+2922& 0.046& 587741532251685053&  178.903& 29.351& 10.4& $-$10.8& 42.1&  215& COS$-$GASS & N & LINER\\
J1214+0825& 0.074& 588017726547361972&  183.630& 8.374& 11.2& \nodata{}& 42.9&  237& HSLA & N & LINER\\
J1404+3353& 0.026& 587739131343929515&  211.122& 33.953& 10.4& $-$10.8& 43.1&  115& HSLA & N & Seyfert\\
J1419+0606& 0.049& 587730022252675197&  214.909& 6.135& 10.9& $-$10.8& 42.9&  180& HSLA & N & ---\\
J1454+3046& 0.031& 587739131885649936&  223.612& 30.909& 10.3& $-$11.3& 42.1&  287& COS$-$GASS & Y & ---\\
J1536+1412& 0.093& 587742551760765148&  234.172& 14.227& 10.7& $-$10.8& 43.3&  154& COS$-$AGN & N & Seyfert\\
J1607+1334& 0.069& 587742614562603047&  241.824& 13.565& 10.6& $-$10.8& 43.2&  161& COS$-$AGN & N & Seyfert\\
J2133$-$0712& 0.064& 587726878878073213&  323.473& $-$7.180& 11.1& $-$10.8& 43.0&  140& COS$-$AGN & N & Seyfert\\
J2322$-$0053& 0.081& 587731185126080831&  350.730& $-$0.893& 10.7& $-$10.5& 43.0& 121& COS$-$AGN & N & ---\\
\hline
\end{tabular}
$^{\star}$The AGN are classified as LINERs or Seyferts using the SDSS emission line diagnostics presented in \citet[Equations 1-15]{Kewley06}.  Cases where the classification is either ambiguous, or the emission line data quality is poor (S/N$<3$ for any emission line) are denoted by blank entries.
\end{center}
\end{table*}

\begin{figure}
\begin{center}
\includegraphics[width=0.4\textwidth]{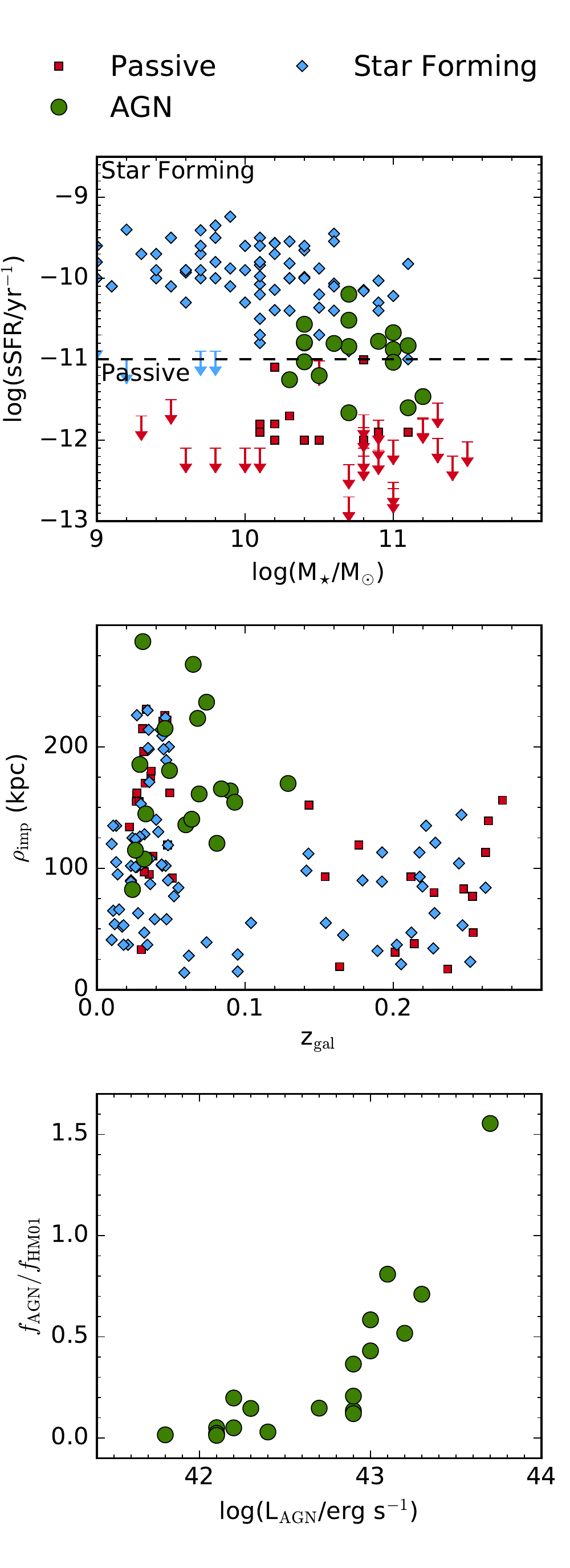}
\caption{Top: sSFR vs \Mstar{} for the COS-AGN sample (green circles) compared to star-forming (blue diamonds) and passive (red squares) galaxies from the literature.  Middle: Impact parameter (\Rimp{}) vs galaxy redshift ($z_{\rm gal}$) for the COS-AGN and literature sample. Bottom:  the strength of Ly$\alpha$ ionizing radition of the AGN relative to the UV background (\RadRat{}) as a function of the bolometric luminosity of the AGN (\Lbol{}) for the COS-AGN sample.}
\label{fig:SampDists}
\end{center}
\end{figure}

\begin{figure}
\begin{center}
\includegraphics[width=0.5\textwidth]{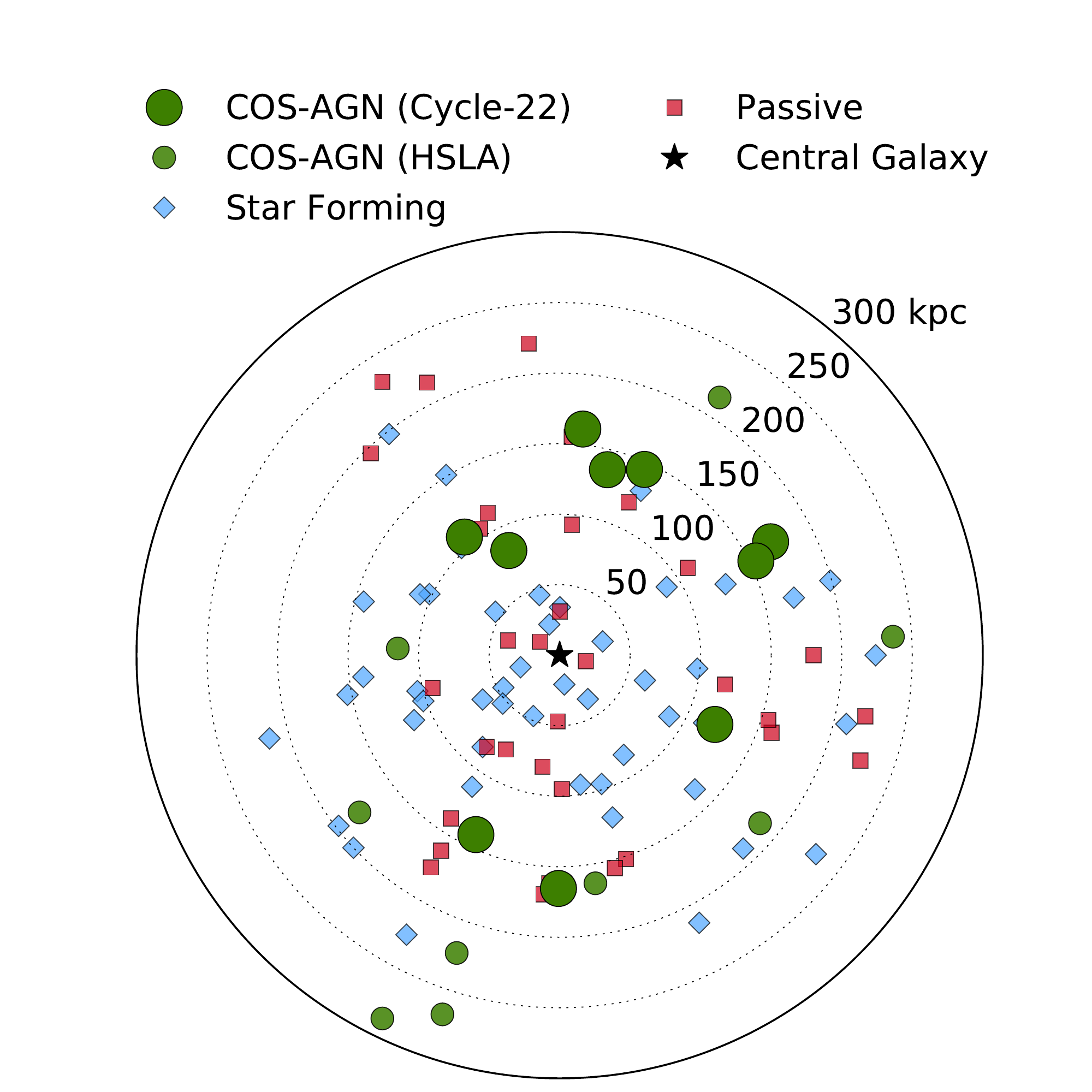}
\caption{The map of QSO sightlines that probe the CGM of AGN-dominated (green circles), star-forming (red squares), and passive galaxies (blue diamonds). All the sightlines are shown relative to the central galaxy being probed (black star). The larger and smaller circles respectively represent sightlines observed in our Cycle 22 program, and previously observed in the HSLA. The passive and star-forming galaxies are from the COS-Halos, COS-Dwarfs, and COS-GASS surveys. The angular position of the points are based on the position angle of the galaxy relative to the background QSO. }
\label{fig:map}
\end{center}
\end{figure}

To provide a systematic comparison between the properties of the CGM of galaxies with and without an AGN, we have compiled a non-AGN literature sample of EWs from all galaxies observed in the COS-Halos \citep{Tumlinson13,Werk13,Werk14}, COS-Dwarfs \citep{Bordoloi14}, and COS-GASS \citep[][]{Borthakur15,Borthakur16} surveys. Figure \ref{fig:SampDists} compares the properties of the non-AGN literature sample to the COS-AGN galaxies (green data) in terms of sSFR as a function of \Mstar{} (top left),  and \Rimp{} vs $z_{gal}$ (bottom left). The non-AGN literature sample is separated into star-forming (blue diamonds) and passive galaxies (red squares) using a log(sSFR/yr$^{-1}$)$=-11$ cut.  We note that the measured SFR values in this literature sample have been derived using different methods for COS-Halos \citep{Werk12}, COS-GASS \citep{Borthakur13}, and COS-Dwarfs \citep{Bordoloi14}. However, small differences (up to a few tenths of dex) will not affect the analyses presented in this paper as sSFR is only used to classify galaxies as star-forming or passive. To demonstrate the properties of the AGN in the COS-AGN sample, the bottom right panel of Figure \ref{fig:SampDists} shows the distributions of \Lbol{} and the strength of the \HI{} ionizing flux  from the AGN \citep[][]{Sazonov04} relative to the UV background \citep{Haardt01}\footnote{We adopt the \cite{Haardt01} UV background to be consistent with what we use in our zoom-in simulations (see Section \ref{sec:Sims}).} at the \Rimp{} of each sightline (\RadRat{}). The distribution of \RadRat{} shows the UV background is the dominant source of ionizing radiation for most of these sightlines except for the sightline towards J0948+5800 (\RadRat{}$=1.56$).  In addition, Figure \ref{fig:map} shows the distribution of QSO sightlines observed for the COS-AGN sample (our Cycle-22 observations are the larger green circles; archival sightlines are the smaller circles) about the central galaxy (black star).

In order to identify additional absorbers that could potentially contribute to the CGM of the AGN hosts, we searched for possible spectroscopic companions in the SDSS for each of the 20 COS-AGN hosts using the catalogue of galaxy companions compiled by \cite{Patton16}. The catalogue from \cite{Patton16} finds the nearest spectroscopic companion in the SDSS DR7 with a stellar mass greater than 10\% of the galaxy in question. In order to flag the possibility of contaminating absorption, we further required that the background QSO be within 300 kpc of the companion, and within $\pm1000$\kms{} of the AGN host such that any contribution from the CGM gas of the companion would be found in our absorption search window. Five of our COS-AGN galaxies have companions that match these requirements (see Table \ref{tab:SightProps}), and are situated between $193\leq$\Rimp{}$\leq274$~kpc from the targetted QSO. We have repeated all of the analysis presented in this paper with and without these five systems, and find that our results do not change qualitatively; we therefore keep these systems with companions in our sample and visually flag their data in relevant figures. We note that 13 of the remaining 15 sightlines still contain a lower mass companion (\Mstar$<10$\% of the AGN host mass) within $\pm1000$\kms{} of the AGN host and 300 kpc projected separation from the background QSO. However, given the low mass of these 13 systems we suspect that the CGM will be dominated by the AGN host.

\subsection{Observations}
\label{sec:Obs}

\begin{table*}
\small
\begin{center}
\caption{Summary of QSO observations}
\label{tab:QSOobs}
\begin{tabular}{lccccccc}
\hline
QSO ID & R.A. & Dec. &  Grating & Central wavelength(s) & Exposure time & S/N$^a$ &Program ID$^b$\\
& [$^\circ$] & [$^\circ$] & & [\AA{}] & [s] & [pixel$^{-1}$] &\\
\hline
\multicolumn{8}{c}{COS-AGN Cycle-22 sightlines}\\
\hline
J0116+1429&  19.096& 14.495& G130M  &  1309  &  5239 & 5--7 &  13774\\
&&& G160M  &  1577  &  11338  & 10&  13774\\

J0851+4243&  132.817& 42.725& G130M  &  1318  &  5408 & 6--12 &  13774\\
&&& G160M  &  1611  &  11696  & 7--12  & 13774\\

J0852+0313&  133.247& 3.222& G130M  &  1291,1327  &  2232  & 3--12 &  12603\\
&&& G160M  &  1600  &  8135  & 6--15 &  13774\\

J0948+5800&  147.167& 58.011& G130M  &  1291  &  8834  & 5--10 &  13774\\
&&& G160M  &  1589  &  18676  & 4--11 &  13774\\

J1127+2654&  171.902& 26.914& G130M  &  1291  &  2255  & 5--18 &  12603\\
&&& G160M  &  1577  &  8193  & 8--15 &  13774\\

J1142+3016&  175.551& 30.270& G130M  &  1291,1327  &  4790  & 4--14 &  12603\\
&&& G160M  &  1577  &  11466  & 8--14 &  13774\\

J1536+1412&  234.187& 14.208& G130M  &  1291  &  4251  & 4--9 &  13774\\
&&& G160M  &  1600  &  6232  & 5--9 &  13774\\

J1607+1334&  241.791& 13.572&G130M  &  1318  &  4265  & 4--7 &  13774\\
&&& G160M  &  1577  &  6218  & 3--7 &  13774\\

J2133$-$0712&  323.491& $-$7.205& G130M  &  1309  &  8245  & 7--9 &  13774\\
&&& G160M  &  1577  &  11322  & 4--9 &  13774\\

J2322$-$0053&  350.750& $-$0.900&G130M  &  1291  &  5242  & 7 &  13774\\
&&& G160M  &  1600  &  11326  & 4--10 &  13774\\
\hline
\multicolumn{7}{c}{COS-AGN HSLA sightlines}\\
\hline
J0042$-$1037& 10.593& $-$10.629&  G130M  &  1291  &  2448  & 5--7  &  11598 \\
&&& G160M  &  1600,1623  &  2781  & 9 &  11598\\

J0843+4117&  130.956& 41.295& G130M  &   1291,1309  & 4359  & 2--9 &  12248\\
&&& G160M  &  1577,1623  &  7010  & 4--9 &  12248\\

J0853+4349&  133.393& 43.817&  G130M  &  1222  &  14809  & 5--10  &  13398\\

J1117+2634&  169.476& 26.571& G160M  &  1577,1600  &  4783  & 4--16 &  12248\\

J1155+2922&  178.970& 29.377& G130M  &  1300,1327  &  10916  & 3--9 &  12603\\

J1214+0825&  183.627& 8.419& G130M  &  1300  &  4813  & 6--8 &  11698\\

J1404+3353&  211.118& 33.895&G130M  &  1309,1327  &  7706  & 4--7 &  12603\\

J1419+0606&  214.956& 6.115&G130M  &  1291  &  11028  & 5--7 &  13473\\
&&& G160M  &  1600  &  8735  & 3--6 &  13473\\

J1454+3046&  223.601& 30.783&  G130M  &  1291,1327  &  7712  & 3--8 &  12603\\

\hline

\end{tabular}
\\
\begin{flushleft}

$^a$ -- Range of continuum S/N measured at the wavelengths of Ly$\alpha$ and metal species listed in Section \ref{sec:DataEW}.\\
$^b$ -- HST MAST archive program ID. Notable program IDs include: Cycle 22 COS-AGN (13774), COS-Halos (11598),  COS-Dwarfs (12248), and COS-GASS (12603).\\
\end{flushleft}
\end{center}
\end{table*}

Observations for our ten newly targeted sightlines were completed with HST/COS during Cycle-22 (program ID 13774; PI S. Ellison).  To probe a range of ionization species, we required wavelength coverage to probe the prominent lines of \HI{} $\lambda$ 1215~\AA{}, Si\ion{ii} $\lambda$1260~\AA{}, C\ion{iv} $\lambda$1548~\AA{}, Si\ion{iv} $\lambda$1393~\AA{}, and N\ion{v} $\lambda$1238~\AA{}. This required observations using both the G130M and G160M gratings on COS. We note that the G130M data used for three of these ten sightlines were previously obtained by the COS-GASS program \citep{Borthakur15}, and were not re-observed in our program. For each individual sightline, the central wavelengths were selected to ensure coverage of these lines, and all four FP-POS offsets were used to minimize gaps in the wavelength coverage and reduce the fixed pattern noise. Exposure times were selected to obtain a similar signal-to-noise ratio (S/N) of $\approx10$ to the COS-Halos survey across the entire wavelength range \citep{Tumlinson13}. To ensure homogeneity in S/N amongst the archival HSLA sample and our Cycle-22 observations, we required a S/N $\gtrsim4$ near absorption lines of interest for the archival sightlines. It is important to note that the HSLA sightlines do not have the same wavelength coverage as our Cycle-22 sample. A summary of the observational details of all 19 COS-AGN sightlines is provided in Table \ref{tab:QSOobs}.

\subsection{Data reduction and equivalent width measurements}
\label{sec:DataEW}
To provide a systematic comparison to the EWs measured in COS-Halos and COS-GASS, we use the same data reduction technique as COS-Halos \citep{Tumlinson13}. In brief, the final \textsc{calcos} (version 3.1) extracted 1D spectra (x1d files)  are taken from the HST archive\footnote{\url{https://archive.stsci.edu/hst/}}, and are coadded and rebinned to contain $\approx6$ pixels per resolution element. The QSO continuum is fitted locally ($\pm1500$ pixels) around each absorption feature using fifth-order Legendre polynomials.

Following the COS-Halos methodology, all absorption located within $\pm500$ \kms{} of the systemic redshift of the galaxy is assumed to be associated with the CGM of the galaxy. Within the $\pm500$ \kms{} window, the integration limits of the equivalent width derivation are chosen on a line by line basis to avoid any regions of contamination, whilst limiting the amount of clean continuum within the integration bounds.

To determine if the absorption within this $\pm500$ \kms{} window is indeed associated with the CGM of the host galaxy, we confirmed  that there was neither contamination from the Galaxy's ISM nor from intervening absorbers at other redshifts more than $1000$ \kms{} from the AGN. In cases where multiple lines are covered for a given species, contamination was also flagged by comparing the strengths of the lines relative to the ratio of oscillator strengths ($f$), as well as requiring similar velocity profiles.

For five of the COS-AGN sightlines, we are uncertain if the absorption detected at the expected location of Ly$\alpha$ or Si\ion{iii} is associated with the CGM of the COS-AGN host galaxy as there are no other absorption lines to confirm its velocity or structure. We have conservatively not adopted the corresponding EW measurements in our analysis for these lines, but note (when relevant) how our results change if these EW measurements are adopted. Appendix \ref{App:EWs} presents the measured EWs for these line and the justification for why these cases are not included.

When no absorption is detected without signs of blending, 3$\sigma$ EW upper limits on these undetected lines are derived using the error spectrum within a $\pm 50$\kms{} interval of clean continuum near the systemic redshift of the AGN-dominated galaxy. The EW for lines that are visually detected but were $<3\sigma$ above the noise were automatically set to the $3\sigma$ noise threshold and flagged as upper limits. All EW upper limits adopted from the literature are converted to 3$\sigma$ values for consistency.

\begin{figure*}
\begin{center}
\includegraphics[width=\textwidth]{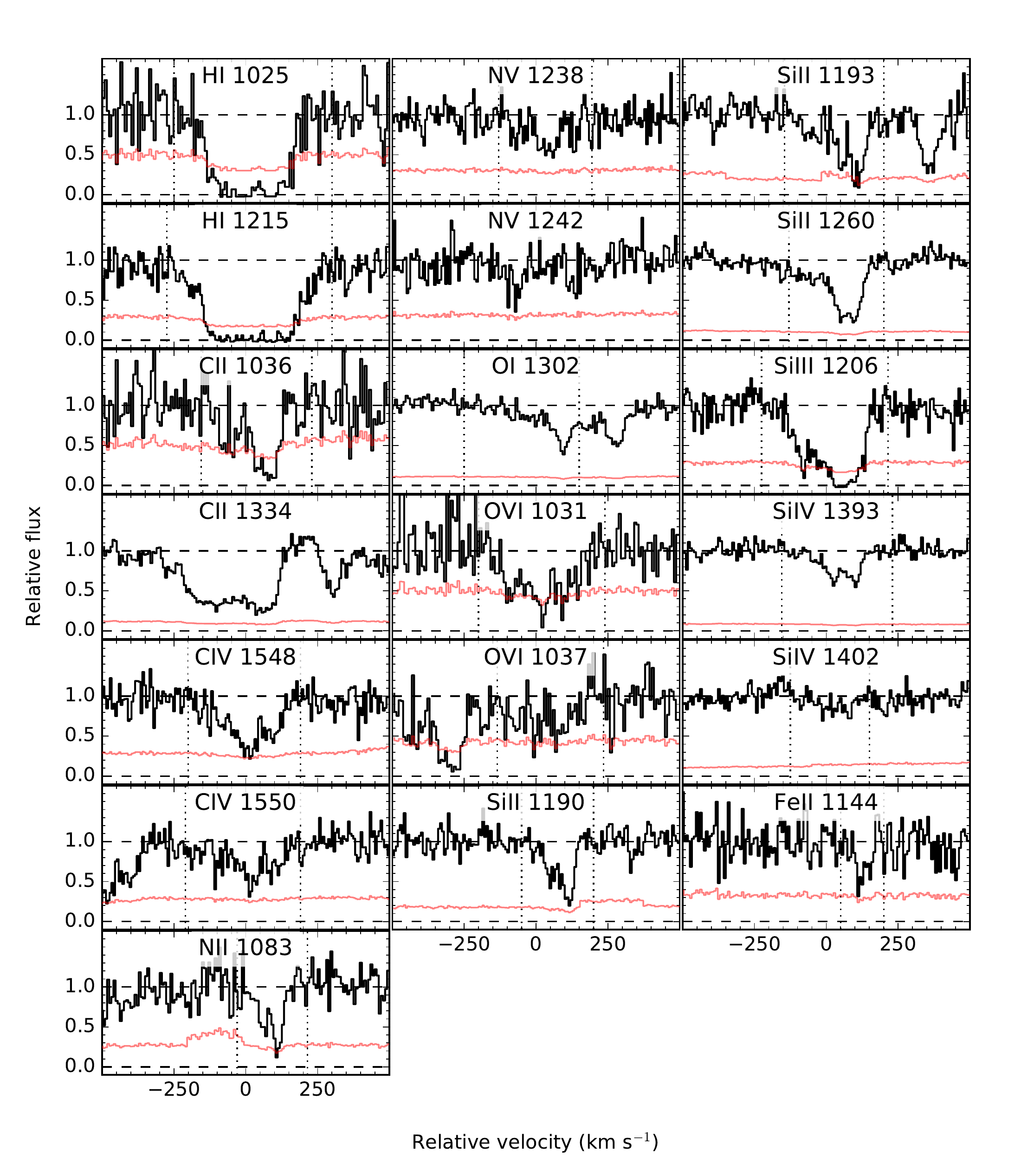}
\caption{Velocity profiles for the sightline towards J0852+0313 (z$_{\rm gal}$=0.129). The vertical dotted lines show the integration limits for determining the EW for adopted lines only. The red line shows the associated error spectrum.}
\label{fig:J0852}
\end{center}
\end{figure*}

The derived EWs and velocity profiles are given in Appendix \ref{App:EWs}. An example is given for the sightline towards J0852+0313 (the most gas-rich sightline); Table \ref{tab:J0852+0313,26_71} gives the EW and the associated data quality flags for each absorption line shown in Figure \ref{fig:J0852}. The data quality flags represent a sum of whether or not the line is adopted ($+1$), blended ($+2$), undetected ($+4$), or saturated ($+8$). Table \ref{tab:all_ews} contains a summary of adopted EWs for all the COS-AGN sightlines.

\begin{table}
\scriptsize
\begin{center}
\caption{Measured EWs for J0852+0313 (z=0.129)}
\label{tab:J0852+0313,26_71}
\begin{tabular}{lcccccc}
\hline
Ion& $\lambda$& $f$& $v_{\rm min}$& $v_{\rm max}$& EW& flag$^{\star}$\\
& [\AA{}]& & [\kms{}]& [\kms{}]& [m\AA{}]& \\
\hline
H\ion{i}& 1025.722& 7.912E-02& -250& 300& $752\pm91$& \textbf{9}\\
H\ion{i}& 1215.670& 4.164E-01& -275& 300& $1356\pm56$& \textbf{9}\\
C\ion{ii}& 1036.337& 1.231E-01& -155& 230& $<366$& \textbf{11}\\
C\ion{ii}& 1334.532& 1.278E-01& -315& 215& \nodata{}& 2\\
C\ion{iv}& 1548.195& 1.908E-01& -200& 190& $657\pm66$& \textbf{1}\\
C\ion{iv}& 1550.770& 9.522E-02& -210& 190& $474\pm71$& \textbf{1}\\
N\ion{ii}& 1083.990& 1.031E-01& -30& 215& $200\pm41$& \textbf{1}\\
N\ion{v}& 1238.821& 1.570E-01& -130& 195& $260\pm56$& \textbf{1}\\
N\ion{v}& 1242.804& 7.823E-02& 0& 100& \nodata{}& 2\\
O\ion{i}& 1302.168& 4.887E-02& -250& 150& $<264$& \textbf{3}\\
O\ion{vi}& 1031.926& 1.329E-01& -200& 240& $<451$& \textbf{11}\\
O\ion{vi}& 1037.617& 6.609E-02& -135& 235& $273\pm80$& \textbf{1}\\
Si\ion{ii}& 1190.416& 2.502E-01& -50& 200& $234\pm30$& \textbf{1}\\
Si\ion{ii}& 1193.290& 4.991E-01& -145& 200& $378\pm40$& \textbf{9}\\
Si\ion{ii}& 1260.422& 1.007E+00& -130& 200& $458\pm18$& \textbf{1}\\
Si\ion{iii}& 1206.500& 1.660E+00& -225& 215& $804\pm52$& \textbf{9}\\
Si\ion{iv}& 1393.755& 5.280E-01& -155& 230& $235\pm19$& \textbf{1}\\
Si\ion{iv}& 1402.770& 2.620E-01& -125& 150& $115\pm28$& \textbf{1}\\
Fe\ion{ii}& 1144.938& 1.060E-01& 50& 200& $87\pm39$& \textbf{1}\\
\hline
\end{tabular}
\\$^{\star}$The data quality flags represent a sum of whether or not the line is:\\ adopted ($+1$), blended ($+2$), undetected ($+4$), or saturated ($+8$).\\\end{center}
\end{table}

\subsection{Control matching}
\label{sec:ControlMatching}

 To provide a fair comparison between the CGM properties of AGN and non-AGN host galaxies, we implement a control-matching scheme that matches galaxies from our literature sample (COS-Halos, COS-GASS and COS-Dwarf galaxies) to each COS-AGN galaxy.  By adopting the literature sample as a pool for our control galaxies, we assume that these galaxies are representative of galaxies that do not host AGN. As the EW of a given species in the CGM has tentatively been shown to scale with a combination of \Rimp{} and the host's \Mstar{} \citep{Chen10,Werk14,Borthakur16}, we use \Mstar{} and \Rimp{} as our control matching parameters, thus simultaneously removing the effects from these observed scaling relations in our EW analysis.

The control matching scheme in this work uses a fixed maximum tolerated offset in both \Mstar{} and \Rimp{} between a given control sightline and the AGN sightline ($\Delta$\logMstar{} and $\Delta$\logRimp{}; where $\Delta X=X_{control} - X_{AGN}$ for the matched parameter $X$). These two offsets are combined into a single matching parameter $r^2$, which is given by the sum of the squares of $\Delta$\logMstar{} and $\Delta$\logRimp{}, i.e.
\begin{equation}
r^2 = \Delta $\logMstar$^2 + \Delta $\logRimp$^2,
\end{equation}
 such that $r$ provides a single factor of how far off each control match is from the AGN value. Adopting a tolerance of 0.2 dex in each $\Delta$\logMstar{} and $\Delta$\logRimp{} (and thus a tolerance in $r=0.28$ dex) provides matching within a factor of two of the COS-AGN sightline's \Rimp{} and \Mstar{}. Given that the COS-AGN sample tends to probe higher \Rimp{} and \logMstar{} relative to the literature sample (e.g. see Figures \ref{fig:SampDists} and \ref{fig:map}), we computed the skewness for both $\Delta$\logMstar{} and $\Delta$\logRimp{} (independently) to confirm that the adopted tolerances do not lead to a biased control sample. The skewness tests reveal that our selected $r$ tolerance is similar to a Gaussian distribution at $\geq95$\% confidence, suggesting that the initial tolerance does not select a significantly skewed sample. Thus we adopt our cut of $r=0.28$ as our control matching tolerance, which selects at least five different control sightlines for over 80\% of the COS-AGN sample. We point out that this broad tolerance allows for any of the literature galaxies to be matched to multiple COS-AGN galaxies.  As one COS-AGN absorber does not have a single control match within the adopted tolerance range (J0042-1037; z=0.036, \logMstar{}=9.5, and \Rimp{}$=299$ kpc); we do not include this sightline in the our analysis. We note that the adopted tolerance in $r$ is similar to the one required to match the scatter of the Mg\ion{ii}-\Rimp{} relation computed by \cite{Chen10}.

In addition to \Mstar{} and \Rimp{}, the comparison sample could potentially benefit from controlling additional parameters. For example, the specific star formation rate (sSFR) appears to play a key role in the EW distribution of O\ion{vi} \citep{Tumlinson11,Werk14},  however an alternative suggestion for the paucity of metals around passive galaxies is related to the halo mass rather than the sSFR of a galaxy \citep[][]{Oppenheimer16}. Figure \ref{fig:SampDists} demonstrates that many of the COS-AGN galaxies have a sSFR that is intermediate to the star-forming and passive galaxies \citep[these lower SFR for Seyfert AGN relative to star-forming galaxies have previously been seen in the SDSS;][]{Ellison16b,Leslie16}. The lack of overlap between the COS-AGN sSFR distribution and that of the control pool prevents a matching within meaningful tolerances of sSFR. However, we can simply distinguish between star-forming and passive galaxies in the control sample based on the \logsSFR{}$=-11.0$ cut adopted by COS-Halos, and draw comparisons between AGN and star-forming or passive galaxies independently. Later, we will implement this sSFR cut into our analysis.

Environment may also influence the CGM of galaxies \citep{Bordoloi11,Stocke14,Burchett16}. Environment can be quantified using a metric such as $\delta_{5}$, a measure
of the surface density of galaxies ($\Sigma_{5}$) within the distance to the
5\ts{th} nearest neighbour ($d_{5}$) relative to the average surface density
across the sky at that redshift, i.e.
\begin{equation}
\delta_{5} \equiv \frac{\Sigma_{5,gal}}{{\Sigma_{5,sky}}}=\left(\frac{d_{5,sky}}{d_{5,gal}}\right)^{2},
\end{equation}
where $\Sigma_{5,gal}$ and $d_{5,gal}$ are measured for the host galaxy (within $\pm1000$ \kms{} of the host's redshift) and  $\Sigma_{5,sky}$ and $d_{5,sky}$ are the average values measured across the sky at the host's redshift. Since we do not have a robust measurement of environment consistently quantified across our control and COS-AGN samples, we are unable to control for environment. However, we have compared the distribution of $\delta_{5}$ measured in the SDSS \citep{Baldry06} for all AGN galaxies to the distribution for star-forming galaxies with matching stellar mass and redshift distributions and find that the resulting $\delta_{5}$ distributions between these two galaxy populations are similar to within a couple of per cent at all values of \Mstar{}. The consistent distributions suggests that, on average, the star-forming and the SDSS AGN galaxies occupy similar environments, thus we do not need to match for environment in our control sample.

\section{Results}

\subsection{Kinematics}

To study the kinematics of the CGM gas, we are limited to using the saturated Ly$\alpha$ absorption as many of the typically unsaturated metal lines are frequently undetected in the COS-AGN sample (see Table \ref{tab:all_ews}; the covering fractions of metal lines will be further discussed in the following section). We first assess whether or not the CGM gas is bound to the host galaxy halo by comparing the relative velocity of the Ly$\alpha$ absorption ($v_{\rm Ly\alpha}$) to the systemic velocity of the galaxy ($v_{\rm Gal}$). As the dark matter halo mass (${\rm M_{Halo}}$) is required to assess the escape velocity of the galaxy halo, we use the \Mstar{}-${\rm M_{Halo}}$ relation provided by \cite{Moster10} to calculate the halo mass of each galaxy, namely
\begin{equation}
\frac{\rm{M_{\star}}}{\rm{M_{Halo}}} = 2\left( \frac{\rm{M_{\star}}}{\rm{M_{Halo}}} \right) _{0}\left[ \left( \frac{\rm{M_{Halo}}}{M_{1}} \right) ^{-\beta} + \left( \frac{\rm{M_{Halo}}}{M_{1}} \right) ^{\gamma} \right] ^{-1},
\end{equation}
where $M_{1}=10^{11.884} M_{\odot}$, $\left(\frac{\rm{M_{\star}}}{\rm{M_{Halo}}}\right)_{0}=0.0282$, $\beta=1.057$, and $\gamma=0.556$ \cite[i.e. the best fit parameters from table 1 in][]{Moster10}. The escape velocity of the halo (calculated from \Rimp{} of the QSO sightline) is found using
\begin{equation}
v_{\rm{Esc}}=\left[ 2\rm{\int_{\rho_{imp}}^{\infty}\frac{GM_{Halo}(<\rho)}{\rho^{2}}d\rho} \right]^{0.5},
\end{equation}
assuming the halo mass is distributed following a \cite{NFW} dark matter profile with a concentration parameter of 15.

\begin{figure*}
\begin{center}
\includegraphics[width=0.9\textwidth]{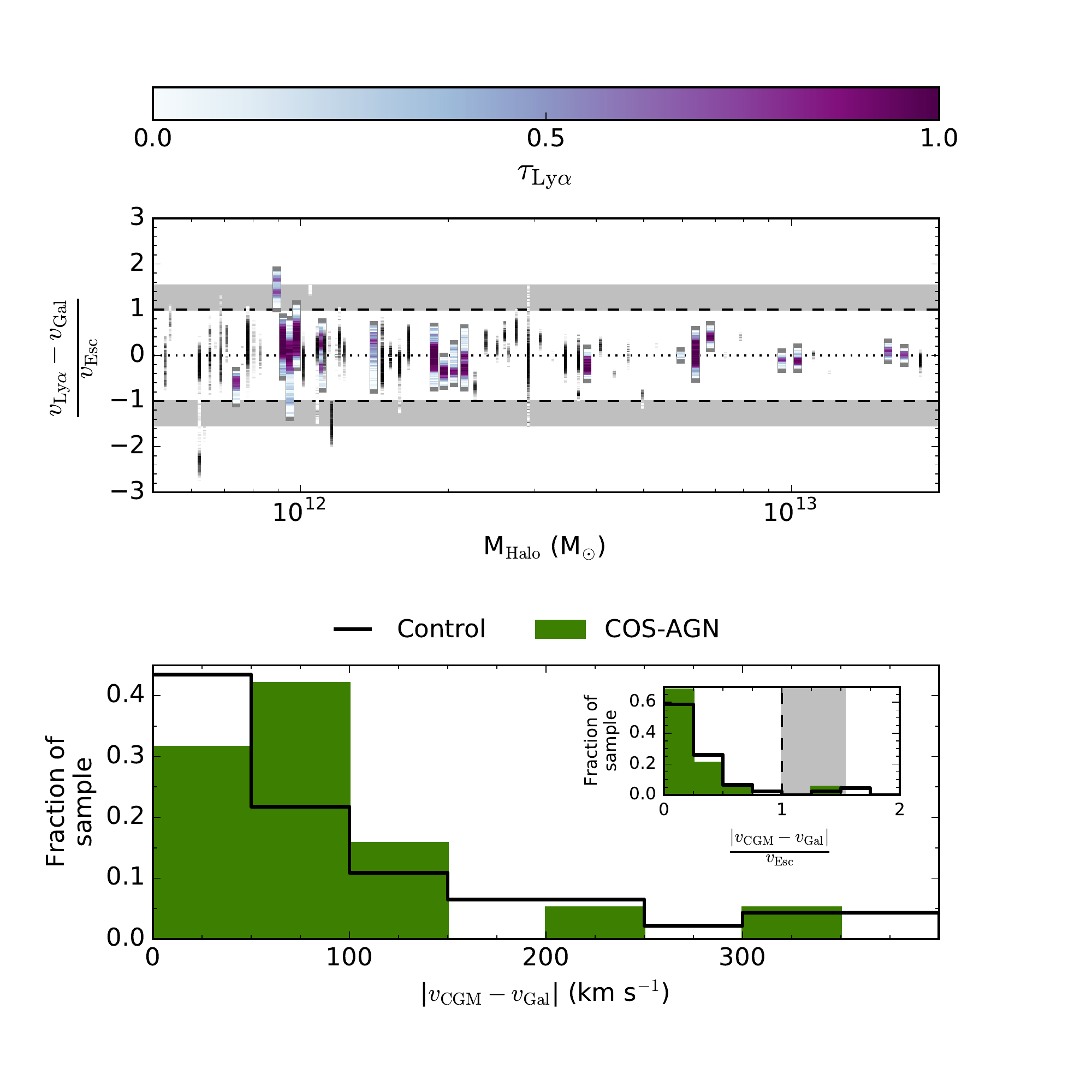}
\caption{The top panel shows the velocity span of the Ly$\alpha$ profiles (within $\pm500$ \kms{} of the host galaxy) from COS-AGN and the control sample as a function of halo mass (M$_{\rm Halo}$). The COS-AGN sightlines are denoted by the thicker coloured lines, while the control sample are shown as thin grey lines. The velocity is normalized by the escape velocity from the halo at the observed \Rimp{}. The horizontal dashed lines denotes the relative velocity for which gas at a distance of \Rimp{} from the galaxy becomes unbound.  As we do not know the precise location of the gas along the line of sight, the grey band shows where gas for the median COS-AGN halo becomes unbounded within a $\pm$\Rimp{} region along the sightline from the perpendicular. Each line is colour-coded by the optical depth of the Ly$\alpha$ profile to indicate the velocity of all the observed components. For visualization, all galaxies of the same halo mass are offset horizontally by a small amount. The bottom panel displays the distribution of the velocity centroid offsets of the Ly$\alpha$ absorption profile relative to the systemic velocity of the galaxy for the COS-AGN (green bars) and control (black line) samples. The inset panel shows the distribution of $|v_{\rm CGM} - v_{\rm Gal}|$ data normalized by the escape velocity of the halo. As in the top panel, the shaded region denotes the approximate value where the gas may become unbound from the host galaxy. A KS test suggests there is no difference between the AGN and control sample distributions.}
\label{fig:Kinematics}
\end{center}
\end{figure*}

The top panel of Figure \ref{fig:Kinematics} compares the kinematic extent of the Ly$\alpha$ profile to the escape velocity of the halo for both the COS-AGN (thick bars) and control samples (thin lines). The horizontal lines denote the escape velocity of the halo, such that any gas located beyond these lines is likely not bound to the host galaxy. The colour coding of the vertical bars in the top panel of Figure \ref{fig:Kinematics} represents the optical depth of the velocity profile of the Ly$\alpha$ line, such that dark colours show the strongest components of the line. The bulk of the gas in the COS-AGN is within the escape velocity of hosts' haloes, and is likely bound to the host galaxy\footnote{We note that the gas still appears bound to the galaxy when only matter interior to \Rimp{} is considered in the calculation of $v_{\rm{Esc}}$.}. We note that the gas probed by the control sightlines also appears bound \citep[][]{Werk13,Borthakur16}

The only AGN host that shows gas likely moving at speeds greater than the escape velocity is J1117+2634 (absorber at z$=0.029$; see Figure \ref{fig:J1117,351}). We note that in this system the $\pm500$ \kms{} absorption search window for Ly$\alpha$ contains two absorption features: Galactic S\ion{ii} $\lambda$ 1250~\AA{} absorption at $\approx-100$ \kms{}, and the unidentified absorption assumed to be Ly$\alpha$ between 250 \kms{} and 400 \kms{} (see Appendix \ref{sec:J1117}). Without additional detected metal lines to attempt to confirm the absorption at 250--400 \kms{}, it is possible the proposed Ly$\alpha$ absorption towards J1117+2634 at z$=0.029$ is not actually associated with the AGN host.

To compare the bulk motion of the CGM surrounding AGN hosts to their control match sample, we compute the velocity centroid of the Ly$\alpha$ absorption profiles ($v_{\rm CGM}$) to look for any kinematic differences between the AGN and non-AGN populations. The bottom panel of Figure \ref{fig:Kinematics} shows the distributions of kinematic offset between the CGM gas and the systemic redshift of the host ($|v_{\rm CGM} - v_{\rm Gal}|$) for the COS-AGN (green shaded region) and control samples (black line). The inset panel shows the distribution of the same $|v_{\rm CGM} - v_{\rm Gal}|$ data, but normalized by $v_{\rm{Esc}}$. Note that the majority of the COS-AGN sightlines probe gas within 100 \kms{} of their host galaxy that is likely bound to the host galaxies. A Kolmogorov-Smirnov (KS) test rejects the null hypothesis that the two distributions of $|v_{\rm CGM} - v_{\rm Gal}|$ are similar at 75\% confidence (94\% for the $v_{\rm{Esc}}$-normalized distribution of the inset panel), suggesting there is likely no difference in the bulk motions of the AGN gas compared to their control-matched counterparts. Therefore the kinematics of the gas traced by Ly$\alpha$ around AGN hosts suggests the material is likely bound and has bulk motion properties similar to the CGM surrounding non-AGN hosts in the control sample.

\subsection{EW analysis}
Since many of the detected absorption lines are potentially saturated, our analysis is limited to using EWs, rather than column densities.  In this sub-section, we present a number of complementary analyses to investigate both the strength and frequency of absorption features in the COS-AGN sample, relative to the matched control sightlines.

\subsubsection{Covering Fractions}
\begin{figure*}
\begin{center}
\includegraphics[width=\textwidth]{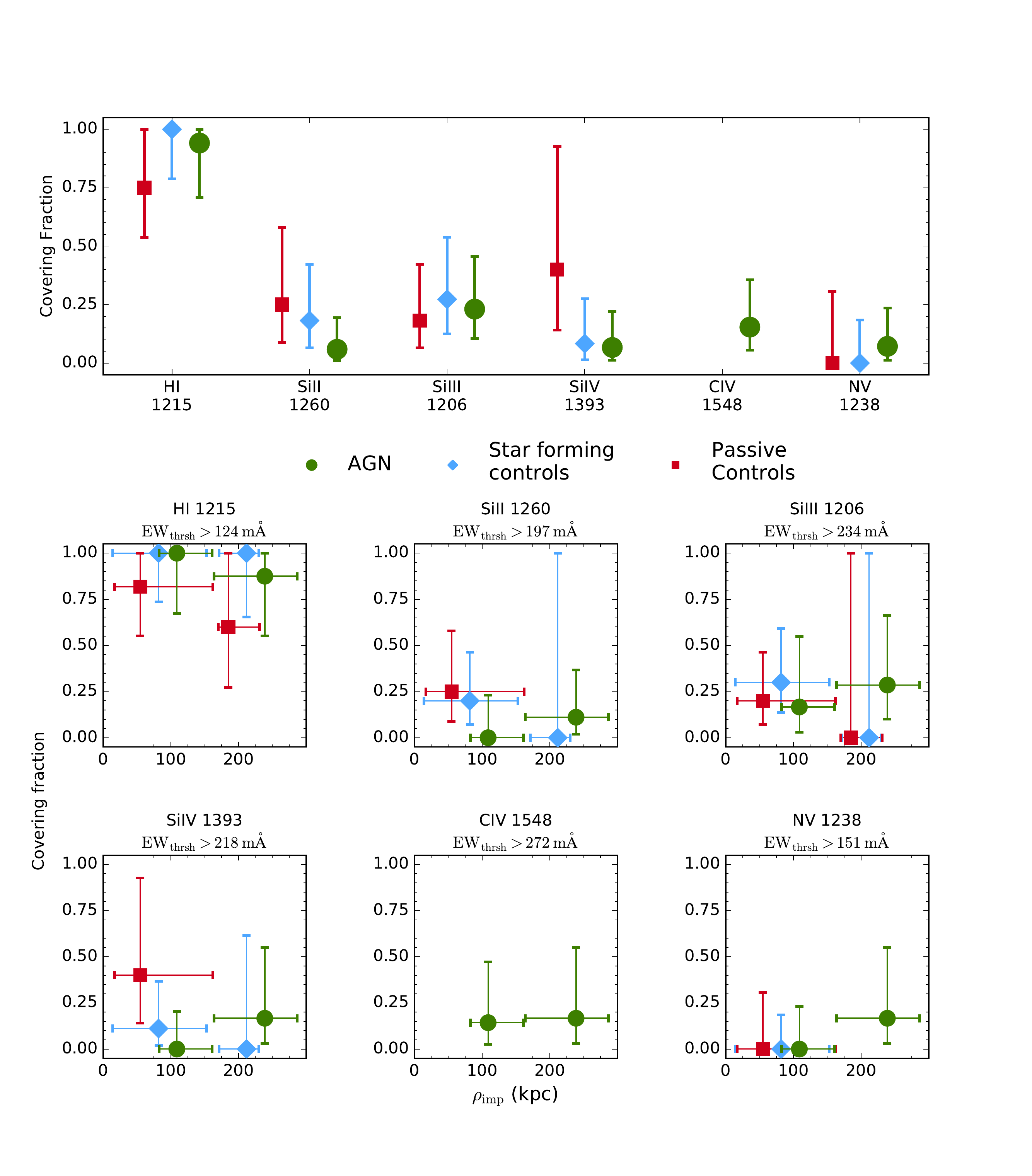}
\caption{Covering fractions (with 1$\sigma$ errors) of gas within $\pm500$ \kms{} of the host galaxy for a variety of species measured in COS-AGN and control-matched samples. Top panel: The global covering fractions measured across all impact parameters from Table \ref{tab:covfracs} are shown for the COS-AGN galaxies (green circles), and the control-matched passive (red squares) and star-forming (blue diamonds) galaxies. Bottom panels: The covering fractions of an individual species as a function of impact parameter (\Rimp{}), split into two bins by the median \Rimp{} of the COS-AGN sample (164 kpc). The EW thresholds (EW$_{\rm thrsh}$) used to determine the covering fraction for each species (including the measurements presented in the top panel) are given above the corresponding species panel in the bottom two rows. The horizontal error bars represent the entire range of \Rimp{} probed by each sample within the respective bin. The three points are offset from the centre of each bin by a small amount for clarity. No points are shown for a species that do not have spectral coverage of the corresponding absorption line. }
\label{fig:CovFracR}
\end{center}
\end{figure*}

Covering fractions are determined as the number of CGM sightlines  that have a detection of a given species greater than some EW threshold (EW$_{\rm thrsh}$) relative to the total number of sightlines in a given sample with adopted EW measurements presented in Table \ref{tab:all_ews}. Similar to \cite{Werk13}, the EW$_{\rm thrsh}$ are selected based on the lowest S/N near the species of interest such that all $3\sigma$ non-detections from COS-AGN are below this threshold. As the S/N of some of our spectra is lower than for COS-Halos, our EW$_{\rm thrsh}$ are different than those used in \cite{Werk13}. Table \ref{tab:covfracs} lists the covering fractions calculated for the entire COS-AGN and control-matched sample (split by sSFR into star-forming [SF] and passive [P] controls; blue diamonds and red squares respectively in Figure \ref{fig:CovFracR}) using our EW$_{\rm thrsh}$ as well as the threshold adopted by COS-Halos for reference (note that when using the COS-Halos thresholds, upper limits above the threshold are removed from the covering fraction calculations). We elect to use our EW$_{\rm thrsh}$ for consistency with our data quality. Due to the low number of sightlines, we use the 1$\sigma$ binomial confidence intervals in the Poisson regime \citep[tabulated in][]{Gehrels86} for our covering fraction errors. These covering fractions using our EW$_{\rm thrsh}$ are plotted in the top panel of Figure \ref{fig:CovFracR} for a variety of species, spanning a range of ionization states.

\begin{table}
\begin{center}
\caption{Covering Fractions}
\label{tab:covfracs}
\begin{tabular}{lccccc}
\hline
Ion& $\lambda$& EW$_{\rm thrsh}$& \multicolumn{3}{c}{Covering Fraction}\\
& [\AA]& [m\AA]& AGN& SF controls& P controls\\
\hline
\HI{}& 1215& 124& 0.94 $_{-0.23}^{+0.06}$& 1.00 $_{-0.21}^{+0.00}$& 0.75 $_{-0.21}^{+0.25}$\\
& & 200$^{\star}$& 0.88 $_{-0.23}^{+0.12}$& 0.86 $_{-0.20}^{+0.14}$& 0.69 $_{-0.20}^{+0.28}$\\
\hline
Si\ion{ii}& 1260& 197& 0.06 $_{-0.05}^{+0.14}$& 0.18 $_{-0.12}^{+0.24}$& 0.25 $_{-0.16}^{+0.33}$\\
& & 150$^{\star}$& 0.12 $_{-0.08}^{+0.16}$& 0.27 $_{-0.15}^{+0.27}$& 0.25 $_{-0.16}^{+0.33}$\\
\hline
Si\ion{iii}& 1206& 234& 0.23 $_{-0.13}^{+0.22}$& 0.27 $_{-0.15}^{+0.27}$& 0.18 $_{-0.12}^{+0.24}$\\
& & 100$^{\star}$& 0.31 $_{-0.15}^{+0.24}$& 0.55 $_{-0.22}^{+0.33}$& 0.36 $_{-0.17}^{+0.29}$\\
\hline
Si\ion{iv}& 1393& 218& 0.07 $_{-0.06}^{+0.15}$& 0.08 $_{-0.07}^{+0.19}$& 0.40 $_{-0.26}^{+0.53}$\\
& & 100$^{\star}$& 0.13 $_{-0.09}^{+0.18}$& 0.33 $_{-0.16}^{+0.26}$& 0.40 $_{-0.26}^{+0.53}$\\
\hline
C\ion{iv}& 1548& 272& 0.15 $_{-0.10}^{+0.20}$& \nodata{}& \nodata{}\\
\hline
N\ion{v}& 1238& 151& 0.07 $_{-0.06}^{+0.16}$& 0.00 $_{-0.00}^{+0.18}$& 0.00 $_{-0.00}^{+0.31}$\\
\hline
\end{tabular}
\\$^{\star}$ COS-Halos EW threshold.\\\end{center}
\end{table}

The bottom six panels of Figure \ref{fig:CovFracR} present the covering fractions in bins of \Rimp{} (split by the median \Rimp{} of the COS-AGN sample; 164 kpc). The covering fractions of the COS-AGN sample are the green circles. For reference, the covering fractions of the literature galaxies that were matched to the COS-AGN galaxies are shown as blue diamonds (for star-forming controls) and red squares (passive controls).

The error bars on all of the data points overlap, indicating no significant difference between the AGN sightlines and the controls for any of the species.  Nonetheless, we note three tentative differences. The first is that the \HI{} covering fraction of the COS-AGN sightlines ($94^{+6}_{-23}$\%) is most similar to the covering fraction of the star-forming control galaxies ($100^{+0}_{-21}$\%), and higher that the control-matched passive galaxies ($75^{+25}_{-21}$\%). Secondly, the covering fraction for Si\ion{iii} 1206~\AA{} is $29^{+38}_{-18}$\% for the COS-AGN sample at high impact parameters, which is larger than the covering fraction of the control samples at the same distance (0\%; $164\leq$\Rimp{}$<300$~kpc). Lastly, the inner 164 kpc bin shows no detections of Si\ion{ii} and Si\ion{iv}  in the COS-AGN sample while the control sample has non-zero covering fractions at those impact parameters ($\approx25$\%, and 10--40\%; respectively).

Although the covering fraction of N\ion{v} in the COS-AGN sample is $7^{+16}_{-6}$\% for the entire range of \Rimp{}, the sightline towards J0852+0313 (see Figure \ref{fig:J0852}) represents the only detection of N\ion{v} in COS-AGN (1/20) and our controlled matched galaxies from COS-Halos (0/16 sightlines) and COS-GASS (0/30; S. Borthakur, private communication). We note however that COS-Halos detected N\ion{v} in four of their 44 sightlines.

The $94^{+6}_{-23}$\% covering fraction measured for Ly$\alpha$ in the COS-AGN galaxies may initially appear to be in tension with the conclusions of \cite{Kacprzak15}, who used Mg\ion{ii} as a neutral gas tracer and found much smaller 10\% covering fraction (EW$>300$m\AA{}) between 100 and 200 kpc of the AGN. The Mg\ion{ii} EW threshold adopted by \cite{Kacprzak15} is typical of that used to select strong \HI{} absorbers at low redshifts \citep[log(N(\HI{})/cm$^{-2}$)$\gtrsim18.5$;][]{Rao06}. Translating this Mg\ion{ii} EW threshold into a corresponding Ly$\alpha$ EW theshold for these column densities of gas yields a much higher than the threshold used in this work (EW $\gtrsim 1300$ m\AA{} --- assuming a minimum broadening parameter of 5 \kms{} ---  compared to EW $>124$ m\AA{} for COS-AGN). Using this larger threshold, only one of the COS-AGN sightlines has an EW $\gtrsim 1300$ m\AA{}, giving a consistent result with the observations from \cite{Kacprzak15}.

\subsubsection{Relative EW analysis}

\begin{figure*}
\begin{center}
\includegraphics[width=\textwidth]{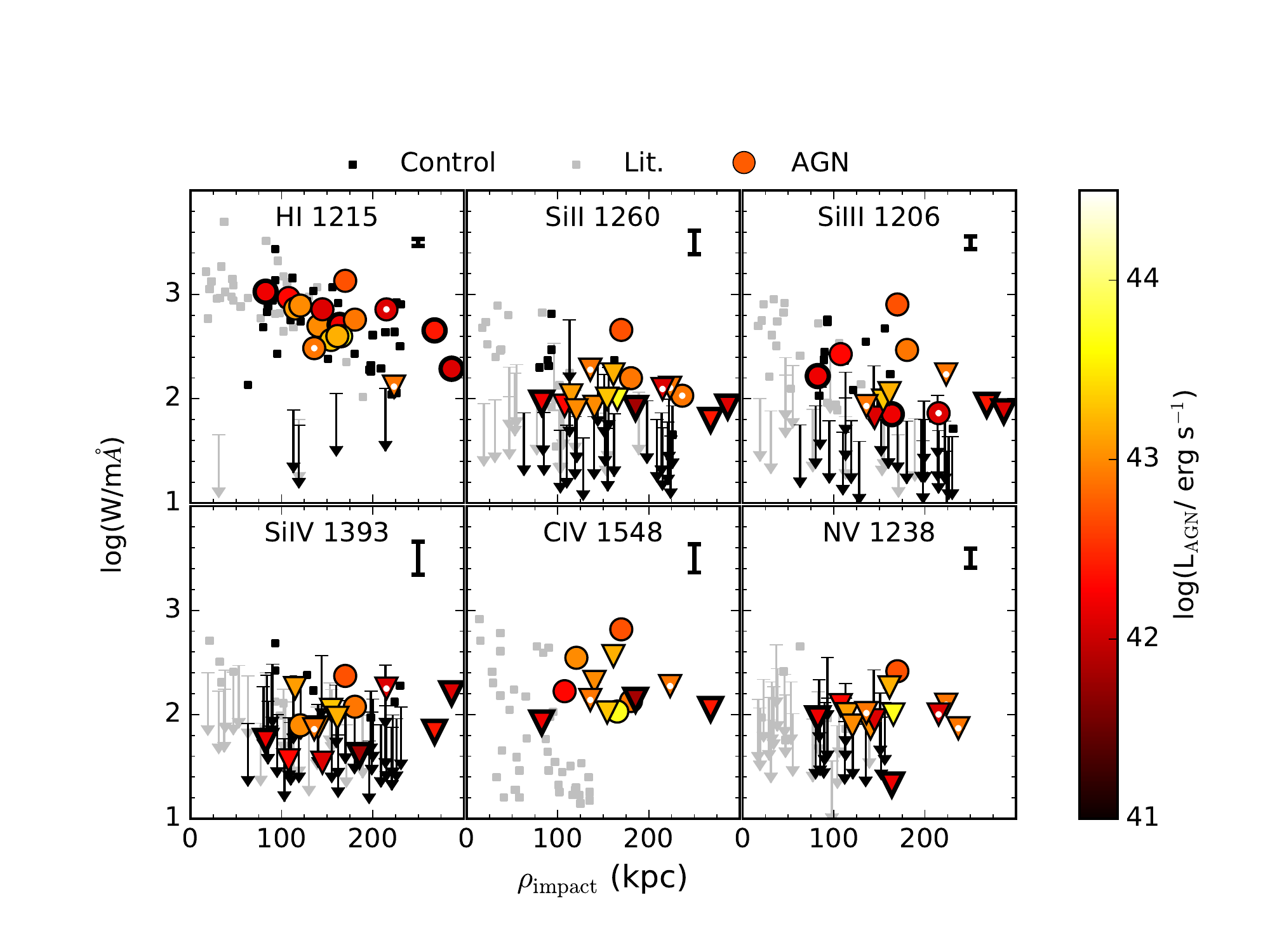}
\caption{The rest-frame equivalent widths (within $\pm500$ \kms{} of the host galaxy) as a function of impact parameter. The coloured circles (EW detections) and triangles (EW upper limits) represent the COS-AGN sample, and are colour coded by the bolometric luminosity of the AGN (\Lbol{}). The median error bar on the COS-AGN EW measurements is given by the black error bar in the top right region of each panel. The black squares show the EW values of the control-matched galaxies, while the grey squares denote the rest of the literature comparison sample. Data outlined with a thick black line represent COS-AGN systems with nearby galaxies. COS-AGN galaxies flagged as LINERs are indicated by small white dots on top of the respective  data points.}
\label{fig:EWRho}
\end{center}
\end{figure*}

Figure \ref{fig:EWRho} shows the raw EW values for a variety of ionic species as a function of \Rimp{}. The COS-AGN points are colour-coded by their \Lbol{}. The control-matched sample is shown as black points, while the grey points are the remaining un-matched literature sightlines. The bold COS-AGN points are CGM sightlines that have spectroscopic companions (see Section \ref{sec:SampProps}). The top left panel demonstrates that the Ly$\alpha$ EWs for the COS-AGN follow the general trend of decreasing EW as a function of \Rimp{} seen in previous low redshift studies \citep[][]{Chen10, Werk14,Borthakur16}. For metal species, the lack of detections in both the COS-AGN and control sample makes a comparative analysis difficult.  For the remainder of this section, we focus only on the Ly$\alpha$ EWs.

\begin{figure*}
\begin{center}
\includegraphics[width=\textwidth]{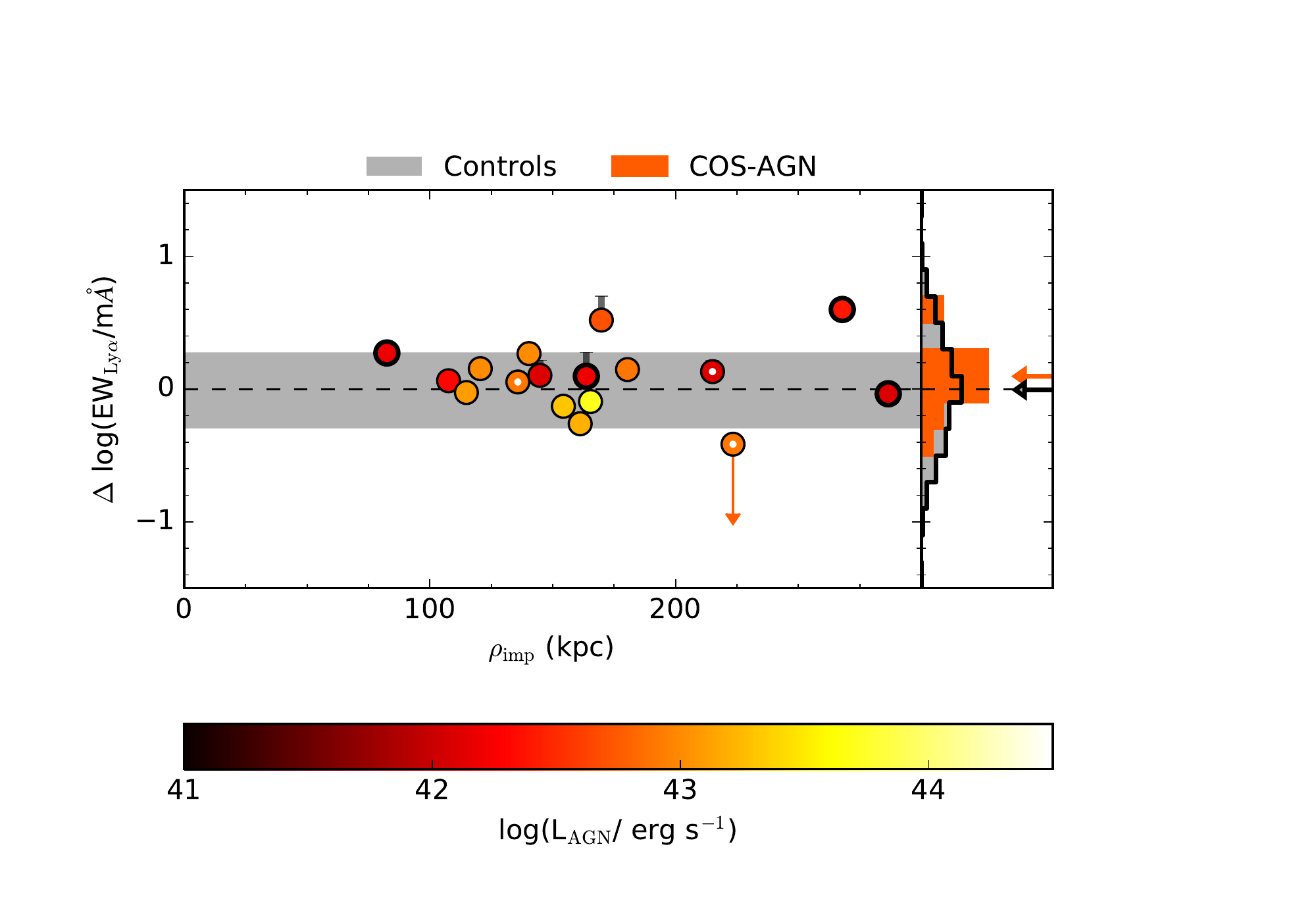}
\caption{The difference in the Ly$\alpha$ equivalent width of the COS-AGN sightlines relative to their control matched counterparts  ($\Delta$log(EW/m\AA{})) as a function of impact parameter (\Rimp{}). The EW measure material within $\pm500$ \kms{} of the host galaxy. The points are colour-coded by the bolometric luminosity of the AGN (\Lbol{}). The errorbars denote how the maximal shift on including control matched EW upper limits in the calculation of $\Delta$log(EW/m\AA{}).  For reference, the horizontal grey band represents the interquartile range of $\Delta$EW of the literature sample control matched with itself. The normalized distributions of $\Delta$EW for both the COS-AGN and control galaxies are shown in the right panel, with the median of each histogram given by an arrow. Data outlined by a thick black line are COS-AGN sightlines flagged as having nearby galaxies. COS-AGN galaxies flagged as LINERs are indicated by small white dots on top of the respective  data points.}
\label{fig:deltaEWRho}
\end{center}
\end{figure*}

In order to quantify any difference between the COS-AGN and control samples, we calculate $\Delta$log(EW/m\AA), which is defined as ${\rm \Delta log(EW/m\AA) = log[EW_{AGN}/median(EW_{Controls})]}$, such that a positive \deltaEW{} would imply that the CGM surrounding the AGN has a larger EW than the median of its control-matched galaxies. The left panel of Figure \ref{fig:deltaEWRho} shows $\Delta$log(EW) for Ly$\alpha$ as a function of \Rimp{} for the COS-AGN galaxies. For reference, the grey band represents the interquartile range of $\Delta$log(EW) for the entire literature sample matched to itself. The right panel shows the distributions of $\Delta$log(EW) for the COS-AGN (orange) and literature (grey) galaxies, with medians of the distributions indicated by the arrows. 

To include the non-detections of the controls in the analysis, we calculate the median EW of the controls twice: once including limits as if they were detections, and once setting the non-detected EWs to 0~m\AA{}. These median EWs span the range of true median EW if the absorption lines were actually detected. For this calculation, we only include non-detections when the upper limits are more sensitive (i.e. smaller) than the largest detected EW as these limits  are constraining enough to affect the median value. The corresponding $\Delta$log(EW) range is shown on Figure \ref{fig:deltaEWRho} as the thick grey errorbars. The 1$\sigma$ jackknife errors on $\Delta$log(EW) are typically smaller than the size of the points.

The median $\Delta$log(EW) of the COS-AGN sample is enhanced by $+0.10\pm0.13$  dex relative to the controls. Repeating this control-matching experiment for the literature sample yields a median $\Delta$log(EW) of $0.00\pm0.28$. Note that the errors on these median $\Delta$log(EW) represent the median absolute deviation (MAD) of the distribution. A KS test rejects the null hypothesis that the distributions of $\Delta$log(EW) for the COS-AGN and control samples are same at 20\% confidence. When the LINER galaxies are removed from the COS-AGN sample, the median $\Delta$log(EW) changes to $+0.10\pm0.15$, and the KS test yields a rejection of the null hypothesis at 14\% confidence.

To check if there is any effect from splitting the control sample into star-forming and passive galaxies, we measured the $\Delta$log(EW) for all star-forming galaxies in the non-AGN literature sample to their control-matched passive counterparts. The median $\Delta$log(EW) obtained for star-forming galaxies relative to the matched passive controls is $+0.03\pm0.26$, while a KS-test reveals  the null hypothesis of the $\Delta$log(EW) distributions for the star-forming and passive galaxies are the same is rejected at 38\% confidence. Therefore there is no significant difference in the offset of the AGN hosts from the controls that would be caused by the star formation rate.

\subsubsection{Stacked Spectra}

The results in the previous sub-section indicate a possible (but not significant) difference in the Ly$\alpha$ absorption properties of the COS-AGN sample, and possibly in some of the metal species as well.  However, that analysis is limited by the modest S/N of the data, the small sample size and lack of detections of metal species.  Therefore, in this section we consolidate the data by stacking all of the COS-AGN spectra and comparing it to a stack of the control sample. In brief, all spectra are shifted to the rest-frame (using the redshift of the strongest component of the \HI{} absorption profile) and rebinned to a linear dispersion of 0.064 \AA{} pixel$^{-1}$ (similar to the resolution of the COS-AGN spectra). These rebinned spectra are mean combined without any weighting. We note that either using the systemic redshifts of the galaxies or using different weighting schemes does not significantly change the results.

Table \ref{tab:EWStack} gives the measured EWs of the various absorption lines of interest from the final stacked COS-AGN spectrum. We require that the absorption line be detected at $>3\sigma$, otherwise the EW is set to a $3\sigma$ upper limit.  The stacked EW errors are calculated using a standard jackknife approach by removing each COS-AGN from the stacked spectrum and recalculating the EW. The numbers in brackets in each column give the EW offsets from removing the `strongest' (${jack,min}$) and `weakest' (${jack,max}$) absorber sightline from the stack in a jackknife fashion (i.e. these are the maximal variations in the EW from the jackknife), as well as the number of sightlines that contributed to the stack (N$_{spec}$). The four columns on the right are the measured EWs when the stacking process is only applied to COS-AGN sightlines split into two bins of log(\Lbol{}) and \Rimp{} at the median value of the COS-AGN sample (log(\Lbol{}/erg s$^{-1}$)$=42.9$ and \Rimp{}$=164$ kpc; respectively).

\begin{table*}
\begin{center}
\caption{Measured EW of stacked spectra}
\label{tab:EWStack}
\begin{tabular}{lc|c|cc|cc|}
\hline
Ion & Line & \multicolumn{5}{c}{Stacked EW ($_{jack,min}^{jack,max}$; N$_{spec}$) [m\AA{}]} \\
 & [\AA] & All sight-lines&	log(\Lbol{}/erg s$^{-1}$)$\leq$42.9 & log(\Lbol{}/erg s$^{-1}$)$>$42.9& \Rimp{}$\leq$164 kpc& \Rimp$>$164 kpc \\
\hline
\HI{}	 & 1215	 & 739$\pm$25 ($_{-69}^{+44}$; 14)	 & 937$\pm$53 ($_{-116}^{+77}$; 7)	 & 539$\pm$38 ($_{-66}^{+70}$; 7)	 & 799$\pm$41 ($_{-68}^{+73}$; 7)	 & 680$\pm$66 ($_{-176}^{+87}$; 7)	\\
C\ion{ii}	 & 1036	 & $<$80 (2)	 & $<$80 (2)	 & \nodata & \nodata & $<$80 (2)	\\
C\ion{ii}	 &	1334& $<$37 (11)	 & $<$53 (6)	 & $<$50 (5)	 & $<$54 (5)	 & $<$51 (6)	\\
C\ion{iv}	 & 1548	 & 194$\pm$13 ($_{-40}^{+20}$; 11)	 & 236$\pm$38 ($_{-89}^{+61}$; 5)	 & $<$194 (6)	 & 185$\pm$17 ($_{-38}^{+29}$; 6)	 & 205$\pm$39 ($_{-96}^{+53}$; 5)	\\
C\ion{iv}	 & 1550	 & 131$\pm$9 ($_{-24}^{+15}$; 13)	 & 132$\pm$21 ($_{-56}^{+21}$; 6)	 & $<$167 (7)	 & $<$132 (7)	 & 159$\pm$24 ($_{-52}^{+33}$; 6)	\\
N\ion{v}	 & 1238	 & $<$32 (13)	 & $<$42 (6)	 & $<$47 (7)	 & $<$47 (7)	 & $<$43 (6)	\\
N\ion{v}	 & 1242	 & $<$38 (10)	 & $<$46 (5)	 & $<$60 (5)	 & $<$51 (6)	 & $<$54 (4)	\\
O\ion{i}	 &	1302& $<$31 (12)	 & $<$45 (5)	 & $<$42 (7)	 & $<$47 (6)	 & $<$39 (6)	\\
O\ion{vi}	 &	1037& $<$379 (1)	 & $<$379 (1)	 & \nodata & \nodata & $<$379 (1)	\\
SiII	 & 1190	 & $<$29 (10)	 & $<$36 (5)	 & $<$46 (5)	 & $<$61 (3)	 & $<$32 (7)	\\
Si\ion{ii}	 & 1190	 & $<$32 (9)	 & $<$44 (4)	 & $<$46 (5)	 & $<$61 (3)	 & $<$37 (6)	\\
Si\ion{ii}	 & 1260	 & 88$\pm$8 ($_{-28}^{+11}$; 15)	 & 94$\pm$28 ($_{-75}^{+30}$; 6)	 & 84$\pm$10 ($_{-17}^{+15}$; 9)	 & $<$101 (6)	 & 103$\pm$18 ($_{-47}^{+21}$; 9)	\\
Si\ion{iii}	 & 1206	 & 152$\pm$16 ($_{-62}^{+18}$; 12)	 & 211$\pm$37 ($_{-98}^{+33}$; 8)	 & $<$120 (4)	 & $<$98 (6)	 & 238$\pm$40 ($_{-112}^{+32}$; 6)	\\
Si\ion{iv}	 & 1393	 & $<$44 (11)	 & $<$69 (4)	 & $<$57 (7)	 & $<$59 (6)	 & $<$66 (5)	\\
Fe\ion{ii}	 & 1144	 & $<$29 (12)	 & $<$34 (7)	 & $<$52 (5)	 & $<$49 (5)	 & $<$35 (7)	\\
Fe\ion{ii}	 & 1608& $<$102 (8)	 & $<$94 (2)	 & $<$134 (6)	 & $<$129 (5)	 & $<$168 (3)	\\
\hline

\end{tabular}
\end{center}
\end{table*}

\begin{table*}
\begin{center}
\caption{Measured EW of stacked control spectra}
\label{tab:EWStackControl}
\begin{tabular}{lc|c|cc|cc|}
\hline
Ion & Line & \multicolumn{5}{c}{Stacked EW ($_{jack,min}^{jack,max}$; N$_{spec}$) [m\AA{}]} \\
 & [\AA] & All sight-lines&	 log(sSFR/yr$^{-1}$)$<-11$ & log(sSFR/yr$^{-1}$)$\geq-11$ & \Rimp{}$\leq$164 kpc& \Rimp$>$164 kpc \\
\hline
\HI{}	 & 1215	 & 577$\pm$12 ($_{-39}^{+24}$; 43)	 & 485$\pm$30 ($_{-84}^{+52}$; 19)	 & 649$\pm$21 ($_{-65}^{+40}$; 24)	 & 781$\pm$20 ($_{-50}^{+50}$; 25)	 & 239$\pm$15 ($_{-37}^{+30}$; 18)	\\
C\ion{ii}	 & 1334	 & 36$\pm$3 ($_{-11}^{+6}$; 43)	 & $<$29 (19)	 & 46$\pm$5 ($_{-20}^{+9}$; 24)	 & 62$\pm$5 ($_{-18}^{+10}$; 25)	 & $<$31 (18)	\\
Si\ion{ii}	 & 1190	 & $<$22 (15)	 & $<$39 (5)	 & $<$26 (10)	 & $<$22 (15)	 & \nodata \\
Si\ion{ii}	 & 1260	 & 56$\pm$4 ($_{-14}^{+7}$; 33)	 & 85$\pm$10 ($_{-28}^{+17}$; 16)	 & $<$42 (17)	 & 103$\pm$6 ($_{-20}^{+8}$; 23)	 & $<$55 (10)	\\
Si\ion{iii}	 & 1206	 & 98$\pm$4 ($_{-15}^{+9}$; 39)	 & 77$\pm$9 ($_{-28}^{+14}$; 21)	 & 126$\pm$8 ($_{-24}^{+15}$; 18)	 & 156$\pm$7 ($_{-20}^{+14}$; 25)	 & $<$47 (14)	\\
Si\ion{iv}	 & 1393	 & $<$35 (37)	 & $<$44 (17)	 & $<$55 (20)	 & 58$\pm$5 ($_{-15}^{+14}$; 23)	 & $<$65 (14)	\\
Si\ion{iv}	 & 1402	 & $<$27 (37)	 & $<$34 (17)	 & $<$41 (20)	 & $<$33 (21)	 & $<$44 (16)	\\

\hline
\end{tabular}
\end{center}
\end{table*}

A similar procedure was completed for all the spectra in the literature sample that were used in the control sample. Each control sightline in the stacked spectrum was weighted by the number of times the sightline was matched to a unique AGN host (such that the most frequently matched control sightline was given a higher weighting). Although there is very little difference in the measured EW without such a weighting, we elect to use this weighting scheme such that the derived jackknife errors represent the true range in EWs when a given sightline is excluded. The EWs measured from the stacked spectrum of the control galaxies is given in Table \ref{tab:EWStackControl}. We repeat the stacked EW calculations of splitting the controls by the median \Rimp{} of the COS-AGN sample and log(sSFR/yr$^{-1}$)$=-11$ (i.e. whether the controls are star-forming or passive; see Table \ref{tab:EWStackControl}). Note that the COS-GASS results \citep{Borthakur15,Borthakur16} focus on a smaller subset of species (\HI{}, C\ion{ii}, Si\ion{ii}, Si\ion{iii}, and Si\ion{iv}), thus we are only able to provide a stacked spectrum for these species. We point out that the Si\ion{iv} EW of the control stacked spectrum is poorly constrained due to an uncertain continuum in some of the literature sample spectra.

\begin{figure*}
\begin{center}
\includegraphics[width=\textwidth]{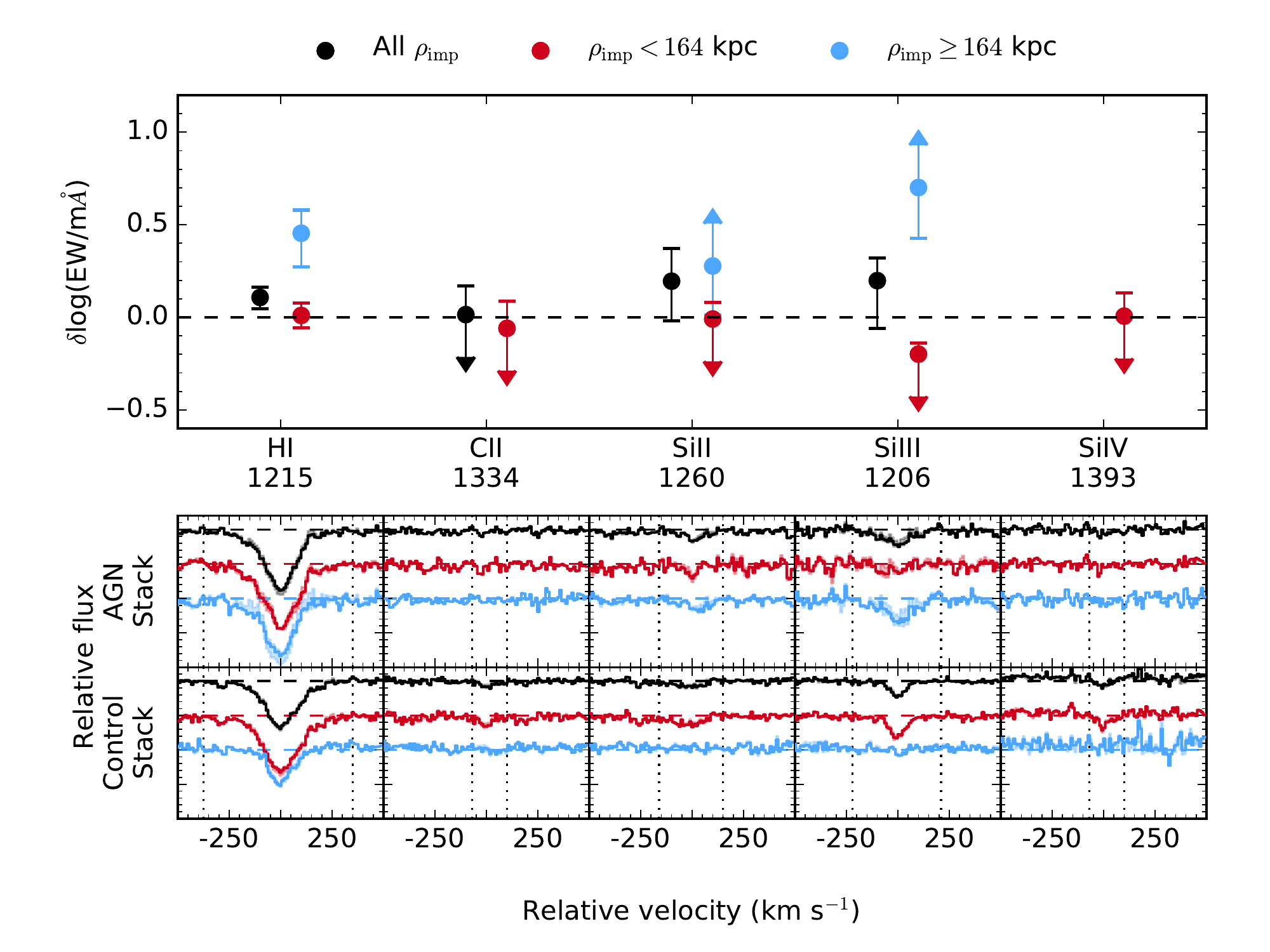}
\caption{$\delta$log(EW) measured from the stacked spectra for a variety of detected species. Top panel: The $\delta$log(EW) is shown for the COS-AGN sightlines relative to: all controls (black circles), AGN and controls for small (\Rimp$<164$ kpc; red circles) and large (\Rimp{}$\geq164$ kpc; blue circles) impact parameters. Errorbars on all points represent the possible range in $\delta$log(EW) spanned by the maximal jackknife errors (i.e.~$_{-jack,min}^{+jack,max}$). Upper and lower limits are plotted when an absorption line is detected in either the COS-AGN or control stack (respectively), but not the other. We note that the lower errorbar on the $>164$ kpc $\delta$log(EW) for Si\ion{iii} 1206~\AA{} line represents how shallow the lower limit becomes upon including the jackknife errors. The bottom ten panels show the absorption profiles of the stacked spectra (the COS-AGN spectrum in the middle row; the control spectrum in the bottom row) for each element. The colour coding of which sightlines are included is the same as in the top panel. The absorption lines are offset vertically by 0.5  in relative flux for clarity.}
\label{fig:StackdEW}
\end{center}
\end{figure*}

We repeat a similar differential EW analysis as above, where we calculate $\delta$log(EW)$={\rm log(EW_{AGN}) - log(EW_{Control})}$ using the EWs derived from the respective stacked spectra (Tables \ref{tab:EWStack} and \ref{tab:EWStackControl}). The top panel of Figure \ref{fig:StackdEW} provides a comparison of EW between all the AGN sightlines and all controls (black points). The errorbars represent the combination of the maximal jackknife errors, providing the entire range of possible $\delta$log(EW) from removing a single sightline from each sample. The stacking confirms the results from Figure \ref{fig:deltaEWRho}, where the COS-AGN sightlines have an enhanced Ly$\alpha$ relative to the controls, with an enhancement of $\delta$log(EW)$\approx0.17$ dex. However, there is a negligible difference in $\delta$log(EW) for the metal species detected at all impact parameters. Removing the absorption surrounding the LINER galaxies does not change the qualitative picture presented in Figure \ref{fig:deltaEWRho}; the measured $\delta$EW from the LINER-free stacks shift $\leq0.08$ dex (in either direction). 

The additional red and blue points in the top panel of Figure \ref{fig:StackdEW} show the values of $\delta$log(EW) calculated only when including only sightlines with \Rimp{} smaller or larger than the median value of the COS-AGN sample (164 kpc; respectively). Splitting the stacking into the `inner' and `outer' CGM around AGN hosts uncovers that the enhancement seen in the Ly$\alpha$ $\delta$log(EW) is driven by the the COS-AGN sightlines that probe \Rimp{}$\geq164$ kpc ($\delta$log(EW)$=+0.45\pm0.05$~dex; using standard jackknife EW errors). This enhancement in the outer CGM gas is also seen by the detections of the cool gas tracers Si\ion{ii} ($\delta$log(EW)$>0.27$~dex) and Si\ion{iii} ($\delta$log(EW)$>0.75$~dex) in COS-AGN, even after removing the strongest metal absorber (towards QSO J0852+0313; see Figure \ref{fig:J0852}), whilst these species are not detected in the stacked spectrum of the control galaxies. We remind the reader that this enhancement was suggested in our covering fraction analysis of Si\ion{iii} (Figure \ref{fig:CovFracR}). For the inner CGM, the stacked spectrum hints that there is a deficit of metal species around AGN galaxies relative to the control sample. The combination of the excess Ly$\alpha$ EW and tentative Si\ion{ii} and Si\ion{iii} EW enhancements at high \Rimp{} is suggestive that these EW enhancement are tracing the cool gas phase of the CGM, rather than just the \HI{} gas kinematics. We note that the inner CGM distribution has a slightly higher median \Mstar{} than the outer CGM  bin (\logMstar{} of 10.8 dex relative to 10.5 dex), and the inner and outer CGM bins contain approximately the same ratio of passive and star-forming galaxies. We remind the reader that this discrepancy in the median \Mstar{} of each bins is on the order of the size of our control matching tolerance.

\section{Discussion}

In the previous section, we demonstrated that the CGM around AGN hosts is not much different than the control-matched non-AGN hosts. Statistically significant differences are found in the analysis of the stacked spectrum; the COS-AGN systems have a higher EW of Ly$\alpha$ (and potentially Si\ion{ii} and Si\ion{iii} as well) relative to their non-AGN host counterparts at high impact parameters (\Rimp{} $\geq 164$ kpc; $\delta$log(EW/m\AA{})$=+0.45\pm0.05$~dex). The kinematics of the gas traced by the absorption show the gas is likely bound to the halo, whilst no strong kinematic offsets relative to their host are present. We now consider whether these observations are a result of the AGN directly influencing the CGM of the host galaxy, or an effect of the environments (either internal or external to the host galaxy) in which AGN are typically found.

\subsection{Are we seeing the effects of AGN feedback?}

\subsubsection*{Mock COS-AGN simulation}
\label{sec:Sims}
The radiation field of the AGN may be expected to have a profound effect on the ionization structure of the CGM, by enhancing the ionizing radiation field to which the surrounding gas is subjected, and in the case of metals even long after the AGN has turned off \citep{Oppenheimer13,Segers17,Oppenheimer18}. To quantify the expected effects of turning on and off the AGN ionizing radiation spectrum on the CGM of a COS-AGN galaxy, we created a mock COS-AGN survey using previously run cosmological zoom-in hydrodynamical simulations with an additional AGN ionizing source following previous work on non-equilibrium ionization effects \citep{Oppenheimer13,Oppenheimer16}. The purpose of these simulations is to isolate the effect of adding a constant AGN ionizing radiation source versus a control sample without AGN ionizing radiation to study the effect of the changed ionization structure of the CGM.  This exploration is not meant to model or consider the effect of AGN feedback mechanically transforming the CGM via superwinds as the inclusion of the AGN radiation is not tied to the accretion onto the central super massive black hole. The AGN radiation is added to these simulations at the position of the central super massive black hole and only alters the ionization states of the CGM.

For our simulation suite we selected three representative galaxy haloes from \cite{Oppenheimer16} that were chosen from the EAGLE volume \citep{Crain15,Schaye15}. These three halos are representative of the properties of the COS-AGN sample at $z=0.075$ [\logMstar$=10.3$, $10.9$, $11.0$, \logsSFR$=-10.3$, $-10.7$, $-11.0$, residing in haloes log(M$_{200}$/M$_{\odot}$)$=12.1$, $12.8$, $13.3$; respectively]. The haloes were ran at EAGLE HiRes resolution (gas particle mass of $2.3\times10^{5}~{\rm M_{\odot}}$, dark matter particle mass of $1.2\times10^{6}~{\rm M_{\odot}}$, and softening length of 350 proper pc) using the Recal feedback prescription \citep{Schaye15} and are zooms Gal001, Grp003, and Grp008 listed in Table 1 of \citet{Oppenheimer16} from the initial conditions ($z=127$).

To include the effects of the AGN ionizing spectrum on the CGM at different luminosities, we inserted an additional AGN ionizing source \citep[with an ionizing spectrum from][]{Sazonov04}  placed at the centre of the galaxy, instantaneously affecting the radiation field from $z=0.1$ onwards and reaches equilibrium at $z=0.075$. We note that this AGN radiation model has no dynamical effect on the gas accretion or outflows from the AGN, and is not tied to the accretion of material onto the central super-massive black hole.  A range of AGN luminosities  was used to match the COS-AGN luminosities (\logLbol{}=42--44 dex, in increments of 0.5 dex), and a control run with no AGN radiation was included for creating the control sample. The time-dependent ionization from the AGN in addition to the \cite{Haardt01}\footnote{The \cite{Haardt01} background is adopted as it better reproduces the statistics of the Ly$\alpha$ forest in the EAGLE volumes and other simulations \citep{Rahmati15,Kollmeier14}.} ultra-violet background is followed using the \cite{Richings14a} ionization network.

A mock COS-AGN sample was then generated from this simulation suite to match the properties of the COS-AGN sample. For each observed galaxy in COS-AGN, a simulated galaxy was selected by matching the stellar mass (with a \logMstar{}$=\pm0.5$~dex tolerance), star formation rate (\logsSFR{}$=\pm0.5$~dex tolerance), and AGN luminosity (\logLbol{}$=\pm0.25$~dex) of the simulated haloes to each COS-AGN galaxy. If multiple matches were identified, a random simulated halo was selected such that a mock sample contains the same number of galaxies as the actual COS-AGN sample (20 galaxies). A mock spectrum was generated for each galaxy  using the \textsc{SpecWizard} package \citep[][]{Theuns98,Schaye03} by placing a randomly oriented quasar sightline through the CGM at the corresponding impact parameter of the matched COS-AGN galaxy sightline. A control sample was generated in a similar fashion, but all AGN ionizing sources in the simulated galaxies were set to \Lbol{}$=0$~erg~s$^{-1}$. The above procedure was repeated 200 times (including choosing another random halo if multiple haloes were matched), in order to produce 200 mock survey samples for both the AGN and control samples.  These 200 realizations of the COS-AGN survey were needed to reproduce the range in measured EW variations for the metal species between the mock samples, and is used to quantify the expected spread in EWs from different orientations and from halo to halo variations. Further details on how these mock spectra were generated will be presented in a forthcoming paper (Horton et al., in prep.).

\begin{figure*}
\begin{center}
\includegraphics[width=\textwidth]{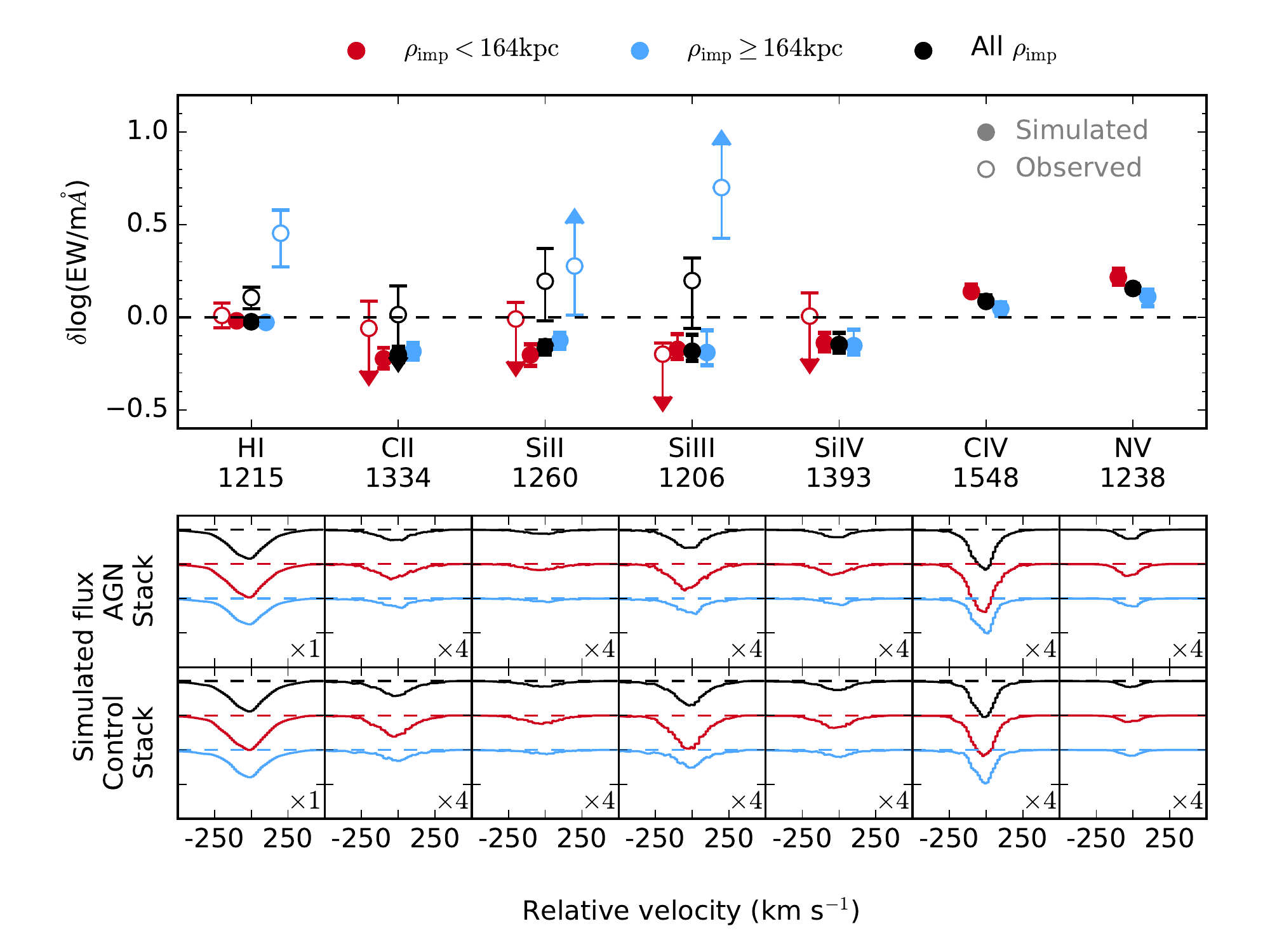}
\caption{The top panel shows the measured $\delta$EW using a stacked spectrum from the zoom-in simulations of AGN hosts, split into bins of \Rimp{} (all \Rimp{}, $<164$ kpc and $\geq164$ kpc; black, red and blue points, respectively). The respective open symbols show the $\delta$EW measured in the observations (i.e.~identical points presented in the top panel of Figure \ref{fig:StackdEW}). Errorbars were calculated using the jackknife approach identical to that used for the observations. The bottom panels show the simulated stacked velocity profiles for each species for the AGN (middle row) and non-AGN control run (bottom row). The dashed lines represent the continuum levels of the absorption profiles. The Ly$\alpha$ are staggered by 0.5 in relative flux for clarity. The relative flux scale has been enlarged by a factor of four for the metal line profiles (indicated by a $\times4$ in the bottom right of each panel), with the dashed continuum lines separated by a relative flux of 0.125. For the outer CGM, the simulations predict little change in the EW of \HI{} and Si\ion{iii} due to the ionizing radiation from the AGN, which is in stark contrast to the COS-AGN observations similarly presented in Figure \ref{fig:StackdEW}.}
\label{fig:Sims}
\end{center}
\end{figure*}

\subsubsection*{Stacked mock COS-AGN spectra}
We assessed the relative impact of the AGN ionizing radiation on the measured simulated EWs by repeating the stacking procedure above, but with the simulated data. We computed the relative EW ratio $\delta$log(EW) as with the observations. Note that this relative EW analysis removes any systematic differences between the observed and simulated EWs caused by assumptions in the physics models (e.g. feedback) or ability to resolve small clouds in the simulations \citep{Schaye07,Stinson12,Crighton15,Gutcke17,Nelson17}. A quantitative comparison between the simulated and observed EWs will be presented in Horton et al.~(in prep.), although we note that the simulations produce a systematically weaker \HI{} EW from the observations, while metal lines show better agreement with the observed data \citep[see][]{Oppenheimer16,Oppenheimer17}. 

The simulated stacked sightline spectra were created from the mocks by centering the velocity profile on the redshift of the simulated galaxy, rebinning the mock spectra to 15 \kms{}, and mean stacking these rebinned spectra. Three simulated stacked spectra were created for the same three bins of impact parameters that were previously used for the observed data: all \Rimp{}, \Rimp{}$<164$ kpc, and \Rimp{}$\geq164$ kpc. Jackknife errors on the measured EWs were computed by removing all 200 mock spectra associated with the matched COS-AGN sightline which contributed the most and least to the derived EW. This approach for calculating the simulated jackknife error is identical to that used for the observations. Using the same analysis for our COS-AGN sightlines, Figure \ref{fig:Sims} presents the $\delta$log(EW) analysis for these simulated stacked spectra (solid points in the top panel) to quantify the effect of including AGN radiation on the CGM relative to the control simulations.  The inclusion of an AGN ionizing spectrum results in a negligible change in the EW as a function of impact parameter for Ly$\alpha$. At high impact parameters (\Rimp{}$\geq 164$~kpc), the simulated results are in contrast to the observations (open points on Figure \ref{fig:Sims}) where the COS-AGN sample shows a significant enhancement in the Ly$\alpha$ EW relative to the control sample, as well as a potential enhancement for Si\ion{ii} and Si\ion{iii}. We do note that for the inner impact parameter bin, the results from the simulation are consistent with the lack of metal species detected of our COS-AGN stacked spectrum (\Rimp{}$<164$ kpc; Figure \ref{fig:StackdEW}).

Compounded by the fact that the \HI{} ionizing radiation from a \cite{Sazonov04} AGN at the observed \Rimp{} of the COS-AGN sample is weaker ($\lesssim25\%$ ) than the ionizing radiation from the UV background \citep{Haardt01} at most of the impact parameters probed by COS-AGN (see the bottom right panel of Figure \ref{fig:SampDists}), it is unlikely that photoionizing radiation from the AGN is responsible for the observed difference in EWs between the COS-AGN and the control samples. We note that at the impact parameters of COS-AGN, the light travel time of radiation from the AGN ($\sim5\times10^{5}$ yr) is comparable to the AGN's lifetime \citep[$\lesssim10^{6}$ yr;][]{Schirber04,Goncalves08,Furlanetto11,Keel12}, implying that photoionization events caused by an AGN likely require previous AGN cycle(s) to have ionized the CGM gas, and have remained in the predicted long-lived fossil zone around the AGN \citep{Oppenheimer13,Segers17} provided an AGN has been active within the last several Myrs. Unfortunately, our simulations predict that low ionization species cannot distinguish the presence of fossil zones in COS-AGN due to the low intensity of the photo-ionizing radiation.

Despite being unable to probe the effects of the lower ionization species,  ions such as C\ion{iv}, N\ion{v}, and O\ion{vi} are better indicators of these proximity zone fossil. We highlight that in our simulated COS-AGN haloes, C\ion{iv} and N\ion{v} are still sensitive to the ionizing radiation of the harder AGN ionizing spectrum relative to the \cite{Haardt01} UV background \citep[see figure 3 in][]{Segers17}, as demonstrated by the enhanced EWs for these ionization species out to 300 kpc. However, to observe such an excess with N\ion{v} 1238~\AA{} would require spectra with S/N of $\sim30$ to detect an absorption line of the predicted strength displayed in Figure \ref{fig:Sims}. The excess EW of C\ion{iv} 1548~\AA{} would be an excellent test of the effects of the AGN ionizing field as we have already detected absorption in our stacked COS-AGN spectra (Table \ref{tab:EWStack}). Such a test would required observing the C\ion{iv} 1548~\AA{} covering fraction in the CGM of non-AGN galaxies from the control-matched sample.

\subsubsection*{AGN-driven winds and outflows}
An alternative form of feedback is AGN-driven winds or outflows \citep[][]{Concas17, Fiore17,Woo17}. As the typical lifetime of an AGN ($\lesssim10^6$ yr) is much smaller than the expected travel time of winds out to the impact parameters probed by COS-AGN \citep[$\sim10^8$ yr, assuming a constant, maximum velocity of $1000$ \kms{}; e.g.][]{Tremonti07,Veilleux13,Ishibashi15}, any signatures of winds or outflowing material would likely trace material expelled from a previous cycles of AGN activity or star-formation in the host galaxy \citep[e.g.][]{Nedelchev17,Kauffmann17,Woo17}. The typically observed signatures of outflowing winds from a galaxy manifest as kinematic offsets of ionized emission or absorption lines from the host galaxy \citep[e.g.][]{Bordoloi14b,Rubin14,Woo16,Heckman17,Perna17}, but as demonstrated in Figure \ref{fig:Kinematics}, the Ly$\alpha$ gas that we are probing in the COS-AGN sightlines does not have any strong bulk motion away from the host galaxy relative to the control matched sample.  If such winds or outflows were driven by previous AGN or star-forming activity, the lack of kinematic offsets from the host galaxy suggests that these winds have dissipated over time, and should have deposited metals into the CGM \citep[e.g.][]{Muzahid15,Turner15}. The low metal covering fraction at low impact parameter ($\leq164$ kpc; Figure \ref{fig:CovFracR}) suggests an absence of outflowing material polluting the CGM with metals, rejecting the notion that any recent (within 160~Myr) AGN-driven winds  have enhanced the CGM. We note that the Ly$\alpha$ EW at these column densities is more sensitive to the kinematics of the gas than to the amount of gas. Although we have rejected the possibility that AGN-driven winds are responsible for our results, it is possible that the gas is more turbulent in the CGM of AGN hosts compared to their control matches. To estimate the effects of turbulence, we would require observations of unsaturated low ionization metal absorption lines to verify if the EW enhancements are from enhanced column densities or kinematic broadening.

Given that AGN-driven winds can affect the absorption line profile by up to $\approx \pm 1000$ \kms{}, our adopted search window of $\pm500$ \kms{} (Section \ref{sec:DataEW}) could potentially miss gas present in a wind.  We searched for the Ly$\alpha$ profiles within $\pm1000$ \kms{} and found two systems (J1214+0825 and J2133$-$0712) with minimal absorption outside the original window not associated with other absorption systems or the Galaxy. These two additional absorption components do not show any associated metal line absorption. As these missed components are small and narrow, the calculated flux-weighted velocity centroids presented in Figure \ref{fig:Kinematics} would still be contained within the already identified absorption component. However, we remind the reader that our adopted search window of $\pm500$ \kms{} is adopted to be consistent with methods used in the literature and control samples.

Rather than AGN feedback, it is possible that the effects we are seeing are from a different process co-eval or prior to the onset of AGN accretion. Several works have pointed out that AGN activity coincide with a recent starburst; with the AGN having significant accretion events at least $\sim200$ Myr after the starburst has occurred \citep{Wild07,Davies07,Wild10,Yesuf14} giving the neutral material time to propagate out to the impact parameters probed by COS-AGN \citep{Heckman17}. With a sample of QSO sightlines probing the CGM around 17 low-redshift starburst and post-starburst galaxies, \cite{Heckman17} have observed a similar signature of enhanced EWs of Ly$\alpha$, Si\ion{iii}, and C\ion{iv} (the latter of which is not measured in our control sample) relative to a control-matched sample (matched in stellar mass and impact parameter). In the range of impact parameters and stellar masses probed by COS-AGN, the strength of our enhanced EW signature is consistent with the values probed by \cite{Heckman17}. However, the results of \cite{Heckman17} show strong offsets in the kinematics of the gas from the host galaxy \citep[$\approx 100$ \kms{}; see figure 5 from][]{Heckman17}, whereas the COS-AGN sightlines do not (bottom panel of Figure \ref{fig:Kinematics}). Assuming the AGN activity was triggered by the starburst, a minimum delay time of 200 Myr could allow for any starburst-driven winds to dissipate and kinematic offsets to no longer be present at the impact parameters of the COS-AGN sample.  Although this starburst picture provides a possible explanation of our observations, we caution that starbursts are not the only astrophysical event linked to AGN accretion activity. For example, mergers that trigger the AGN \citep{Ellison11,Ellison13,Satyapal14,Silverman14,Goulding17} could potentially affect the surrounding CGM gas. Past and future work focussing on the CGM of galaxy mergers can further test this result \citep[][Bordoloi et al. in prep.]{Johnson14,Hani17}.

\subsection{Are we seeing the effects from environment or other galaxy properties?}
If the AGN (or host galaxy) is not responsible for the observed differences in the CGM, an alternative is that the circumgalactic environment in which an AGN host is found is different. Results from \emph{Quasars probing Quasars} \citep{QPQ6}, \cite{Farina14}, and \cite{Johnson15} have all suggested that the excess of cool gas seen in the CGM out to 1 Mpc around $z\approx1$ quasars is a result of residing in group environments. Although the excess of cool gas around quasars goes in the same direction as the 0.1 dex enhancement we find in the outer CGM of $z\approx0.1$ COS-AGN galaxies (though we \emph{do not} see an excess in the Ly$\alpha$ EW in the inner CGM of AGN hosts, as seen for QSOs), the dark matter haloes of the AGN in our sample are typically an order of magnitude smaller than group dark matter haloes that host quasars. As stated in Section \ref{sec:ControlMatching}, the $\delta_{5}$ parameter provides an estimate of the environment. Given that differences in the distributions of $\delta_{5}$ between AGN and star-forming galaxies in the SDSS vary on the order of a percent for a given stellar mass, it is likely that the enhanced EW of cool gas is not due to the contribution from a difference in the galaxy environment.

Given that we find an excess in the \HI{} content of the outer CGM around AGN hosts, the results from \cite{Borthakur13} which find a connection between the ISM and CGM gas properties would imply that the ISM would also host a large reservoir of \HI{} gas. Such an enhancement in the ISM gas mass (relative to non-AGN galaxies) has been previously seen in AGN hosts \citep[e.g.][]{Vito14}, where an excess of $\sim0.2$ dex in gas mass is seen for AGN hosts similar to those probed in our sample.  If such large gas reservoirs do exist in the ISM of AGN hosts, the AGN could be fuelled by the excess cool gas in the ISM, which in turn is fed by the cool CGM  gas surrounding the AGN host. We note that other works such as \cite{Fabello11} have found that the ISM of optically-selected AGN hosts contain the same \HI{} gas mass as their star-forming counterparts. However, $\sim$50\% of the \cite{Fabello11} AGN sample are so-called `composite' AGN (galaxies whose emission lines contain contributions from both star formation and AGN activity), which are not representative of the AGN hosts selected in our COS-AGN sample.

A further test of this accretion picture would be to look for any orientation effects. If we are probing the gas reservoirs that are fuelling the AGN, we are likely to find the accreting gas along the major axis of the galaxy \citep[e.g.][]{Kacprzak12,Nielsen15,Ho17}. Unfortunately, we do not have the ability to measure robust inclinations for many of our COS-AGN galaxies from SDSS imaging (see Figure \ref{fig:postage}), and have too small of a sample size to produce a significant statistic. In addition, the \cite{Ho17} sample are at much closer impact parameters ($\lesssim 50$ kpc) whereas \cite{Borthakur15} showed that at higher impact parameters (such as those probed by COS-AGN) there is no evidence of orientation effects for galaxies in COS-GASS. However, we do note that for the 4--5 sightlines that are probing along the edge of the disc relative to the 2--3 that are perpendicular to the disc, there is no significant difference in the median $\Delta$log(EW). A larger sample would be required to test this explicitly.

\section{Summary}
Using a sample of 19 quasar sightlines through the  circumgalactic medium (CGM) of 20 Type II Seyfert  AGN and LINERs, we have demonstrated that there are mild differences in the rest-frame equivalent widths (EWs) of cool  CGM gas around AGN hosts relative to their non-AGN counterparts. After matching in stellar mass and impact parameter, we find:
\begin{itemize}

\item[1.] The covering fraction of Ly$\alpha$ gas for the AGN is 94$^{+6}_{-23}$\%, which is comparable to the star-forming control galaxies (100$^{+0}_{-21}$\%) and consistent with passive galaxies (75$^{+25}_{-21}$\%). The covering fractions of metal species (C\ion{ii}, Si\ion{ii}, Si\ion{iii}, C\ion{iv}, Si\ion{iv}, and N\ion{v}) are consistent with the control-matched galaxies (Figure \ref{fig:CovFracR}).

\item[2.] An insignificant increase in the Ly$\alpha$ EW for AGN relative to control-matched galaxies on a sightline by sightline basis. The measured median EW offset between these two population is $+0.10\pm0.13$ dex (Figure \ref{fig:deltaEWRho}). 

\item[3.] After stacking the spectra, the observed EW enhancement of low ionization species for AGN is seen only at high impact parameters (\Rimp{}$\geq164$ kpc; the median impact parameter of COS-AGN) for both Ly$\alpha$ ($\delta$log(EW)$=+0.45\pm0.05$) and cool metal line tracers (Si\ion{ii} 1260~\AA{} [$\delta$log(EW)$>0.27$~dex] and Si\ion{iii} 1206~\AA{} [$\delta$log(EW)$>0.75$~dex]; Figure \ref{fig:StackdEW}). These results are inconsistent with the expected effects from AGN feedback seen in our zoom-in simulations at high impact parameters (Figure \ref{fig:Sims}). At lower impact parameters (\Rimp{}$<164$ kpc), our results are consistent with the simulations.

\item[4.] The Ly$\alpha$ line kinematics for the COS-AGN sightlines does not differ significantly from what is observed around the control-matched sample, suggesting there is no strong bulk motion in the CGM due to the presence of an AGN (Figure \ref{fig:Kinematics}). As all but one system show gas within the escape velocity of the host halo, the probed material is likely bound to the AGN host.

\end{itemize}

These results suggest that the circumgalactic environments that host AGN show little difference than their non-AGN hosts on a sightline-by-sightline basis, likely attributed to our small sample size. We only detect a significant difference in the amount of cool gas in our stacked spectrum at high impact parameters (\Rimp{}$\geq164$~kpc), which we use to interpret our results. Given the lack of signatures (in both EW and kinematic diagnostics) of recent AGN feedback on the CGM from winds and ionizing radiation, we speculate that possible causes for these stacked spectrum results could be from the accretion of cool gas to feed the AGN, or a remnant effect from previous evolutionary activity of the host galaxy (such as starbursts) prior to the AGN accretion phase.

\section*{Acknowledgments}
We thank the anonymous referee for their comments to improve the clarity of this manuscript. We are grateful for Sanchayeeta Borthakur providing the reduced spectra and all Si species equivalent widths from the COS-GASS survey. This manuscript benefited greatly from discussions with H.W. Chen, T. Heckman,  C. Martin, J.X. Prochaska, and J. Werk. BDO's contribution was made possible by the HST observing grant HST-GO-13774.

\bibliography{bibref}

\begin{thebibliography}{123}
\expandafter\ifx\csname natexlab\endcsname\relax\def\natexlab#1{#1}\fi

\bibitem[{{Abazajian} {et~al}\mbox{.}(2009){Abazajian}, {Adelman-McCarthy},
  {Ag{\"u}eros}, {Allam}, {Allende Prieto}, {An}, {Anderson}, {Anderson},
  {Annis}, {Bahcall}, \& et~al.}]{Abazajian09}
{Abazajian} K.~N. {et~al.}, 2009, \apjs, 182, 543

\bibitem[{{Adelberger} {et~al}\mbox{.}(2005){Adelberger}, {Shapley}, {Steidel},
  {Pettini}, {Erb}, \& {Reddy}}]{Adelberger05}
{Adelberger} K.~L., {Shapley} A.~E., {Steidel} C.~C., {Pettini} M., {Erb}
  D.~K., {Reddy} N.~A., 2005, \apj, 629, 636

\bibitem[{{Baldry} {et~al}\mbox{.}(2006){Baldry}, {Balogh}, {Bower},
  {Glazebrook}, {Nichol}, {Bamford}, \& {Budavari}}]{Baldry06}
{Baldry} I.~K., {Balogh} M.~L., {Bower} R.~G., {Glazebrook} K., {Nichol} R.~C.,
  {Bamford} S.~P., {Budavari} T., 2006, \mnras, 373, 469

\bibitem[{{Baldwin} {et~al}\mbox{.}(1981){Baldwin}, {Phillips}, \&
  {Terlevich}}]{Baldwin81}
{Baldwin} J.~A., {Phillips} M.~M., {Terlevich} R., 1981, \pasp, 93, 5

\bibitem[{{Belfiore} {et~al}\mbox{.}(2016){Belfiore}, {Maiolino}, {Maraston},
  {Emsellem}, {Bershady}, {Masters}, {Yan}, {Bizyaev}, {Boquien}, {Brownstein},
  {Bundy}, {Drory}, {Heckman}, {Law}, {Roman-Lopes}, {Pan}, {Stanghellini},
  {Thomas}, {Weijmans}, \& {Westfall}}]{Belfiore16}
{Belfiore} F. {et~al.}, 2016, \mnras, 461, 3111

\bibitem[{{Bergeron}(1986)}]{Bergeron86}
{Bergeron} J., 1986, \aap, 155, L8

\bibitem[{{Bluck} {et~al}\mbox{.}(2014){Bluck}, {Mendel}, {Ellison}, {Moreno},
  {Simard}, {Patton}, \& {Starkenburg}}]{Bluck14}
{Bluck} A.~F.~L., {Mendel} J.~T., {Ellison} S.~L., {Moreno} J., {Simard} L.,
  {Patton} D.~R., {Starkenburg} E., 2014, \mnras, 441, 599

\bibitem[{{Bluck} {et~al}\mbox{.}(2016){Bluck}, {Mendel}, {Ellison}, {Patton},
  {Simard}, {Henriques}, {Torrey}, {Teimoorinia}, {Moreno}, \&
  {Starkenburg}}]{Bluck16}
{Bluck} A.~F.~L. {et~al.}, 2016, \mnras, 462, 2559

\bibitem[{{Bordoloi} {et~al}\mbox{.}(2014{\natexlab{a}}){Bordoloi}, {Lilly},
  {Hardmeier}, {Contini}, {Kneib}, {Le Fevre}, {Mainieri}, {Renzini},
  {Scodeggio}, {Zamorani}, {Bardelli}, {Bolzonella}, {Bongiorno}, {Caputi},
  {Carollo}, {Cucciati}, {de la Torre}, {de Ravel}, {Garilli}, {Iovino},
  {Kampczyk}, {Kova{\v c}}, {Knobel}, {Lamareille}, {Le Borgne}, {Le Brun},
  {Maier}, {Mignoli}, {Oesch}, {Pello}, {Peng}, {Perez Montero}, {Presotto},
  {Silverman}, {Tanaka}, {Tasca}, {Tresse}, {Vergani}, {Zucca}, {Cappi},
  {Cimatti}, {Coppa}, {Franzetti}, {Koekemoer}, {Moresco}, {Nair}, \&
  {Pozzetti}}]{Bordoloi14b}
{Bordoloi} R. {et~al.}, 2014{\natexlab{a}}, \apj, 794, 130

\bibitem[{{Bordoloi} {et~al}\mbox{.}(2011){Bordoloi}, {Lilly}, {Knobel},
  {Bolzonella}, {Kampczyk}, {Carollo}, {Iovino}, {Zucca}, {Contini}, {Kneib},
  {Le Fevre}, {Mainieri}, {Renzini}, {Scodeggio}, {Zamorani}, {Balestra},
  {Bardelli}, {Bongiorno}, {Caputi}, {Cucciati}, {de la Torre}, {de Ravel},
  {Garilli}, {Kova{\v c}}, {Lamareille}, {Le Borgne}, {Le Brun}, {Maier},
  {Mignoli}, {Pello}, {Peng}, {Perez Montero}, {Presotto}, {Scarlata},
  {Silverman}, {Tanaka}, {Tasca}, {Tresse}, {Vergani}, {Barnes}, {Cappi},
  {Cimatti}, {Coppa}, {Diener}, {Franzetti}, {Koekemoer}, {L{\'o}pez-Sanjuan},
  {McCracken}, {Moresco}, {Nair}, {Oesch}, {Pozzetti}, \&
  {Welikala}}]{Bordoloi11}
{Bordoloi} R. {et~al.}, 2011, \apj, 743, 10

\bibitem[{{Bordoloi} {et~al}\mbox{.}(2014{\natexlab{b}}){Bordoloi},
  {Tumlinson}, {Werk}, {Oppenheimer}, {Peeples}, {Prochaska}, {Tripp}, {Katz},
  {Dav{\'e}}, {Fox}, {Thom}, {Ford}, {Weinberg}, {Burchett}, \&
  {Kollmeier}}]{Bordoloi14}
{Bordoloi} R. {et~al.}, 2014{\natexlab{b}}, \apj, 796, 136

\bibitem[{{Borthakur} {et~al}\mbox{.}(2013){Borthakur}, {Heckman},
  {Strickland}, {Wild}, \& {Schiminovich}}]{Borthakur13}
{Borthakur} S., {Heckman} T., {Strickland} D., {Wild} V., {Schiminovich} D.,
  2013, \apj, 768, 18

\bibitem[{{Borthakur} {et~al}\mbox{.}(2016){Borthakur}, {Heckman}, {Tumlinson},
  {Bordoloi}, {Kauffmann}, {Catinella}, {Schiminovich}, {Dav{\'e}}, {Moran}, \&
  {Saintonge}}]{Borthakur16}
{Borthakur} S. {et~al.}, 2016, \apj, 833, 259

\bibitem[{{Borthakur} {et~al}\mbox{.}(2015){Borthakur}, {Heckman}, {Tumlinson},
  {Bordoloi}, {Thom}, {Catinella}, {Schiminovich}, {Dav{\'e}}, {Kauffmann},
  {Moran}, \& {Saintonge}}]{Borthakur15}
{Borthakur} S. {et~al.}, 2015, \apj, 813, 46

\bibitem[{{Bowen} {et~al}\mbox{.}(1995){Bowen}, {Blades}, \&
  {Pettini}}]{Bowen95}
{Bowen} D.~V., {Blades} J.~C., {Pettini} M., 1995, \apj, 448, 634

\bibitem[{{Bowen} {et~al}\mbox{.}(2006){Bowen}, {Hennawi}, {M{\'e}nard},
  {Chelouche}, {Inada}, {Oguri}, {Richards}, {Strauss}, {Vanden Berk}, \&
  {York}}]{Bowen06}
{Bowen} D.~V. {et~al.}, 2006, \apjl, 645, L105

\bibitem[{{Bower} {et~al}\mbox{.}(2017){Bower}, {Schaye}, {Frenk}, {Theuns},
  {Schaller}, {Crain}, \& {McAlpine}}]{Bower17}
{Bower} R.~G., {Schaye} J., {Frenk} C.~S., {Theuns} T., {Schaller} M., {Crain}
  R.~A., {McAlpine} S., 2017, \mnras, 465, 32

\bibitem[{{Brinchmann} {et~al}\mbox{.}(2004){Brinchmann}, {Charlot}, {White},
  {Tremonti}, {Kauffmann}, {Heckman}, \& {Brinkmann}}]{Brinchmann04}
{Brinchmann} J., {Charlot} S., {White} S.~D.~M., {Tremonti} C., {Kauffmann} G.,
  {Heckman} T., {Brinkmann} J., 2004, \mnras, 351, 1151

\bibitem[{{Burchett} {et~al}\mbox{.}(2016){Burchett}, {Tripp}, {Bordoloi},
  {Werk}, {Prochaska}, {Tumlinson}, {Willmer}, {O'Meara}, \&
  {Katz}}]{Burchett16}
{Burchett} J.~N. {et~al.}, 2016, \apj, 832, 124

\bibitem[{{Chabrier}(2003)}]{Chabrier03}
{Chabrier} G., 2003, \pasp, 115, 763

\bibitem[{{Chen} {et~al}\mbox{.}(2010){Chen}, {Helsby}, {Gauthier}, {Shectman},
  {Thompson}, \& {Tinker}}]{Chen10}
{Chen} H.-W., {Helsby} J.~E., {Gauthier} J.-R., {Shectman} S.~A., {Thompson}
  I.~B., {Tinker} J.~L., 2010, \apj, 714, 1521

\bibitem[{{Concas} {et~al}\mbox{.}(2017){Concas}, {Popesso}, {Brusa},
  {Mainieri}, {Erfanianfar}, \& {Morselli}}]{Concas17}
{Concas} A., {Popesso} P., {Brusa} M., {Mainieri} V., {Erfanianfar} G.,
  {Morselli} L., 2017, ArXiv e-prints

\bibitem[{{Crain} {et~al}\mbox{.}(2015){Crain}, {Schaye}, {Bower}, {Furlong},
  {Schaller}, {Theuns}, {Dalla Vecchia}, {Frenk}, {McCarthy}, {Helly},
  {Jenkins}, {Rosas-Guevara}, {White}, \& {Trayford}}]{Crain15}
{Crain} R.~A. {et~al.}, 2015, \mnras, 450, 1937

\bibitem[{{Crighton} {et~al}\mbox{.}(2015){Crighton}, {Hennawi}, {Simcoe},
  {Cooksey}, {Murphy}, {Fumagalli}, {Prochaska}, \& {Shanks}}]{Crighton15}
{Crighton} N.~H.~M., {Hennawi} J.~F., {Simcoe} R.~A., {Cooksey} K.~L., {Murphy}
  M.~T., {Fumagalli} M., {Prochaska} J.~X., {Shanks} T., 2015, \mnras, 446, 18

\bibitem[{{Davies} {et~al}\mbox{.}(2007){Davies}, {M{\"u}ller S{\'a}nchez},
  {Genzel}, {Tacconi}, {Hicks}, {Friedrich}, \& {Sternberg}}]{Davies07}
{Davies} R.~I., {M{\"u}ller S{\'a}nchez} F., {Genzel} R., {Tacconi} L.~J.,
  {Hicks} E.~K.~S., {Friedrich} S., {Sternberg} A., 2007, \apj, 671, 1388

\bibitem[{{Ellison} {et~al}\mbox{.}(2013){Ellison}, {Mendel}, {Patton}, \&
  {Scudder}}]{Ellison13}
{Ellison} S.~L., {Mendel} J.~T., {Patton} D.~R., {Scudder} J.~M., 2013, \mnras,
  435, 3627

\bibitem[{{Ellison} {et~al}\mbox{.}(2011){Ellison}, {Patton}, {Mendel}, \&
  {Scudder}}]{Ellison11}
{Ellison} S.~L., {Patton} D.~R., {Mendel} J.~T., {Scudder} J.~M., 2011, \mnras,
  418, 2043

\bibitem[{{Ellison} {et~al}\mbox{.}(2016{\natexlab{a}}){Ellison},
  {Teimoorinia}, {Rosario}, \& {Mendel}}]{Ellison16}
{Ellison} S.~L., {Teimoorinia} H., {Rosario} D.~J., {Mendel} J.~T.,
  2016{\natexlab{a}}, \mnras, 455, 370

\bibitem[{{Ellison} {et~al}\mbox{.}(2016{\natexlab{b}}){Ellison},
  {Teimoorinia}, {Rosario}, \& {Mendel}}]{Ellison16b}
{Ellison} S.~L., {Teimoorinia} H., {Rosario} D.~J., {Mendel} J.~T.,
  2016{\natexlab{b}}, \mnras, 458, L34

\bibitem[{{Fabello} {et~al}\mbox{.}(2011){Fabello}, {Kauffmann}, {Catinella},
  {Giovanelli}, {Haynes}, {Heckman}, \& {Schiminovich}}]{Fabello11}
{Fabello} S., {Kauffmann} G., {Catinella} B., {Giovanelli} R., {Haynes} M.~P.,
  {Heckman} T.~M., {Schiminovich} D., 2011, \mnras, 416, 1739

\bibitem[{{Fabian}(2012)}]{Fabian12}
{Fabian} A.~C., 2012, \araa, 50, 455

\bibitem[{{Farina} {et~al}\mbox{.}(2013){Farina}, {Falomo}, {Decarli},
  {Treves}, \& {Kotilainen}}]{Farina13}
{Farina} E.~P., {Falomo} R., {Decarli} R., {Treves} A., {Kotilainen} J.~K.,
  2013, \mnras, 429, 1267

\bibitem[{{Farina} {et~al}\mbox{.}(2014){Farina}, {Falomo}, {Scarpa},
  {Decarli}, {Treves}, \& {Kotilainen}}]{Farina14}
{Farina} E.~P., {Falomo} R., {Scarpa} R., {Decarli} R., {Treves} A.,
  {Kotilainen} J.~K., 2014, \mnras, 441, 886

\bibitem[{{Fiore} {et~al}\mbox{.}(2017){Fiore}, {Feruglio}, {Shankar},
  {Bischetti}, {Bongiorno}, {Brusa}, {Carniani}, {Cicone}, {Duras}, {Lamastra},
  {Mainieri}, {Marconi}, {Menci}, {Maiolino}, {Piconcelli}, {Vietri}, \&
  {Zappacosta}}]{Fiore17}
{Fiore} F. {et~al.}, 2017, \aap, 601, A143

\bibitem[{{Furlanetto} \& {Lidz}(2011)}]{Furlanetto11}
{Furlanetto} S.~R., {Lidz} A., 2011, \apj, 735, 117

\bibitem[{{Gehrels}(1986)}]{Gehrels86}
{Gehrels} N., 1986, \apj, 303, 336

\bibitem[{{Gon{\c c}alves} {et~al}\mbox{.}(2008){Gon{\c c}alves}, {Steidel}, \&
  {Pettini}}]{Goncalves08}
{Gon{\c c}alves} T.~S., {Steidel} C.~C., {Pettini} M., 2008, \apj, 676, 816

\bibitem[{{Goulding} {et~al}\mbox{.}(2017){Goulding}, {Greene}, {Bezanson},
  {Greco}, {Johnson}, {Leauthaud}, {Matsuoka}, {Medezinski}, \&
  {Price-Whelan}}]{Goulding17}
{Goulding} A.~D. {et~al.}, 2017, ArXiv e-prints

\bibitem[{{Green} {et~al}\mbox{.}(2012){Green}, {Froning}, {Osterman},
  {Ebbets}, {Heap}, {Leitherer}, {Linsky}, {Savage}, {Sembach}, {Shull},
  {Siegmund}, {Snow}, {Spencer}, {Stern}, {Stocke}, {Welsh}, {B{\'e}land},
  {Burgh}, {Danforth}, {France}, {Keeney}, {McPhate}, {Penton}, {Andrews},
  {Brownsberger}, {Morse}, \& {Wilkinson}}]{Green12}
{Green} J.~C. {et~al.}, 2012, \apj, 744, 60

\bibitem[{{Gutcke} {et~al}\mbox{.}(2017){Gutcke}, {Stinson}, {Macci{\`o}},
  {Wang}, \& {Dutton}}]{Gutcke17}
{Gutcke} T.~A., {Stinson} G.~S., {Macci{\`o}} A.~V., {Wang} L., {Dutton} A.~A.,
  2017, \mnras, 464, 2796

\bibitem[{{Haardt} \& {Madau}(2001)}]{Haardt01}
{Haardt} F., {Madau} P., 2001, in Clusters of Galaxies and the High Redshift
  Universe Observed in X-rays, {Neumann} D.~M., {Tran} J.~T.~V., eds.

\bibitem[{Hani {et~al}\mbox{.}(2017)Hani, Sparre, Ellison, Torrey, \&
  Vogelsberger}]{Hani17}
Hani M.~H., Sparre M., Ellison S.~L., Torrey P., Vogelsberger M., 2017, MNRAS,
  stx3252

\bibitem[{{Heckman} {et~al}\mbox{.}(2017){Heckman}, {Borthakur}, {Wild},
  {Schiminovich}, \& {Bordoloi}}]{Heckman17}
{Heckman} T., {Borthakur} S., {Wild} V., {Schiminovich} D., {Bordoloi} R.,
  2017, ArXiv e-prints

\bibitem[{{Heckman}(1980)}]{Heckman80}
{Heckman} T.~M., 1980, \aap, 87, 152

\bibitem[{{Hennawi} \& {Prochaska}(2007)}]{QPQ2}
{Hennawi} J.~F., {Prochaska} J.~X., 2007, \apj, 655, 735

\bibitem[{{Hennawi} {et~al}\mbox{.}(2006){Hennawi}, {Prochaska}, {Burles},
  {Strauss}, {Richards}, {Schlegel}, {Fan}, {Schneider}, {Zakamska}, {Oguri},
  {Gunn}, {Lupton}, \& {Brinkmann}}]{QPQ1}
{Hennawi} J.~F. {et~al.}, 2006, \apj, 651, 61

\bibitem[{{Ho} {et~al}\mbox{.}(1997){Ho}, {Filippenko}, \& {Sargent}}]{Ho97}
{Ho} L.~C., {Filippenko} A.~V., {Sargent} W.~L.~W., 1997, \apj, 487, 591

\bibitem[{{Ho} {et~al}\mbox{.}(2017){Ho}, {Martin}, {Kacprzak}, \&
  {Churchill}}]{Ho17}
{Ho} S.~H., {Martin} C.~L., {Kacprzak} G.~G., {Churchill} C.~W., 2017, \apj,
  835, 267

\bibitem[{{Ishibashi} \& {Fabian}(2015)}]{Ishibashi15}
{Ishibashi} W., {Fabian} A.~C., 2015, \mnras, 451, 93

\bibitem[{{Johnson} {et~al}\mbox{.}(2015){Johnson}, {Chen}, \&
  {Mulchaey}}]{Johnson15}
{Johnson} S.~D., {Chen} H.-W., {Mulchaey} J.~S., 2015, \mnras, 452, 2553

\bibitem[{{Johnson} {et~al}\mbox{.}(2014){Johnson}, {Chen}, {Mulchaey},
  {Tripp}, {Prochaska}, \& {Werk}}]{Johnson14}
{Johnson} S.~D., {Chen} H.-W., {Mulchaey} J.~S., {Tripp} T.~M., {Prochaska}
  J.~X., {Werk} J.~K., 2014, \mnras, 438, 3039

\bibitem[{{Kacprzak} {et~al}\mbox{.}(2015){Kacprzak}, {Churchill}, {Murphy}, \&
  {Cooke}}]{Kacprzak15}
{Kacprzak} G.~G., {Churchill} C.~W., {Murphy} M.~T., {Cooke} J., 2015, \mnras,
  446, 2861

\bibitem[{{Kacprzak} {et~al}\mbox{.}(2012){Kacprzak}, {Churchill}, \&
  {Nielsen}}]{Kacprzak12}
{Kacprzak} G.~G., {Churchill} C.~W., {Nielsen} N.~M., 2012, \apjl, 760, L7

\bibitem[{{Kauffmann} \& {Heckman}(2009)}]{Kauffmann09}
{Kauffmann} G., {Heckman} T.~M., 2009, \mnras, 397, 135

\bibitem[{{Kauffmann} {et~al}\mbox{.}(2003){Kauffmann}, {Heckman}, {Tremonti},
  {Brinchmann}, {Charlot}, {White}, {Ridgway}, {Brinkmann}, {Fukugita}, {Hall},
  {Ivezi{\'c}}, {Richards}, \& {Schneider}}]{Kauffmann03}
{Kauffmann} G. {et~al.}, 2003, \mnras, 346, 1055

\bibitem[{{Kauffmann} {et~al}\mbox{.}(2017){Kauffmann}, {Nelson}, {M{\'e}nard},
  \& {Zhu}}]{Kauffmann17}
{Kauffmann} G., {Nelson} D., {M{\'e}nard} B., {Zhu} G., 2017, \mnras, 468, 3737

\bibitem[{{Keel} {et~al}\mbox{.}(2012){Keel}, {Chojnowski}, {Bennert},
  {Schawinski}, {Lintott}, {Lynn}, {Pancoast}, {Harris}, {Nierenberg},
  {Sonnenfeld}, \& {Proctor}}]{Keel12}
{Keel} W.~C. {et~al.}, 2012, \mnras, 420, 878

\bibitem[{{Kennicutt}(1998)}]{Kennicutt98}
{Kennicutt}, Jr. R.~C., 1998, \apj, 498, 541

\bibitem[{{Kewley} {et~al}\mbox{.}(2006){Kewley}, {Groves}, {Kauffmann}, \&
  {Heckman}}]{Kewley06}
{Kewley} L.~J., {Groves} B., {Kauffmann} G., {Heckman} T., 2006, \mnras, 372,
  961

\bibitem[{{Kewley} {et~al}\mbox{.}(2001){Kewley}, {Heisler}, {Dopita}, \&
  {Lumsden}}]{Kewley01}
{Kewley} L.~J., {Heisler} C.~A., {Dopita} M.~A., {Lumsden} S., 2001, \apjs,
  132, 37

\bibitem[{{Kollmeier} {et~al}\mbox{.}(2014){Kollmeier}, {Weinberg},
  {Oppenheimer}, {Haardt}, {Katz}, {Dav{\'e}}, {Fardal}, {Madau}, {Danforth},
  {Ford}, {Peeples}, \& {McEwen}}]{Kollmeier14}
{Kollmeier} J.~A. {et~al.}, 2014, \apjl, 789, L32

\bibitem[{{Lanzetta} {et~al}\mbox{.}(1995){Lanzetta}, {Bowen}, {Tytler}, \&
  {Webb}}]{Lanzetta95}
{Lanzetta} K.~M., {Bowen} D.~V., {Tytler} D., {Webb} J.~K., 1995, \apj, 442,
  538

\bibitem[{{Leslie} {et~al}\mbox{.}(2016){Leslie}, {Kewley}, {Sanders}, \&
  {Lee}}]{Leslie16}
{Leslie} S.~K., {Kewley} L.~J., {Sanders} D.~B., {Lee} N., 2016, \mnras, 455,
  L82

\bibitem[{{Martin} {et~al}\mbox{.}(2005){Martin}, {Fanson}, {Schiminovich},
  {Morrissey}, {Friedman}, {Barlow}, {Conrow}, {Grange}, {Jelinsky},
  {Milliard}, {Siegmund}, {Bianchi}, {Byun}, {Donas}, {Forster}, {Heckman},
  {Lee}, {Madore}, {Malina}, {Neff}, {Rich}, {Small}, {Surber}, {Szalay},
  {Welsh}, \& {Wyder}}]{Martin05}
{Martin} D.~C. {et~al.}, 2005, \apjl, 619, L1

\bibitem[{{McNamara} \& {Nulsen}(2007)}]{McNamara07}
{McNamara} B.~R., {Nulsen} P.~E.~J., 2007, \araa, 45, 117

\bibitem[{{Mendel} {et~al}\mbox{.}(2014){Mendel}, {Simard}, {Palmer},
  {Ellison}, \& {Patton}}]{Mendel14}
{Mendel} J.~T., {Simard} L., {Palmer} M., {Ellison} S.~L., {Patton} D.~R.,
  2014, \apjs, 210, 3

\bibitem[{{Moster} {et~al}\mbox{.}(2010){Moster}, {Somerville}, {Maulbetsch},
  {van den Bosch}, {Macci{\`o}}, {Naab}, \& {Oser}}]{Moster10}
{Moster} B.~P., {Somerville} R.~S., {Maulbetsch} C., {van den Bosch} F.~C.,
  {Macci{\`o}} A.~V., {Naab} T., {Oser} L., 2010, \apj, 710, 903

\bibitem[{{Muzahid} {et~al}\mbox{.}(2015){Muzahid}, {Kacprzak}, {Churchill},
  {Charlton}, {Nielsen}, {Mathes}, \& {Trujillo-Gomez}}]{Muzahid15}
{Muzahid} S., {Kacprzak} G.~G., {Churchill} C.~W., {Charlton} J.~C., {Nielsen}
  N.~M., {Mathes} N.~L., {Trujillo-Gomez} S., 2015, \apj, 811, 132

\bibitem[{{Navarro} {et~al}\mbox{.}(1997){Navarro}, {Frenk}, \& {White}}]{NFW}
{Navarro} J.~F., {Frenk} C.~S., {White} S.~D.~M., 1997, \apj, 490, 493

\bibitem[{{Nedelchev} {et~al}\mbox{.}(2017){Nedelchev}, {Sarzi}, \&
  {Kaviraj}}]{Nedelchev17}
{Nedelchev} B., {Sarzi} M., {Kaviraj} S., 2017, ArXiv e-prints

\bibitem[{{Nelson} {et~al}\mbox{.}(2017){Nelson}, {Kauffmann}, {Pillepich},
  {Genel}, {Springel}, {Pakmor}, {Hernquist}, {Weinberger}, {Torrey},
  {Vogelsberger}, \& {Marinacci}}]{Nelson17}
{Nelson} D. {et~al.}, 2017, ArXiv e-prints

\bibitem[{{Nielsen} {et~al}\mbox{.}(2015){Nielsen}, {Churchill}, {Kacprzak},
  {Murphy}, \& {Evans}}]{Nielsen15}
{Nielsen} N.~M., {Churchill} C.~W., {Kacprzak} G.~G., {Murphy} M.~T., {Evans}
  J.~L., 2015, \apj, 812, 83

\bibitem[{{Oppenheimer} {et~al}\mbox{.}(2016){Oppenheimer}, {Crain}, {Schaye},
  {Rahmati}, {Richings}, {Trayford}, {Tumlinson}, {Bower}, {Schaller}, \&
  {Theuns}}]{Oppenheimer16}
{Oppenheimer} B.~D. {et~al.}, 2016, \mnras, 460, 2157

\bibitem[{{Oppenheimer} \& {Schaye}(2013)}]{Oppenheimer13}
{Oppenheimer} B.~D., {Schaye} J., 2013, \mnras, 434, 1063

\bibitem[{{Oppenheimer} {et~al}\mbox{.}(2017){Oppenheimer}, {Schaye}, {Crain},
  {Werk}, \& {Richings}}]{Oppenheimer17}
{Oppenheimer} B.~D., {Schaye} J., {Crain} R.~A., {Werk} J.~K., {Richings}
  A.~J., 2017, ArXiv e-prints

\bibitem[{{Oppenheimer} {et~al}\mbox{.}(2018){Oppenheimer}, {Segers}, {Schaye},
  {Richings}, \& {Crain}}]{Oppenheimer18}
{Oppenheimer} B.~D., {Segers} M., {Schaye} J., {Richings} A.~J., {Crain} R.~A.,
  2018, \mnras, 474, 4740

\bibitem[{{Patton} {et~al}\mbox{.}(2016){Patton}, {Qamar}, {Ellison}, {Bluck},
  {Simard}, {Mendel}, {Moreno}, \& {Torrey}}]{Patton16}
{Patton} D.~R., {Qamar} F.~D., {Ellison} S.~L., {Bluck} A.~F.~L., {Simard} L.,
  {Mendel} J.~T., {Moreno} J., {Torrey} P., 2016, \mnras, 461, 2589

\bibitem[{{Perna} {et~al}\mbox{.}(2017){Perna}, {Lanzuisi}, {Brusa}, {Mignoli},
  \& {Cresci}}]{Perna17}
{Perna} M., {Lanzuisi} G., {Brusa} M., {Mignoli} M., {Cresci} G., 2017, \aap,
  603, A99

\bibitem[{{Planck Collaboration} {et~al}\mbox{.}(2015){Planck Collaboration},
  {Aghanim}, {Arnaud}, {Ashdown}, {Aumont}, {Baccigalupi}, {Banday},
  {Barreiro}, {Bartlett}, {Bartolo}, \& et~al.}]{Planck15}
{Planck Collaboration} {et~al.}, 2015, ArXiv e-prints

\bibitem[{{Prochaska} {et~al}\mbox{.}(2013){Prochaska}, {Hennawi}, {Lee},
  {Cantalupo}, {Bovy}, {Djorgovski}, {Ellison}, {Lau}, {Martin}, {Myers},
  {Rubin}, \& {Simcoe}}]{QPQ6}
{Prochaska} J.~X. {et~al.}, 2013, \apj, 776, 136

\bibitem[{{Prochaska} {et~al}\mbox{.}(2014){Prochaska}, {Lau}, \&
  {Hennawi}}]{QPQ7}
{Prochaska} J.~X., {Lau} M.~W., {Hennawi} J.~F., 2014, \apj, 796, 140

\bibitem[{{Prochaska} {et~al}\mbox{.}(2011){Prochaska}, {Weiner}, {Chen},
  {Mulchaey}, \& {Cooksey}}]{Prochaska11}
{Prochaska} J.~X., {Weiner} B., {Chen} H.-W., {Mulchaey} J., {Cooksey} K.,
  2011, \apj, 740, 91

\bibitem[{{Putman} {et~al}\mbox{.}(2012){Putman}, {Peek}, \&
  {Joung}}]{Putman12}
{Putman} M.~E., {Peek} J.~E.~G., {Joung} M.~R., 2012, \araa, 50, 491

\bibitem[{{Rahmati} {et~al}\mbox{.}(2015){Rahmati}, {Schaye}, {Bower}, {Crain},
  {Furlong}, {Schaller}, \& {Theuns}}]{Rahmati15}
{Rahmati} A., {Schaye} J., {Bower} R.~G., {Crain} R.~A., {Furlong} M.,
  {Schaller} M., {Theuns} T., 2015, \mnras, 452, 2034

\bibitem[{{Rao} {et~al}\mbox{.}(2006){Rao}, {Turnshek}, \& {Nestor}}]{Rao06}
{Rao} S.~M., {Turnshek} D.~A., {Nestor} D.~B., 2006, \apj, 636, 610

\bibitem[{{Richings} {et~al}\mbox{.}(2014){Richings}, {Schaye}, \&
  {Oppenheimer}}]{Richings14a}
{Richings} A.~J., {Schaye} J., {Oppenheimer} B.~D., 2014, \mnras, 440, 3349

\bibitem[{{Rosario} {et~al}\mbox{.}(2016){Rosario}, {Mendel}, {Ellison},
  {Lutz}, \& {Trump}}]{Rosario16}
{Rosario} D.~J., {Mendel} J.~T., {Ellison} S.~L., {Lutz} D., {Trump} J.~R.,
  2016, \mnras, 457, 2703

\bibitem[{{Rubin} {et~al}\mbox{.}(2014){Rubin}, {Prochaska}, {Koo}, {Phillips},
  {Martin}, \& {Winstrom}}]{Rubin14}
{Rubin} K.~H.~R., {Prochaska} J.~X., {Koo} D.~C., {Phillips} A.~C., {Martin}
  C.~L., {Winstrom} L.~O., 2014, \apj, 794, 156

\bibitem[{{Salim} {et~al}\mbox{.}(2007){Salim}, {Rich}, {Charlot},
  {Brinchmann}, {Johnson}, {Schiminovich}, {Seibert}, {Mallery}, {Heckman},
  {Forster}, {Friedman}, {Martin}, {Morrissey}, {Neff}, {Small}, {Wyder},
  {Bianchi}, {Donas}, {Lee}, {Madore}, {Milliard}, {Szalay}, {Welsh}, \&
  {Yi}}]{Salim07}
{Salim} S. {et~al.}, 2007, \apjs, 173, 267

\bibitem[{{Satyapal} {et~al}\mbox{.}(2014){Satyapal}, {Ellison}, {McAlpine},
  {Hickox}, {Patton}, \& {Mendel}}]{Satyapal14}
{Satyapal} S., {Ellison} S.~L., {McAlpine} W., {Hickox} R.~C., {Patton} D.~R.,
  {Mendel} J.~T., 2014, \mnras, 441, 1297

\bibitem[{{Sazonov} {et~al}\mbox{.}(2004){Sazonov}, {Ostriker}, \&
  {Sunyaev}}]{Sazonov04}
{Sazonov} S.~Y., {Ostriker} J.~P., {Sunyaev} R.~A., 2004, \mnras, 347, 144

\bibitem[{{Schawinski} {et~al}\mbox{.}(2007){Schawinski}, {Thomas}, {Sarzi},
  {Maraston}, {Kaviraj}, {Joo}, {Yi}, \& {Silk}}]{Schawinski07}
{Schawinski} K., {Thomas} D., {Sarzi} M., {Maraston} C., {Kaviraj} S., {Joo}
  S.-J., {Yi} S.~K., {Silk} J., 2007, \mnras, 382, 1415

\bibitem[{{Schaye} {et~al}\mbox{.}(2003){Schaye}, {Aguirre}, {Kim}, {Theuns},
  {Rauch}, \& {Sargent}}]{Schaye03}
{Schaye} J., {Aguirre} A., {Kim} T.-S., {Theuns} T., {Rauch} M., {Sargent}
  W.~L.~W., 2003, \apj, 596, 768

\bibitem[{{Schaye} {et~al}\mbox{.}(2007){Schaye}, {Carswell}, \&
  {Kim}}]{Schaye07}
{Schaye} J., {Carswell} R.~F., {Kim} T.-S., 2007, \mnras, 379, 1169

\bibitem[{{Schaye} {et~al}\mbox{.}(2015){Schaye}, {Crain}, {Bower}, {Furlong},
  {Schaller}, {Theuns}, {Dalla Vecchia}, {Frenk}, {McCarthy}, {Helly},
  {Jenkins}, {Rosas-Guevara}, {White}, {Baes}, {Booth}, {Camps}, {Navarro},
  {Qu}, {Rahmati}, {Sawala}, {Thomas}, \& {Trayford}}]{Schaye15}
{Schaye} J. {et~al.}, 2015, \mnras, 446, 521

\bibitem[{{Schirber} {et~al}\mbox{.}(2004){Schirber}, {Miralda-Escud{\'e}}, \&
  {McDonald}}]{Schirber04}
{Schirber} M., {Miralda-Escud{\'e}} J., {McDonald} P., 2004, \apj, 610, 105

\bibitem[{{Segers} {et~al}\mbox{.}(2017){Segers}, {Oppenheimer}, {Schaye}, \&
  {Richings}}]{Segers17}
{Segers} M.~C., {Oppenheimer} B.~D., {Schaye} J., {Richings} A.~J., 2017, ArXiv
  e-prints

\bibitem[{{Silverman} {et~al}\mbox{.}(2014){Silverman}, {Miniati},
  {Finoguenov}, {Carollo}, {Cibinel}, {Lilly}, \& {Schawinski}}]{Silverman14}
{Silverman} J.~D., {Miniati} F., {Finoguenov} A., {Carollo} C.~M., {Cibinel}
  A., {Lilly} S.~J., {Schawinski} K., 2014, \apj, 780, 67

\bibitem[{{Springel} {et~al}\mbox{.}(2005){Springel}, {Di Matteo}, \&
  {Hernquist}}]{Springel05}
{Springel} V., {Di Matteo} T., {Hernquist} L., 2005, \apjl, 620, L79

\bibitem[{{Steidel} {et~al}\mbox{.}(2010){Steidel}, {Erb}, {Shapley},
  {Pettini}, {Reddy}, {Bogosavljevi{\'c}}, {Rudie}, \& {Rakic}}]{Steidel10}
{Steidel} C.~C., {Erb} D.~K., {Shapley} A.~E., {Pettini} M., {Reddy} N.,
  {Bogosavljevi{\'c}} M., {Rudie} G.~C., {Rakic} O., 2010, \apj, 717, 289

\bibitem[{{Stinson} {et~al}\mbox{.}(2012){Stinson}, {Brook}, {Prochaska},
  {Hennawi}, {Shen}, {Wadsley}, {Pontzen}, {Couchman}, {Quinn}, {Macci{\`o}},
  \& {Gibson}}]{Stinson12}
{Stinson} G.~S. {et~al.}, 2012, \mnras, 425, 1270

\bibitem[{{Stocke} {et~al}\mbox{.}(2014){Stocke}, {Keeney}, {Danforth},
  {Syphers}, {Yamamoto}, {Shull}, {Green}, {Froning}, {Savage}, {Wakker},
  {Kim}, {Ryan-Weber}, \& {Kacprzak}}]{Stocke14}
{Stocke} J.~T. {et~al.}, 2014, \apj, 791, 128

\bibitem[{{Sturm} {et~al}\mbox{.}(2011){Sturm}, {Gonz{\'a}lez-Alfonso},
  {Veilleux}, {Fischer}, {Graci{\'a}-Carpio}, {Hailey-Dunsheath}, {Contursi},
  {Poglitsch}, {Sternberg}, {Davies}, {Genzel}, {Lutz}, {Tacconi}, {Verma},
  {Maiolino}, \& {de Jong}}]{Sturm11}
{Sturm} E. {et~al.}, 2011, \apjl, 733, L16

\bibitem[{{Theuns} {et~al}\mbox{.}(1998){Theuns}, {Leonard}, \&
  {Efstathiou}}]{Theuns98}
{Theuns} T., {Leonard} A., {Efstathiou} G., 1998, \mnras, 297, L49

\bibitem[{{Thom} {et~al}\mbox{.}(2012){Thom}, {Tumlinson}, {Werk}, {Prochaska},
  {Oppenheimer}, {Peeples}, {Tripp}, {Katz}, {O'Meara}, {Ford}, {Dav{\'e}},
  {Sembach}, \& {Weinberg}}]{Thom12}
{Thom} C. {et~al.}, 2012, \apjl, 758, L41

\bibitem[{{Tremonti} {et~al}\mbox{.}(2007){Tremonti}, {Moustakas}, \&
  {Diamond-Stanic}}]{Tremonti07}
{Tremonti} C.~A., {Moustakas} J., {Diamond-Stanic} A.~M., 2007, \apjl, 663, L77

\bibitem[{{Tumlinson} {et~al}\mbox{.}(2017){Tumlinson}, {Peeples}, \&
  {Werk}}]{Tumlinson17}
{Tumlinson} J., {Peeples} M.~S., {Werk} J.~K., 2017, \araa, 55, 389

\bibitem[{{Tumlinson} {et~al}\mbox{.}(2013){Tumlinson}, {Thom}, {Werk},
  {Prochaska}, {Tripp}, {Katz}, {Dav{\'e}}, {Oppenheimer}, {Meiring}, {Ford},
  {O'Meara}, {Peeples}, {Sembach}, \& {Weinberg}}]{Tumlinson13}
{Tumlinson} J. {et~al.}, 2013, \apj, 777, 59

\bibitem[{{Tumlinson} {et~al}\mbox{.}(2011){Tumlinson}, {Thom}, {Werk},
  {Prochaska}, {Tripp}, {Weinberg}, {Peeples}, {O'Meara}, {Oppenheimer},
  {Meiring}, {Katz}, {Dav{\'e}}, {Ford}, \& {Sembach}}]{Tumlinson11}
{Tumlinson} J. {et~al.}, 2011, Science, 334, 948

\bibitem[{{Turner} {et~al}\mbox{.}(2014){Turner}, {Schaye}, {Steidel}, {Rudie},
  \& {Strom}}]{Turner14}
{Turner} M.~L., {Schaye} J., {Steidel} C.~C., {Rudie} G.~C., {Strom} A.~L.,
  2014, \mnras, 445, 794

\bibitem[{{Turner} {et~al}\mbox{.}(2015){Turner}, {Schaye}, {Steidel}, {Rudie},
  \& {Strom}}]{Turner15}
{Turner} M.~L., {Schaye} J., {Steidel} C.~C., {Rudie} G.~C., {Strom} A.~L.,
  2015, \mnras, 450, 2067

\bibitem[{{Veilleux} {et~al}\mbox{.}(2005){Veilleux}, {Cecil}, \&
  {Bland-Hawthorn}}]{Veilleux05}
{Veilleux} S., {Cecil} G., {Bland-Hawthorn} J., 2005, \araa, 43, 769

\bibitem[{{Veilleux} {et~al}\mbox{.}(2013){Veilleux}, {Mel{\'e}ndez}, {Sturm},
  {Gracia-Carpio}, {Fischer}, {Gonz{\'a}lez-Alfonso}, {Contursi}, {Lutz},
  {Poglitsch}, {Davies}, {Genzel}, {Tacconi}, {de Jong}, {Sternberg}, {Netzer},
  {Hailey-Dunsheath}, {Verma}, {Rupke}, {Maiolino}, {Teng}, \&
  {Polisensky}}]{Veilleux13}
{Veilleux} S. {et~al.}, 2013, \apj, 776, 27

\bibitem[{{Vito} {et~al}\mbox{.}(2014){Vito}, {Maiolino}, {Santini}, {Brusa},
  {Comastri}, {Cresci}, {Farrah}, {Franceschini}, {Gilli}, {Granato},
  {Gruppioni}, {Lutz}, {Mannucci}, {Pozzi}, {Rosario}, {Scott}, {Viero}, \&
  {Vignali}}]{Vito14}
{Vito} F. {et~al.}, 2014, \mnras, 441, 1059

\bibitem[{{Werk} {et~al}\mbox{.}(2012){Werk}, {Prochaska}, {Thom}, {Tumlinson},
  {Tripp}, {O'Meara}, \& {Meiring}}]{Werk12}
{Werk} J.~K., {Prochaska} J.~X., {Thom} C., {Tumlinson} J., {Tripp} T.~M.,
  {O'Meara} J.~M., {Meiring} J.~D., 2012, \apjs, 198, 3

\bibitem[{{Werk} {et~al}\mbox{.}(2013){Werk}, {Prochaska}, {Thom}, {Tumlinson},
  {Tripp}, {O'Meara}, \& {Peeples}}]{Werk13}
{Werk} J.~K., {Prochaska} J.~X., {Thom} C., {Tumlinson} J., {Tripp} T.~M.,
  {O'Meara} J.~M., {Peeples} M.~S., 2013, \apjs, 204, 17

\bibitem[{{Werk} {et~al}\mbox{.}(2014){Werk}, {Prochaska}, {Tumlinson},
  {Peeples}, {Tripp}, {Fox}, {Lehner}, {Thom}, {O'Meara}, {Ford}, {Bordoloi},
  {Katz}, {Tejos}, {Oppenheimer}, {Dav{\'e}}, \& {Weinberg}}]{Werk14}
{Werk} J.~K. {et~al.}, 2014, \apj, 792, 8

\bibitem[{{Wild} {et~al}\mbox{.}(2010){Wild}, {Heckman}, \& {Charlot}}]{Wild10}
{Wild} V., {Heckman} T., {Charlot} S., 2010, \mnras, 405, 933

\bibitem[{{Wild} {et~al}\mbox{.}(2007){Wild}, {Kauffmann}, {Heckman},
  {Charlot}, {Lemson}, {Brinchmann}, {Reichard}, \& {Pasquali}}]{Wild07}
{Wild} V., {Kauffmann} G., {Heckman} T., {Charlot} S., {Lemson} G.,
  {Brinchmann} J., {Reichard} T., {Pasquali} A., 2007, \mnras, 381, 543

\bibitem[{{Woo} {et~al}\mbox{.}(2016){Woo}, {Bae}, {Son}, \&
  {Karouzos}}]{Woo16}
{Woo} J.-H., {Bae} H.-J., {Son} D., {Karouzos} M., 2016, \apj, 817, 108

\bibitem[{{Woo} {et~al}\mbox{.}(2017){Woo}, {Son}, \& {Bae}}]{Woo17}
{Woo} J.-H., {Son} D., {Bae} H.-J., 2017, \apj, 839, 120

\bibitem[{{Yan} \& {Blanton}(2012)}]{Yan12}
{Yan} R., {Blanton} M.~R., 2012, \apj, 747, 61

\bibitem[{{Yesuf} {et~al}\mbox{.}(2014){Yesuf}, {Faber}, {Trump}, {Koo},
  {Fang}, {Liu}, {Wild}, \& {Hayward}}]{Yesuf14}
{Yesuf} H.~M., {Faber} S.~M., {Trump} J.~R., {Koo} D.~C., {Fang} J.~J., {Liu}
  F.~S., {Wild} V., {Hayward} C.~C., 2014, \apj, 792, 84

\end{thebibliography}
\appendix

\section{Velocity profiles and equivalent widths}
\label{App:EWs}

In this Appendix, we present the velocity profiles and measured EWs for the COS-AGN sample on a sightline by sightline basis. Table \ref{tab:all_ews} contains a summary of all adopted EW values. For the five sightlines with absorption lines that appear marginally contaminated but could very well be real, we have justified why we do not adopt the measured EWs below.

\subsection*{J0042$-$1037 (z$_{\rm gal}$=0.036)}
\begin{figure*}
\begin{center}
\includegraphics[width=\textwidth]{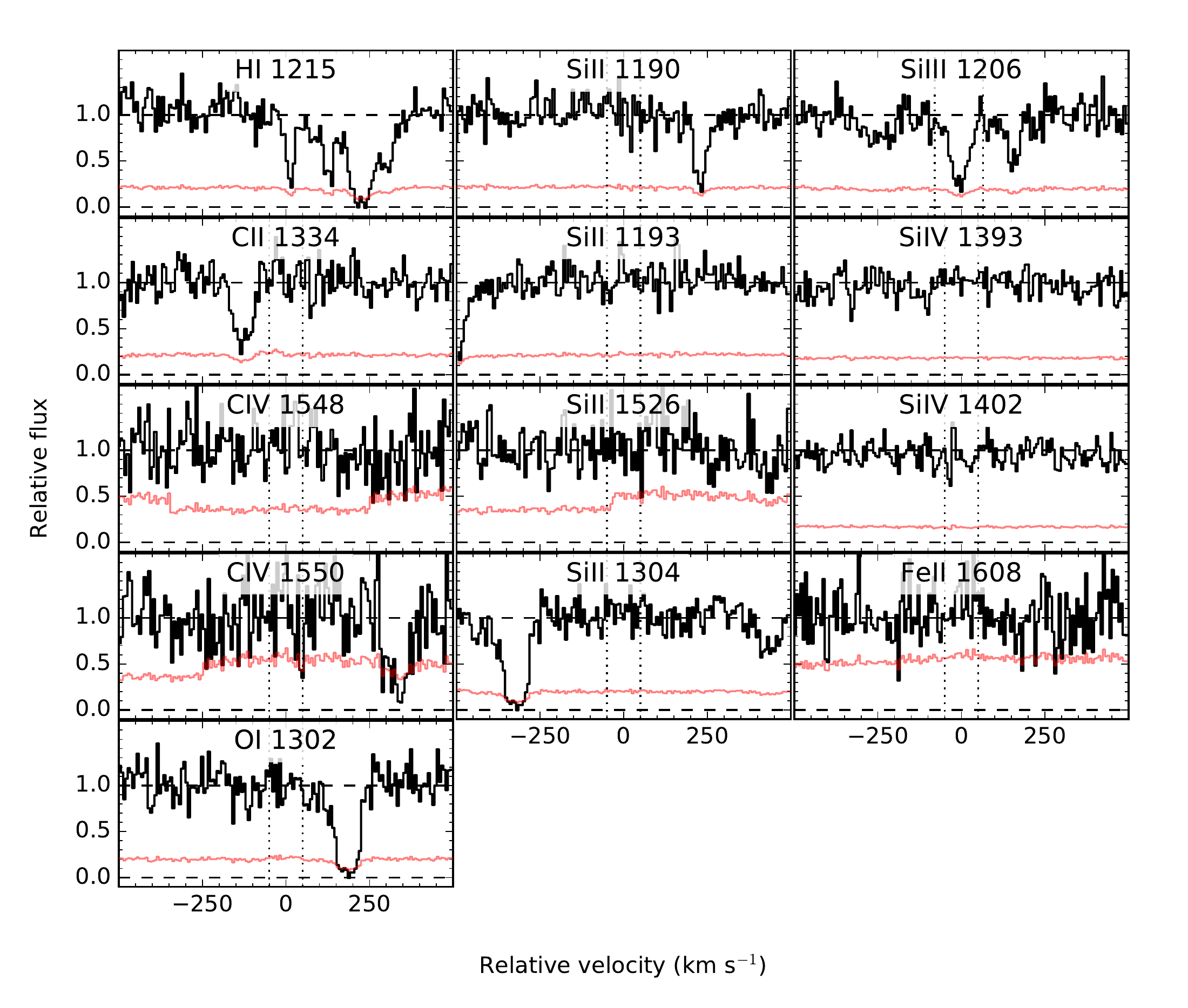}
\caption{Velocity profiles for the sightline towards J0042$-$1037 (z$_{\rm gal}$=0.036).}
\label{fig:J0042}
\end{center}
\end{figure*}

The region of the Ly$\alpha$ absorption  towards J0042$-$1037 at the redshift of the AGN host (z$_{\rm gal}$=0.036) is heavily contaminated by Milky Way S\ion{ii} 1259 ~\AA{} and Si\ion{ii} 1260~\AA{} absorption lines (see Figure \ref{fig:J0042}). As a result, we cannot constrain the amount of Ly$\alpha$ absorption in the CGM around this galaxy. The presence of a strong Si\ion{iii} 1206~\AA{} line with minimal Ly$\alpha$  absorption is extremely unlikely, thus we do not trust the absorption present at the location of Si\ion{iii} 1206~\AA{} (Figure \ref{fig:J0042}) to be associated with the CGM of the AGN host.

\subsection*{J1117+2634 (z$_{\rm gal}$=0.029)}
\label{sec:J1117}
The CGM around the AGN host towards the QSO J1117+2634 (z$_{\rm gal}$=0.029) is the only absorber  whose  Ly$\alpha$ feature is kinematically offset by a large amount ($+350$\kms{}; Figure \ref{fig:J1117,351}) relative to the host galaxy.  This feature at $+350$\kms{} was selected as the associated CGM material as it is the only absorption not associated with the Galaxy (the component at $\approx-100$\kms{} is Galactic S\ion{ii} 1250~\AA{}) or another system at a different redshift. However, we are cautious in adopting the measured EW as this gas is highly offset, with no other absorption line to confirm this offset.

\subsection*{J1155+2922 (z$_{\rm gal}$=0.046)}
We flagged the absorption towards J1155+2922 at the expected location of Fe\ion{ii} 1144~\AA{} of the AGN host (z$_{\rm gal}$=0.046) as tenuous. The strongest component of the Fe\ion{ii} 1144~\AA{} absorption is on the edge of the entire Ly$\alpha$ absorption feature. Given that the neutral CGM gas should be coincidental, and that the feature in question is uniquely very strong compared to other systems; we suspect this absorption is not from the CGM of the AGN host.

\subsection*{J1214+0825 (z$_{\rm gal}$=0.074)}
We note that for the AGN host towards J1214+0825 (z$_{\rm gal}$=0.074) that the Ly$\alpha$ EW is heavily blended with the O\ion{i} 1302~\AA{} emission line from the Galaxy, making it impossible to determine a robust measurement of the EW. As a result, we do not adopt this EW into our final catalogue, even though the absorption feature is present.

\subsection*{J2133$-$0712 (z$_{\rm gal}$=0.064)}
The apparent Si\ion{iii} 1206~\AA{} absorption for the AGN host towards J2133$-$0712 (z$_{\rm gal}$=0.064)  is offset $\approx+200$\kms{} from the Ly$\alpha$ absorption feature. Such strong velocity offsests are rarely seen for Si\ion{iii} from neutral \HI{} in the CGM. We note the possibility that this absorption feature is coincident with an  O\ion{vi} 1037~\AA{} absorber at higher redshift (z$=0.238$), which is confirmed by a tenuously detected O\ion{vi} 1031~\AA{} feature at the same redshift. As a result, we do not associate this absorption with the AGN host.

\begin{table}
\scriptsize
\begin{center}
\caption{Measured EWs for J0042-1037 195\_406 (z=0.036)}
\label{tab:J0042-1037,195_406}
\begin{tabular}{lcccccc}
\hline
Ion& $\lambda$& $f$& $v_{\rm min}$& $v_{\rm max}$& EW& flag$^{\star}$\\
& \AA{}& & \kms{}& \kms{}& m\AA{}& \\
\hline
HI& 1215.670& 4.164E-01& 85& 160& \nodata{}& 2\\
CII& 1334.532& 1.278E-01& -50& 50& $<78$& \textbf{5}\\
CIV& 1548.195& 1.908E-01& -50& 50& $<148$& \textbf{5}\\
CIV& 1550.770& 9.522E-02& -50& 50& $<230$& \textbf{5}\\
OI& 1302.168& 4.887E-02& -50& 50& $<74$& \textbf{5}\\
SiII& 1190.416& 2.502E-01& -50& 50& $<71$& \textbf{5}\\
SiII& 1193.290& 4.991E-01& -50& 50& $<69$& \textbf{5}\\
SiII& 1526.707& 1.270E-01& -50& 50& $<192$& \textbf{5}\\
SiII& 1304.370& 9.400E-02& -50& 50& $<71$& \textbf{5}\\
SiIII& 1206.500& 1.660E+00& -80& 65& $206\pm22$& \textbf{1}\\
SiIV& 1393.755& 5.280E-01& -50& 50& $<64$& \textbf{5}\\
SiIV& 1402.770& 2.620E-01& -50& 50& $<57$& \textbf{5}\\
FeII& 1608.451& 5.800E-02& -50& 50& $<244$& \textbf{5}\\
\end{tabular}
\end{center}
\end{table}

\begin{figure*}
\begin{center}
\includegraphics[width=\textwidth]{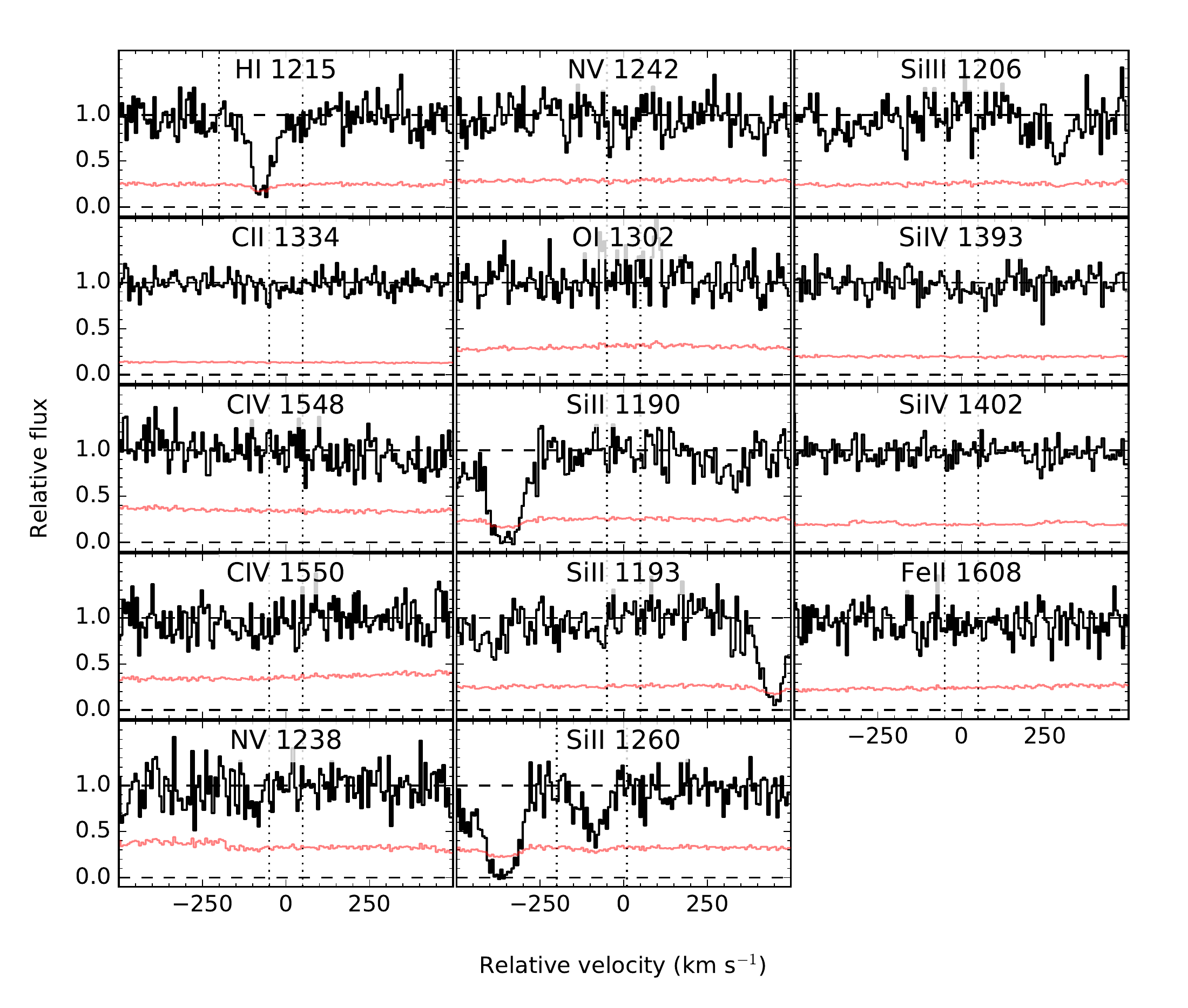}
\caption{Velocity profiles for the sightline towards J0116+1429 (z$_{\rm gal}$=0.060).}
\label{fig:J0116}
\end{center}
\end{figure*}

\begin{table}
\scriptsize
\begin{center}
\caption{Measured EWs for J0116+1429 (z=0.060)}
\label{tab:J0116+1429,114_114}
\begin{tabular}{lcccccc}
\hline
Ion& $\lambda$& $f$& $v_{\rm min}$& $v_{\rm max}$& EW& flag$^{\star}$\\
& [\AA{}]& & [\kms{}]& [\kms{}]& [m\AA{}]& \\
\hline
H\ion{i}& 1215.670& 4.164E-01& -200& 50& $306\pm38$& \textbf{9}\\
C\ion{ii}& 1334.532& 1.278E-01& -50& 50& $<45$& \textbf{5}\\
C\ion{iv}& 1548.195& 1.908E-01& -50& 50& $<138$& \textbf{5}\\
C\ion{iv}& 1550.770& 9.522E-02& -50& 50& $<141$& \textbf{5}\\
N\ion{v}& 1238.821& 1.570E-01& -50& 50& $<103$& \textbf{5}\\
N\ion{v}& 1242.804& 7.823E-02& -50& 50& $<89$& \textbf{5}\\
O\ion{i}& 1302.168& 4.887E-02& -50& 50& $<106$& \textbf{5}\\
Si\ion{ii}& 1190.416& 2.502E-01& -50& 50& $<84$& \textbf{5}\\
Si\ion{ii}& 1193.290& 4.991E-01& -50& 50& $<83$& \textbf{5}\\
Si\ion{ii}& 1260.422& 1.007E+00& -200& 10& $<190$& \textbf{3}\\
Si\ion{iii}& 1206.500& 1.660E+00& -50& 50& $<85$& \textbf{5}\\
Si\ion{iv}& 1393.755& 5.280E-01& -50& 50& $<72$& \textbf{5}\\
Si\ion{iv}& 1402.770& 2.620E-01& -50& 50& $<75$& \textbf{5}\\
Fe\ion{ii}& 1608.451& 5.800E-02& -50& 50& $<100$& \textbf{5}\\
\hline
\end{tabular}
\\$^{\star}$The data quality flags represent a sum of whether or not the line is:\\ adopted ($+1$), blended ($+2$), undetected ($+4$), or saturated ($+8$).\\\end{center}
\end{table}

\begin{figure*}
\begin{center}
\includegraphics[width=\textwidth]{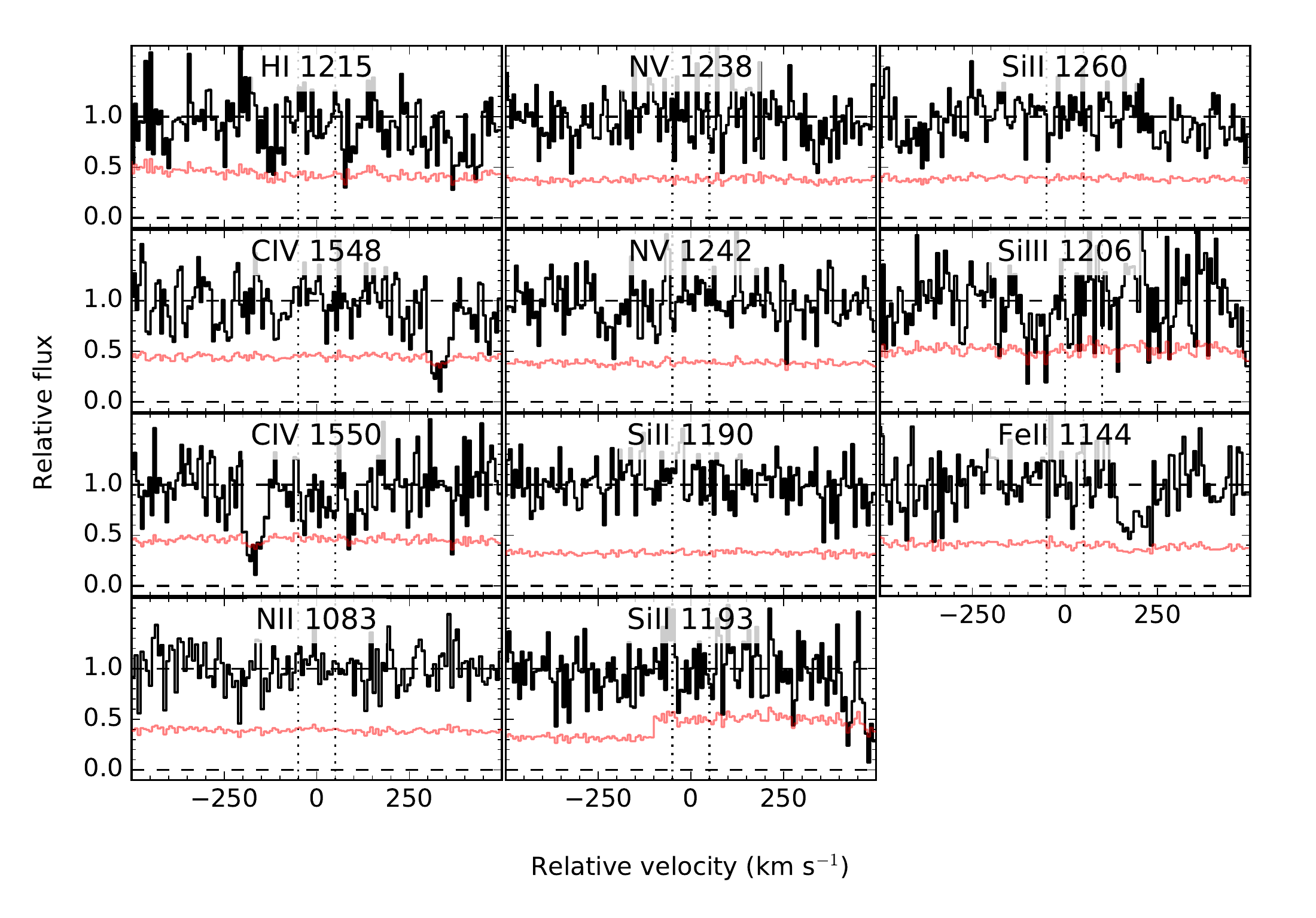}
\caption{Velocity profiles for the sightline towards J0843+4117 (z$_{\rm gal}$=0.068).}
\label{fig:J0843}
\end{center}
\end{figure*}

\begin{table}
\scriptsize
\begin{center}
\caption{Measured EWs for J0843+4117 (z=0.068)}
\label{tab:J0843+4117,287_165}
\begin{tabular}{lcccccc}
\hline
Ion& $\lambda$& $f$& $v_{\rm min}$& $v_{\rm max}$& EW& flag$^{\star}$\\
& [\AA{}]& & [\kms{}]& [\kms{}]& [m\AA{}]& \\
\hline
H\ion{i}& 1215.670& 4.164E-01& -50& 50& $<131$& \textbf{5}\\
C\ion{iv}& 1548.195& 1.908E-01& -50& 50& $<188$& \textbf{5}\\
C\ion{iv}& 1550.770& 9.522E-02& -50& 50& $<188$& \textbf{5}\\
N\ion{ii}& 1083.990& 1.031E-01& -50& 50& $<123$& \textbf{5}\\
N\ion{v}& 1238.821& 1.570E-01& -50& 50& $<124$& \textbf{5}\\
N\ion{v}& 1242.804& 7.823E-02& -50& 50& $<127$& \textbf{5}\\
Si\ion{ii}& 1190.416& 2.502E-01& -50& 50& $<104$& \textbf{5}\\
Si\ion{ii}& 1193.290& 4.991E-01& -50& 50& $<156$& \textbf{5}\\
Si\ion{ii}& 1260.422& 1.007E+00& -50& 50& $<128$& \textbf{5}\\
Si\ion{iii}& 1206.500& 1.660E+00& 0& 100& $<172$& \textbf{5}\\
Fe\ion{ii}& 1144.938& 1.060E-01& -50& 50& $<129$& \textbf{5}\\
\hline
\end{tabular}
\\$^{\star}$The data quality flags represent a sum of whether or not the line is:\\ adopted ($+1$), blended ($+2$), undetected ($+4$), or saturated ($+8$).\\\end{center}
\end{table}

\begin{figure*}
\begin{center}
\includegraphics[width=\textwidth]{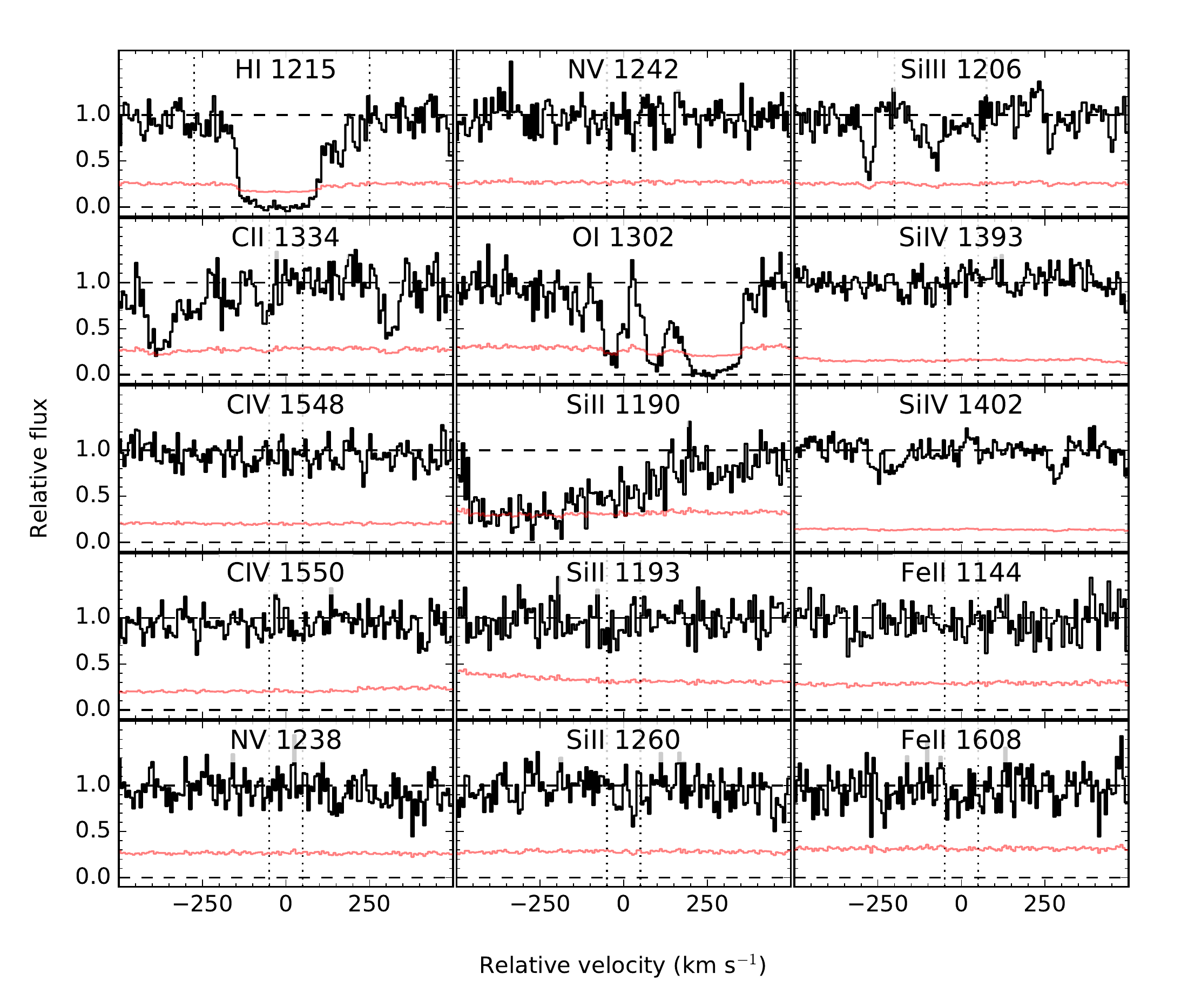}
\caption{Velocity profiles for the sightline towards J0851+4243 (z$_{\rm gal}$=0.024).}
\label{fig:J0851}
\end{center}
\end{figure*}

\begin{table}
\scriptsize
\begin{center}
\caption{Measured EWs for J0851+4243 (z=0.024)}
\label{tab:J0851+4243,285_164}
\begin{tabular}{lcccccc}
\hline
Ion& $\lambda$& $f$& $v_{\rm min}$& $v_{\rm max}$& EW& flag$^{\star}$\\
& [\AA{}]& & [\kms{}]& [\kms{}]& [m\AA{}]& \\
\hline
H\ion{i}& 1215.670& 4.164E-01& -275& 250& $1066\pm53$& \textbf{9}\\
C\ion{ii}& 1334.532& 1.278E-01& -50& 50& $<64$& \textbf{7}\\
C\ion{iv}& 1548.195& 1.908E-01& -50& 50& $<84$& \textbf{5}\\
C\ion{iv}& 1550.770& 9.522E-02& -50& 50& $<85$& \textbf{5}\\
N\ion{v}& 1238.821& 1.570E-01& -50& 50& $<92$& \textbf{5}\\
N\ion{v}& 1242.804& 7.823E-02& -50& 50& $<87$& \textbf{5}\\
O\ion{i}& 1302.168& 4.887E-02& -275& 250& \nodata{}& 2\\
Si\ion{ii}& 1190.416& 2.502E-01& -275& 250& \nodata{}& 2\\
Si\ion{ii}& 1193.290& 4.991E-01& -50& 50& $<98$& \textbf{5}\\
Si\ion{ii}& 1260.422& 1.007E+00& -50& 50& $<92$& \textbf{5}\\
Si\ion{iii}& 1206.500& 1.660E+00& -200& 75& $165\pm44$& \textbf{1}\\
Si\ion{iv}& 1393.755& 5.280E-01& -50& 50& $<56$& \textbf{5}\\
Si\ion{iv}& 1402.770& 2.620E-01& -50& 50& \nodata{}& 6\\
Fe\ion{ii}& 1144.938& 1.060E-01& -50& 50& $<91$& \textbf{5}\\
Fe\ion{ii}& 1608.451& 5.800E-02& -50& 50& $<128$& \textbf{5}\\
\hline
\end{tabular}
\\$^{\star}$The data quality flags represent a sum of whether or not the line is:\\ adopted ($+1$), blended ($+2$), undetected ($+4$), or saturated ($+8$).\\\end{center}
\end{table}

\begin{figure*}
\begin{center}
\includegraphics[width=\textwidth]{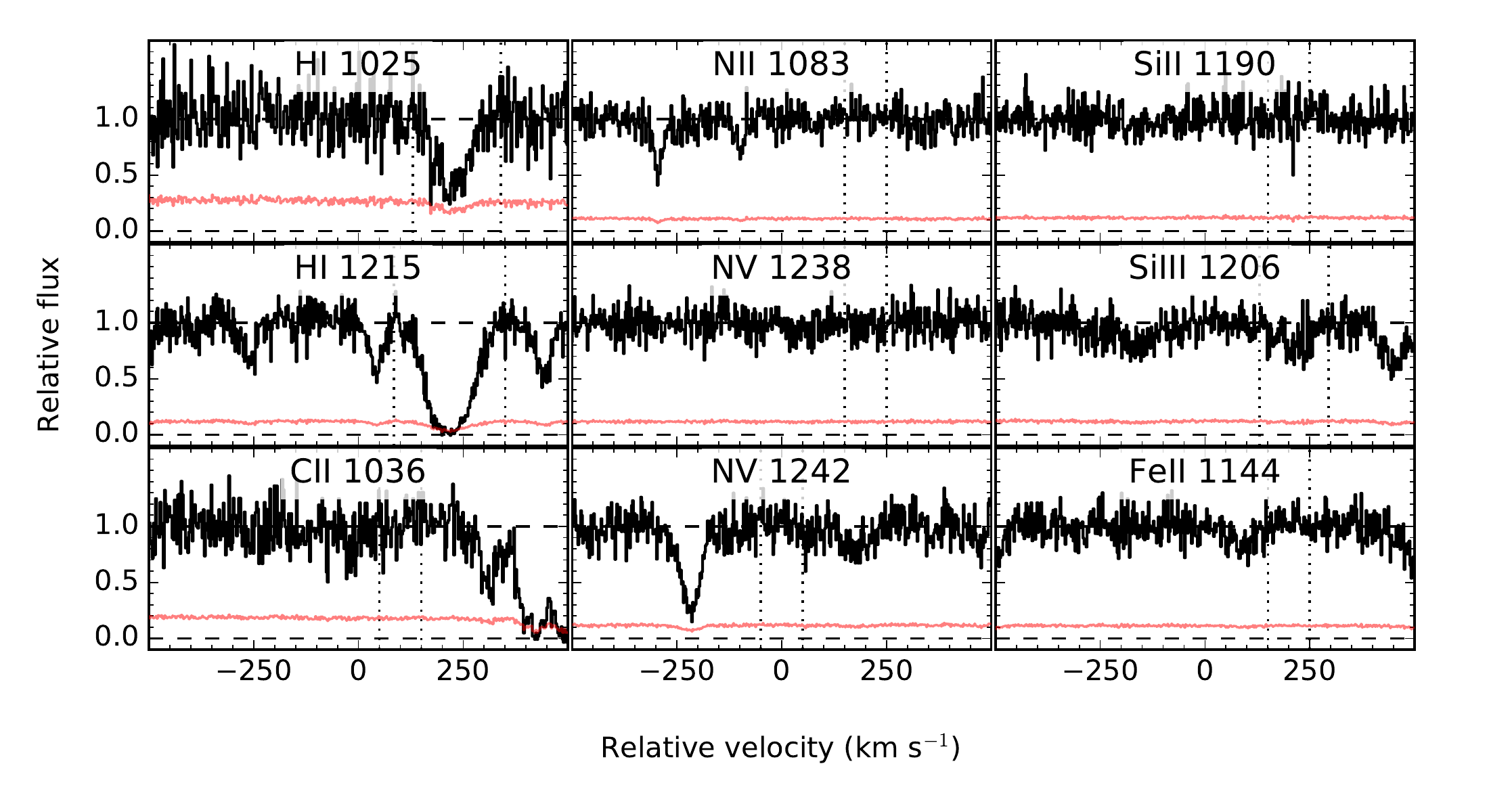}
\caption{Velocity profiles for the sightline towards J0853+4349 (z$_{\rm gal}$=0.090).}
\label{fig:J0853}
\end{center}
\end{figure*}

\begin{table}
\scriptsize
\begin{center}
\caption{Measured EWs for J0853+4349 (z=0.090)}
\label{tab:J0853+4349,275_94}
\begin{tabular}{lcccccc}
\hline
Ion& $\lambda$& $f$& $v_{\rm min}$& $v_{\rm max}$& EW& flag$^{\star}$\\
& [\AA{}]& & [\kms{}]& [\kms{}]& [m\AA{}]& \\
\hline
H\ion{i}& 1025.722& 7.912E-02& 130& 340& $185\pm19$& \textbf{9}\\
H\ion{i}& 1215.670& 4.164E-01& 85& 350& $512\pm9$& \textbf{9}\\
C\ion{ii}& 1036.337& 1.231E-01& 50& 150& $<30$& \textbf{5}\\
N\ion{ii}& 1083.990& 1.031E-01& 150& 250& $<20$& \textbf{5}\\
N\ion{v}& 1238.821& 1.570E-01& 150& 250& $<21$& \textbf{5}\\
N\ion{v}& 1242.804& 7.823E-02& -50& 50& $<22$& \textbf{5}\\
Si\ion{ii}& 1190.416& 2.502E-01& 150& 250& $<22$& \textbf{5}\\
Si\ion{iii}& 1206.500& 1.660E+00& 130& 295& $71\pm9$& \textbf{1}\\
Fe\ion{ii}& 1144.938& 1.060E-01& 150& 250& $<20$& \textbf{5}\\
\hline
\end{tabular}
\\$^{\star}$The data quality flags represent a sum of whether or not the line is:\\ adopted ($+1$), blended ($+2$), undetected ($+4$), or saturated ($+8$).\\\end{center}
\end{table}

\clearpage

\begin{figure*}
\begin{center}
\includegraphics[width=\textwidth]{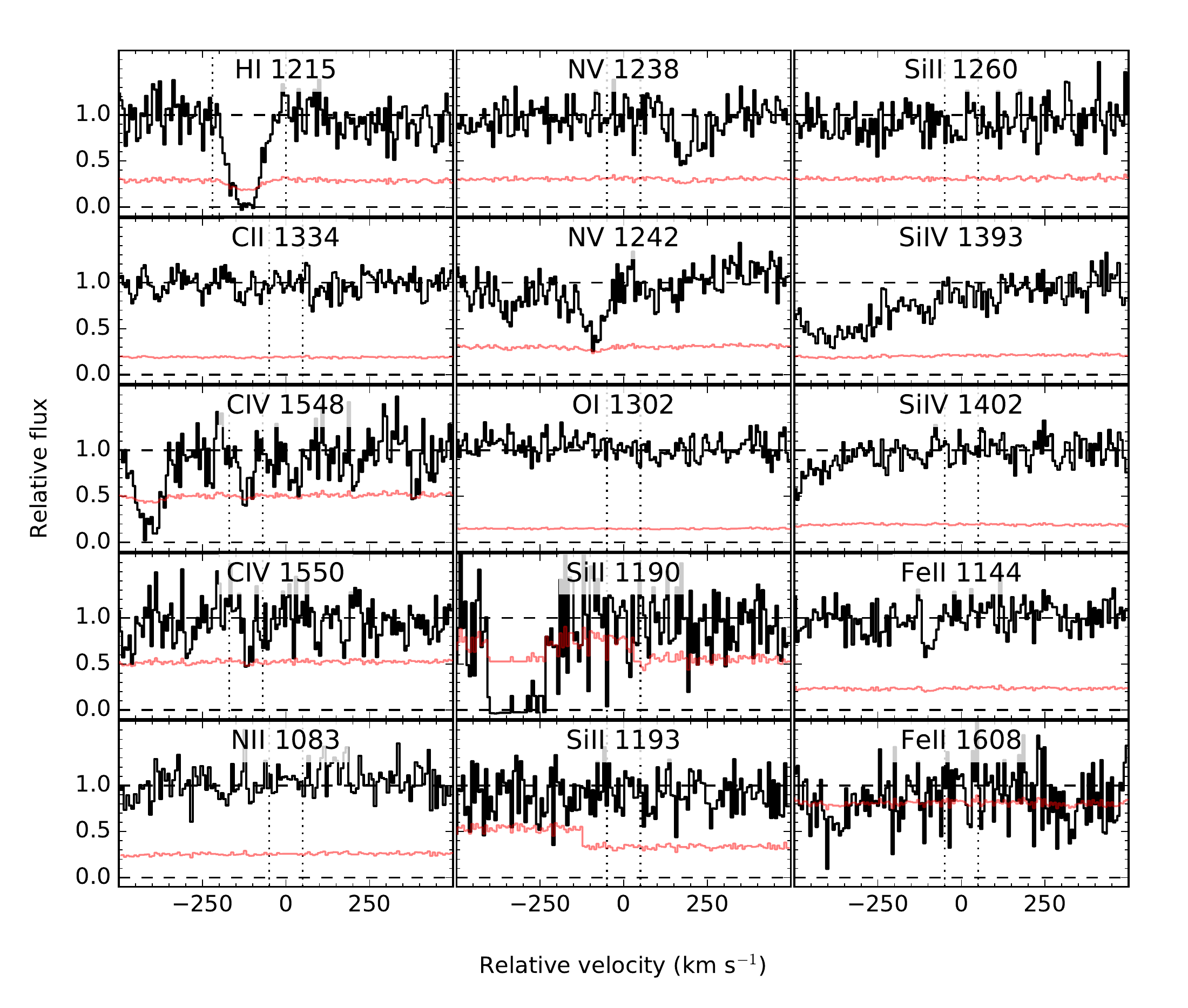}
\caption{Velocity profiles for the sightline towards J0948+5800 (z$_{\rm gal}$=0.084).}
\label{fig:J0948}
\end{center}
\end{figure*}

\begin{table}
\scriptsize
\begin{center}
\caption{Measured EWs for J0948+5800 (z=0.084)}
\label{tab:J0948+5800,294_102}
\begin{tabular}{lcccccc}
\hline
Ion& $\lambda$& $f$& $v_{\rm min}$& $v_{\rm max}$& EW& flag$^{\star}$\\
& [\AA{}]& & [\kms{}]& [\kms{}]& [m\AA{}]& \\
\hline
H\ion{i}& 1215.670& 4.164E-01& -220& 0& $398\pm39$& \textbf{9}\\
C\ion{ii}& 1334.532& 1.278E-01& -50& 50& $<64$& \textbf{5}\\
C\ion{iv}& 1548.195& 1.908E-01& -170& -70& $107\pm67$& \textbf{1}\\
C\ion{iv}& 1550.770& 9.522E-02& -170& -70& $29\pm70$& \textbf{1}\\
N\ion{ii}& 1083.990& 1.031E-01& -50& 50& $<77$& \textbf{5}\\
N\ion{v}& 1238.821& 1.570E-01& -50& 50& $<100$& \textbf{5}\\
N\ion{v}& 1242.804& 7.823E-02& -220& 0& \nodata{}& 2\\
O\ion{i}& 1302.168& 4.887E-02& -50& 50& $<49$& \textbf{5}\\
Si\ion{ii}& 1190.416& 2.502E-01& -50& 50& $<221$& \textbf{5}\\
Si\ion{ii}& 1193.290& 4.991E-01& -50& 50& $<106$& \textbf{5}\\
Si\ion{ii}& 1260.422& 1.007E+00& -50& 50& $<98$& \textbf{5}\\
Si\ion{iv}& 1393.755& 5.280E-01& -220& 0& \nodata{}& 6\\
Si\ion{iv}& 1402.770& 2.620E-01& -50& 50& $<74$& \textbf{5}\\
Fe\ion{ii}& 1144.938& 1.060E-01& -200& -25& \nodata{}& 2\\
Fe\ion{ii}& 1608.451& 5.800E-02& -50& 50& $<335$& \textbf{5}\\
\hline
\end{tabular}
\\$^{\star}$The data quality flags represent a sum of whether or not the line is:\\ adopted ($+1$), blended ($+2$), undetected ($+4$), or saturated ($+8$).\\\end{center}
\end{table}

\begin{figure*}
\begin{center}
\includegraphics[width=\textwidth]{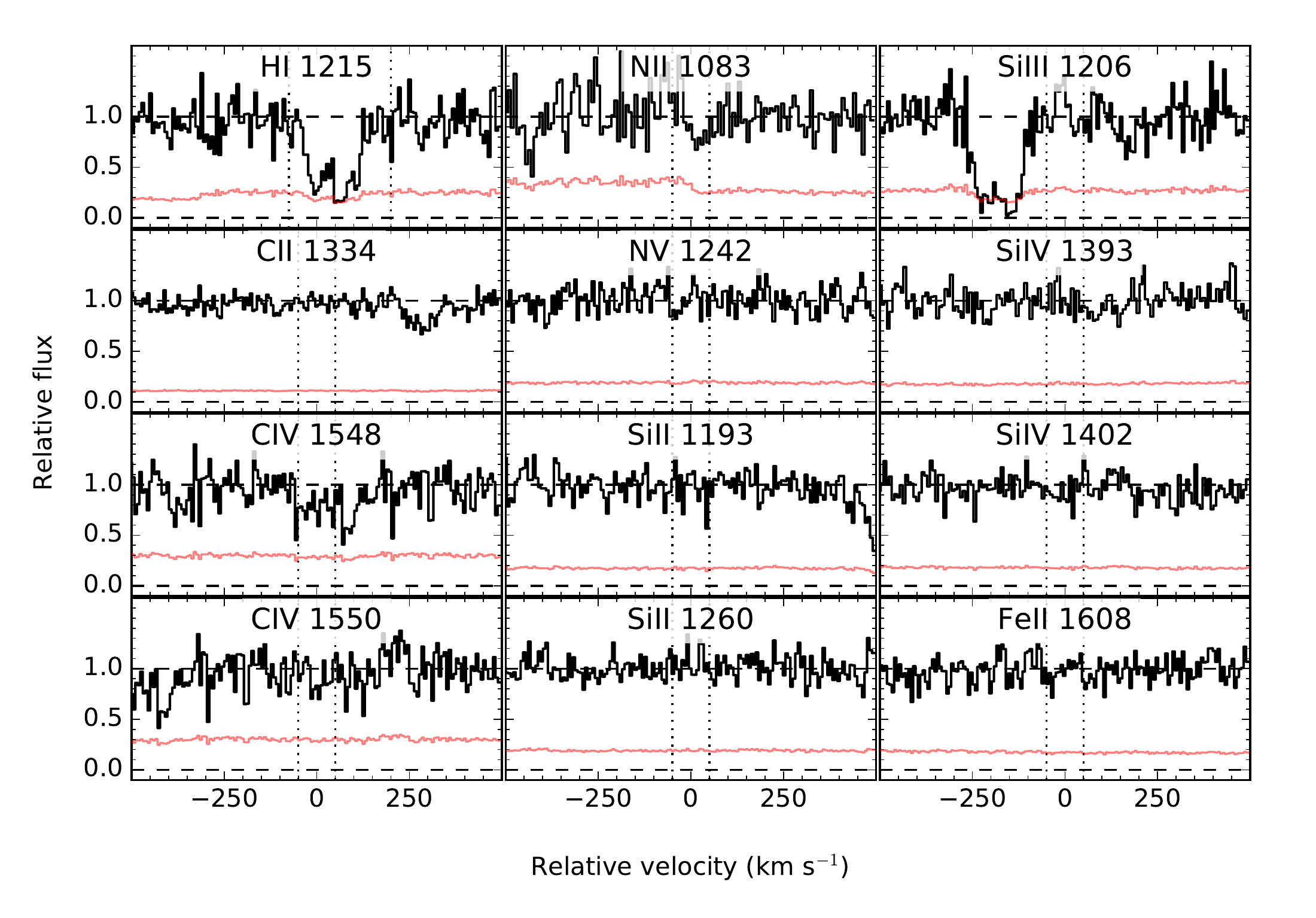}
\caption{Velocity profiles for the sightline towards J1117+2634 (z$_{\rm gal}$=0.065).}
\label{fig:J1117,218}
\end{center}
\end{figure*}

\begin{table}
\scriptsize
\begin{center}
\caption{Measured EWs for J1117+2634 (z=0.065)}
\label{tab:J1117+2634,218_207}
\begin{tabular}{lcccccc}
\hline
Ion& $\lambda$& $f$& $v_{\rm min}$& $v_{\rm max}$& EW& flag$^{\star}$\\
& [\AA{}]& & [\kms{}]& [\kms{}]& [m\AA{}]& \\
\hline
H\ion{i}& 1215.670& 4.164E-01& -75& 200& $453\pm38$& \textbf{9}\\
C\ion{ii}& 1334.532& 1.278E-01& -50& 50& $<38$& \textbf{5}\\
C\ion{iv}& 1548.195& 1.908E-01& -50& 50& $<113$& \textbf{5}\\
C\ion{iv}& 1550.770& 9.522E-02& -50& 50& $<118$& \textbf{5}\\
N\ion{ii}& 1083.990& 1.031E-01& -50& 50& $<96$& \textbf{5}\\
N\ion{v}& 1242.804& 7.823E-02& -50& 50& $<61$& \textbf{5}\\
Si\ion{ii}& 1193.290& 4.991E-01& -50& 50& $<54$& \textbf{5}\\
Si\ion{ii}& 1260.422& 1.007E+00& -50& 50& $<63$& \textbf{5}\\
Si\ion{iii}& 1206.500& 1.660E+00& -50& 50& $<87$& \textbf{5}\\
Si\ion{iv}& 1393.755& 5.280E-01& -50& 50& $<68$& \textbf{5}\\
Si\ion{iv}& 1402.770& 2.620E-01& -50& 50& $<68$& \textbf{5}\\
Fe\ion{ii}& 1608.451& 5.800E-02& -50& 50& $<68$& \textbf{5}\\
\hline
\end{tabular}
\\$^{\star}$The data quality flags represent a sum of whether or not the line is:\\ adopted ($+1$), blended ($+2$), undetected ($+4$), or saturated ($+8$).\\\end{center}
\end{table}

\begin{figure*}
\begin{center}
\includegraphics[width=\textwidth]{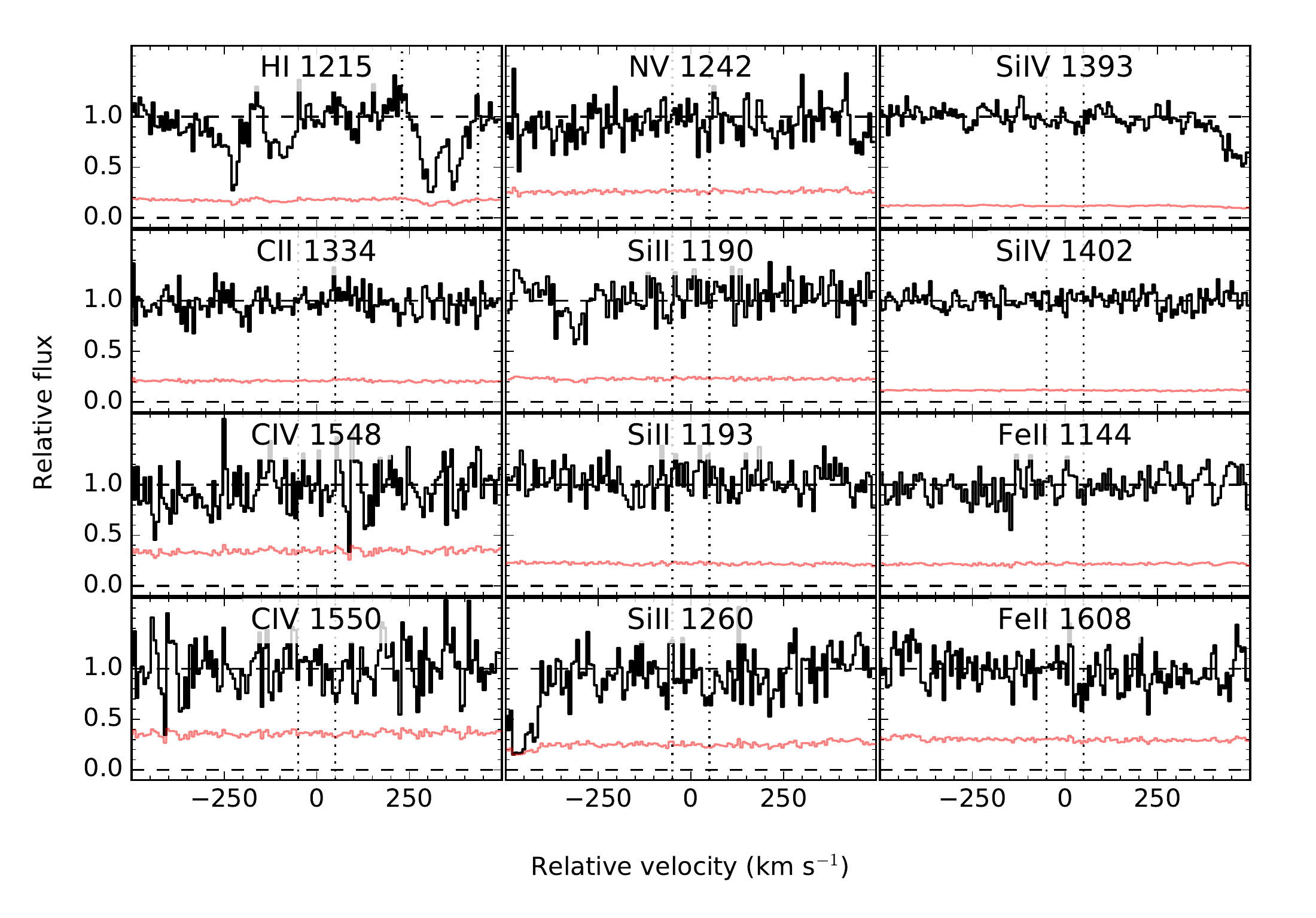}
\caption{Velocity profiles for the sightline towards J1117+2634 (z$_{\rm gal}$=0.029).}
\label{fig:J1117,351}
\end{center}
\end{figure*}

\begin{table}
\scriptsize
\begin{center}
\caption{Measured EWs for J1117+2634 (z=0.029)}
\label{tab:J1117+2634,351_314}
\begin{tabular}{lcccccc}
\hline
Ion& $\lambda$& $f$& $v_{\rm min}$& $v_{\rm max}$& EW& flag$^{\star}$\\
& [\AA{}]& & [\kms{}]& [\kms{}]& [m\AA{}]& \\
\hline
H\ion{i}& 1215.670& 4.164E-01& 230& 435& $256\pm25$& \textbf{9}\\
C\ion{ii}& 1334.532& 1.278E-01& -50& 50& $<73$& \textbf{5}\\
C\ion{iv}& 1548.195& 1.908E-01& -50& 50& $<138$& \textbf{5}\\
C\ion{iv}& 1550.770& 9.522E-02& -50& 50& $<148$& \textbf{5}\\
N\ion{v}& 1242.804& 7.823E-02& -50& 50& $<86$& \textbf{5}\\
Si\ion{ii}& 1190.416& 2.502E-01& -50& 50& $<77$& \textbf{5}\\
Si\ion{ii}& 1193.290& 4.991E-01& -50& 50& $<74$& \textbf{5}\\
Si\ion{ii}& 1260.422& 1.007E+00& -50& 50& $<82$& \textbf{5}\\
Si\ion{iv}& 1393.755& 5.280E-01& -50& 50& $<41$& \textbf{5}\\
Si\ion{iv}& 1402.770& 2.620E-01& -50& 50& $<42$& \textbf{5}\\
Fe\ion{ii}& 1144.938& 1.060E-01& -50& 50& $<68$& \textbf{5}\\
Fe\ion{ii}& 1608.451& 5.800E-02& -50& 50& $<124$& \textbf{5}\\
\hline
\end{tabular}
\\$^{\star}$The data quality flags represent a sum of whether or not the line is:\\ adopted ($+1$), blended ($+2$), undetected ($+4$), or saturated ($+8$).\\\end{center}
\end{table}

\begin{figure*}
\begin{center}
\includegraphics[width=\textwidth]{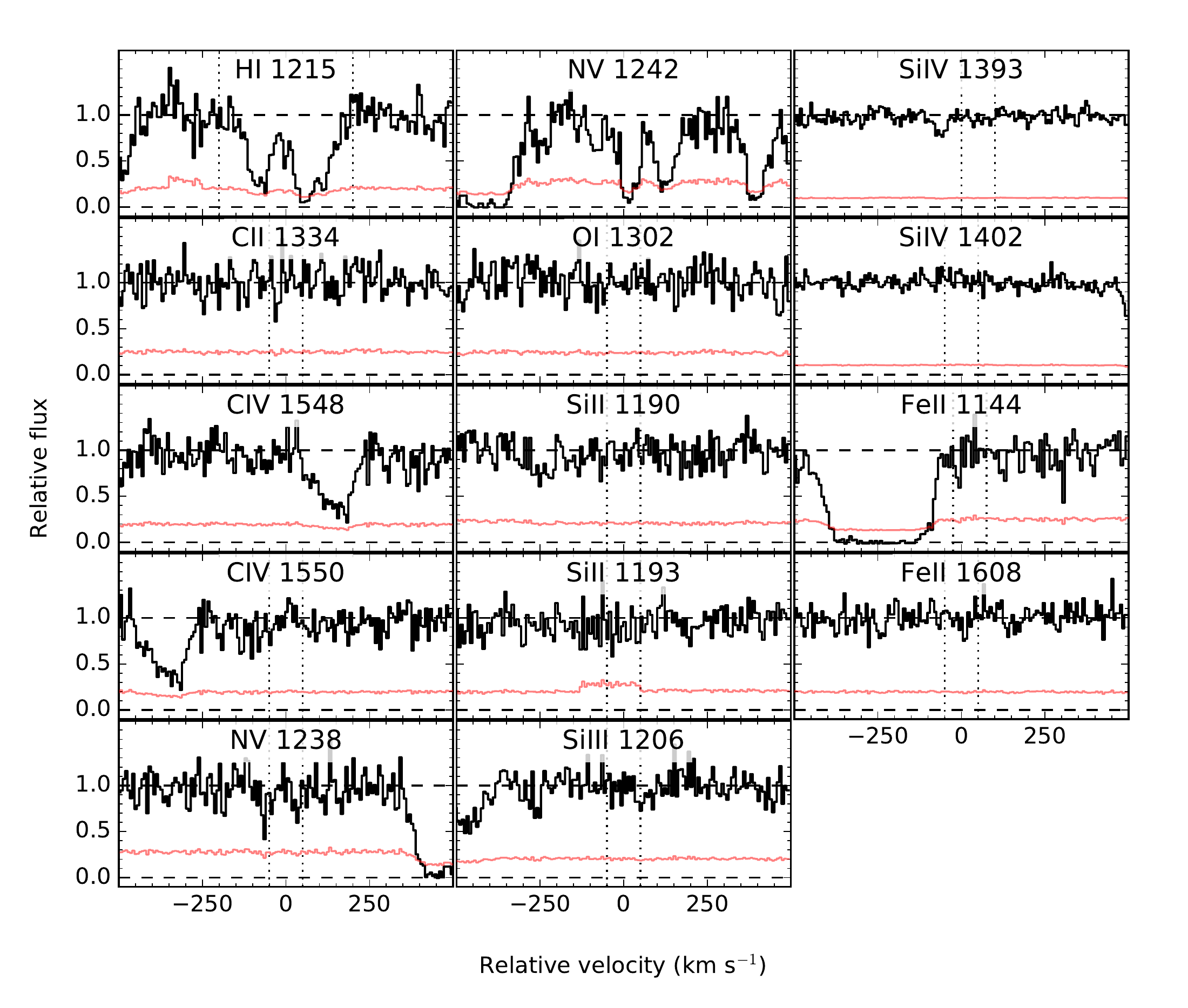}
\caption{Velocity profiles for the sightline towards J1127+2654 (z$_{\rm gal}$=0.033).}
\label{fig:J1127}
\end{center}
\end{figure*}

\begin{table}
\scriptsize
\begin{center}
\caption{Measured EWs for J1127+2654 (z=0.033)}
\label{tab:J1127+2654,39_211}
\begin{tabular}{lcccccc}
\hline
Ion& $\lambda$& $f$& $v_{\rm min}$& $v_{\rm max}$& EW& flag$^{\star}$\\
& [\AA{}]& & [\kms{}]& [\kms{}]& [m\AA{}]& \\
\hline
H\ion{i}& 1215.670& 4.164E-01& -200& 200& $723\pm36$& \textbf{9}\\
C\ion{ii}& 1334.532& 1.278E-01& -50& 50& $<83$& \textbf{5}\\
C\ion{iv}& 1548.195& 1.908E-01& -200& 200& \nodata{}& 2\\
C\ion{iv}& 1550.770& 9.522E-02& -50& 50& $<79$& \textbf{5}\\
N\ion{v}& 1238.821& 1.570E-01& -50& 50& $<88$& \textbf{5}\\
N\ion{v}& 1242.804& 7.823E-02& -200& 200& \nodata{}& 2\\
O\ion{i}& 1302.168& 4.887E-02& -50& 50& $<80$& \textbf{5}\\
Si\ion{ii}& 1190.416& 2.502E-01& -50& 50& $<66$& \textbf{5}\\
Si\ion{ii}& 1193.290& 4.991E-01& -50& 50& $<90$& \textbf{5}\\
Si\ion{iii}& 1206.500& 1.660E+00& -50& 50& $<66$& \textbf{5}\\
Si\ion{iv}& 1393.755& 5.280E-01& 0& 100& $<34$& \textbf{5}\\
Si\ion{iv}& 1402.770& 2.620E-01& -50& 50& $<39$& \textbf{5}\\
Fe\ion{ii}& 1144.938& 1.060E-01& -25& 75& $<83$& \textbf{5}\\
Fe\ion{ii}& 1608.451& 5.800E-02& -50& 50& $<80$& \textbf{5}\\
\hline
\end{tabular}
\\$^{\star}$The data quality flags represent a sum of whether or not the line is:\\ adopted ($+1$), blended ($+2$), undetected ($+4$), or saturated ($+8$).\\\end{center}
\end{table}

\begin{figure*}
\begin{center}
\includegraphics[width=\textwidth]{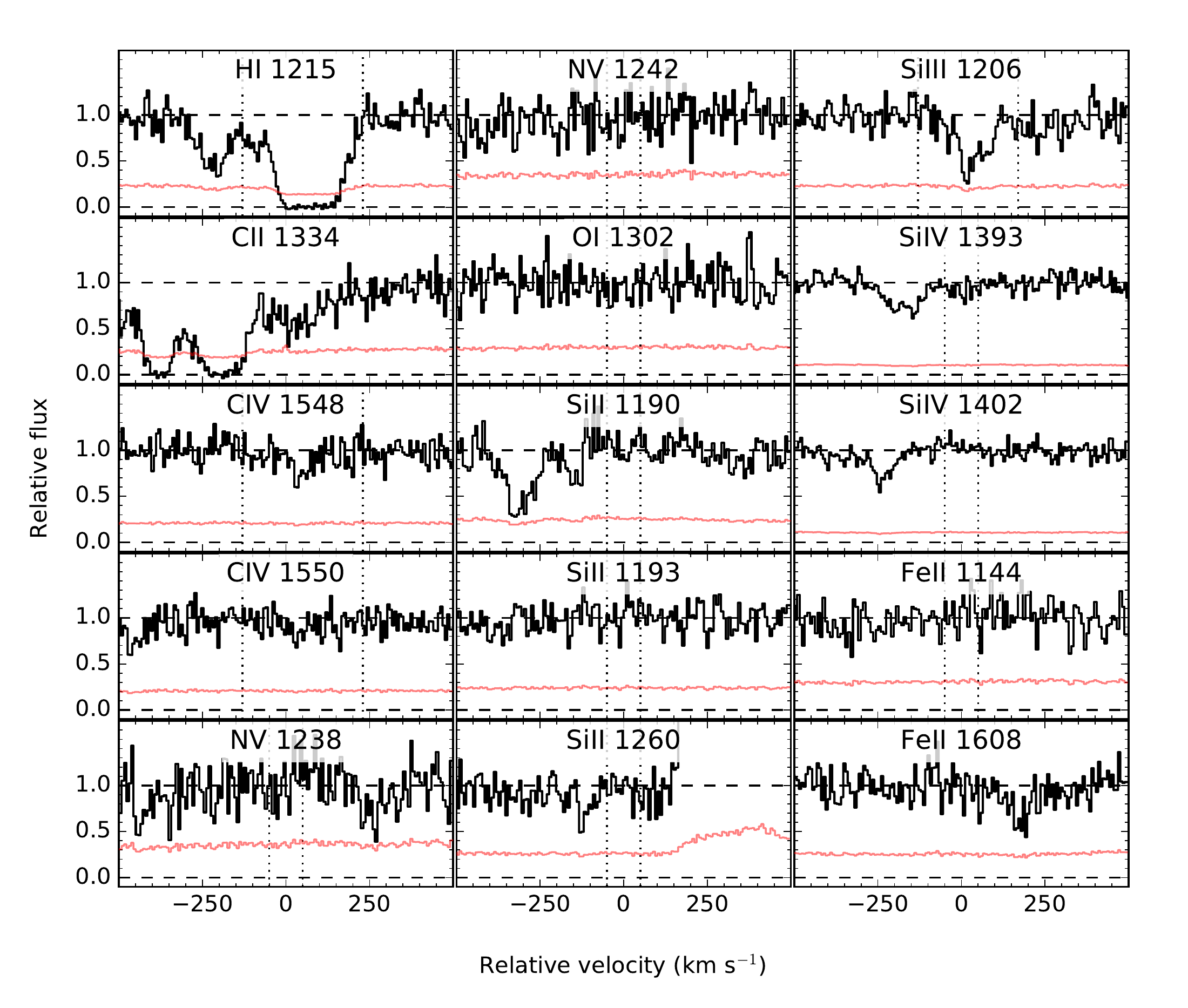}
\caption{Velocity profiles for the sightline towards J1142+3016 (z$_{\rm gal}$=0.032).}
\label{fig:J1142}
\end{center}
\end{figure*}

\begin{table}
\scriptsize
\begin{center}
\caption{Measured EWs for J1142+3016 (z=0.032)}
\label{tab:J1142+3016,153_162}
\begin{tabular}{lcccccc}
\hline
Ion& $\lambda$& $f$& $v_{\rm min}$& $v_{\rm max}$& EW& flag$^{\star}$\\
& [\AA{}]& & [\kms{}]& [\kms{}]& [m\AA{}]& \\
\hline
H\ion{i}& 1215.670& 4.164E-01& -130& 230& $927\pm36$& \textbf{9}\\
C\ion{ii}& 1334.532& 1.278E-01& -230& 130& \nodata{}& 2\\
C\ion{iv}& 1548.195& 1.908E-01& -130& 230& $167\pm53$& \textbf{1}\\
C\ion{iv}& 1550.770& 9.522E-02& -130& 230& $129\pm54$& \textbf{1}\\
N\ion{v}& 1238.821& 1.570E-01& -50& 50& $<123$& \textbf{5}\\
N\ion{v}& 1242.804& 7.823E-02& -50& 50& $<114$& \textbf{5}\\
O\ion{i}& 1302.168& 4.887E-02& -50& 50& $<100$& \textbf{5}\\
Si\ion{ii}& 1190.416& 2.502E-01& -50& 50& $<82$& \textbf{5}\\
Si\ion{ii}& 1193.290& 4.991E-01& -50& 50& $<76$& \textbf{5}\\
Si\ion{ii}& 1260.422& 1.007E+00& -50& 50& $<86$& \textbf{5}\\
Si\ion{iii}& 1206.500& 1.660E+00& -130& 170& $268\pm41$& \textbf{1}\\
Si\ion{iv}& 1393.755& 5.280E-01& -50& 50& $<36$& \textbf{5}\\
Si\ion{iv}& 1402.770& 2.620E-01& -50& 50& $<38$& \textbf{5}\\
Fe\ion{ii}& 1144.938& 1.060E-01& -50& 50& $<98$& \textbf{5}\\
Fe\ion{ii}& 1608.451& 5.800E-02& -50& 50& \nodata{}& 4\\
\hline
\end{tabular}
\\$^{\star}$The data quality flags represent a sum of whether or not the line is:\\ adopted ($+1$), blended ($+2$), undetected ($+4$), or saturated ($+8$).\\\end{center}
\end{table}

\begin{figure*}
\begin{center}
\includegraphics[width=\textwidth]{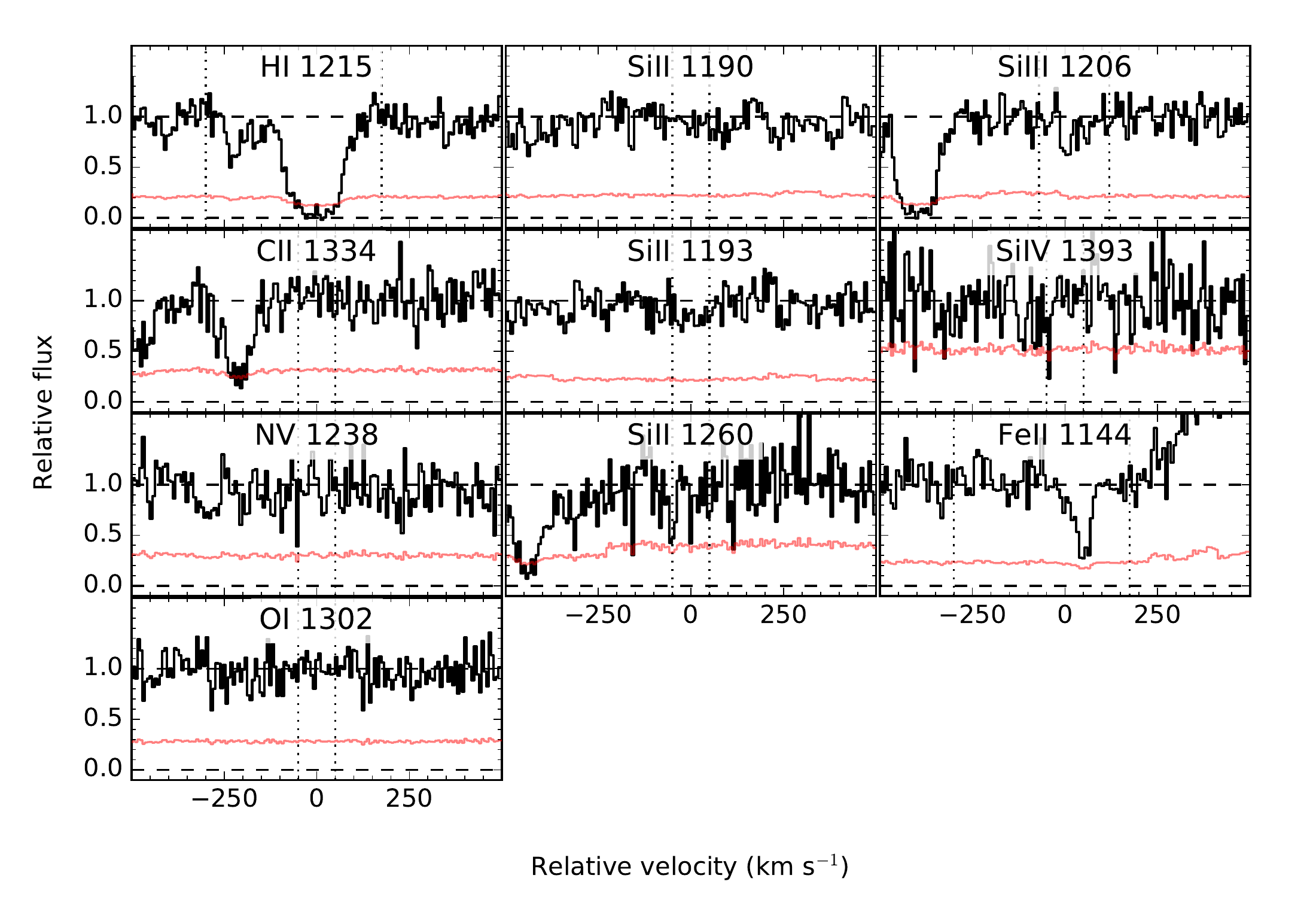}
\caption{Velocity profiles for the sightline towards J1155+2922 (z$_{\rm gal}$=0.046).}
\label{fig:J1155}
\end{center}
\end{figure*}

\begin{table}
\scriptsize
\begin{center}
\caption{Measured EWs for J1155+2922 (z=0.046)}
\label{tab:J1155+2922,246_231}
\begin{tabular}{lcccccc}
\hline
Ion& $\lambda$& $f$& $v_{\rm min}$& $v_{\rm max}$& EW& flag$^{\star}$\\
& [\AA{}]& & [\kms{}]& [\kms{}]& [m\AA{}]& \\
\hline
H\ion{i}& 1215.670& 4.164E-01& -300& 175& $721\pm43$& \textbf{9}\\
C\ion{ii}& 1334.532& 1.278E-01& -50& 50& $<109$& \textbf{5}\\
N\ion{v}& 1238.821& 1.570E-01& -50& 50& $<100$& \textbf{5}\\
O\ion{i}& 1302.168& 4.887E-02& -50& 50& $<97$& \textbf{5}\\
Si\ion{ii}& 1190.416& 2.502E-01& -50& 50& $<71$& \textbf{5}\\
Si\ion{ii}& 1193.290& 4.991E-01& -50& 50& $<69$& \textbf{5}\\
Si\ion{ii}& 1260.422& 1.007E+00& -50& 50& $<124$& \textbf{5}\\
Si\ion{iii}& 1206.500& 1.660E+00& -70& 120& $73\pm33$& \textbf{1}\\
Si\ion{iv}& 1393.755& 5.280E-01& -50& 50& $<177$& \textbf{5}\\
Fe\ion{ii}& 1144.938& 1.060E-01& -300& 175& $85\pm52$& \textbf{1}\\
\hline
\end{tabular}
\\$^{\star}$The data quality flags represent a sum of whether or not the line is:\\ adopted ($+1$), blended ($+2$), undetected ($+4$), or saturated ($+8$).\\\end{center}
\end{table}

\begin{figure*}
\begin{center}
\includegraphics[width=\textwidth]{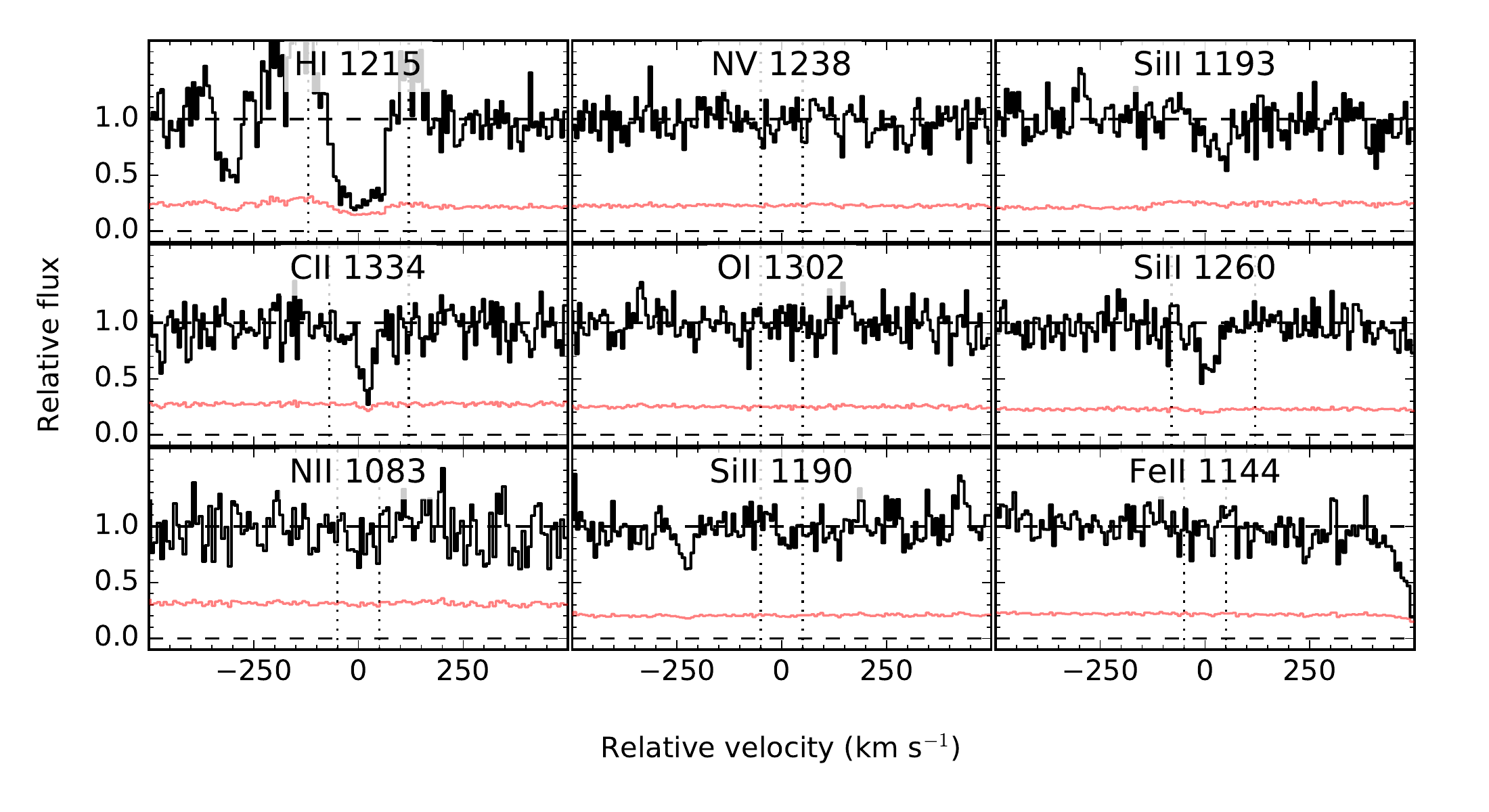}
\caption{Velocity profiles for the sightline towards J1214+0825 (z$_{\rm gal}$=0.074).}
\label{fig:J1214}
\end{center}
\end{figure*}

\begin{table}
\scriptsize
\begin{center}
\caption{Measured EWs for J1214+0825 (z=0.074)}
\label{tab:J1214+0825,176_163}
\begin{tabular}{lcccccc}
\hline
Ion& $\lambda$& $f$& $v_{\rm min}$& $v_{\rm max}$& EW& flag$^{\star}$\\
& [\AA{}]& & [\kms{}]& [\kms{}]& [m\AA{}]& \\
\hline
H\ion{i}& 1215.670& 4.164E-01& -120& 120& $229\pm34$& \textbf{9}\\
C\ion{ii}& 1334.532& 1.278E-01& -70& 120& $154\pm41$& \textbf{1}\\
N\ion{ii}& 1083.990& 1.031E-01& -50& 50& $<92$& \textbf{5}\\
N\ion{v}& 1238.821& 1.570E-01& -50& 50& $<74$& \textbf{5}\\
O\ion{i}& 1302.168& 4.887E-02& -50& 50& $<83$& \textbf{5}\\
Si\ion{ii}& 1190.416& 2.502E-01& -50& 50& $<66$& \textbf{5}\\
Si\ion{ii}& 1193.290& 4.991E-01& -50& 120& \nodata{}& 2\\
Si\ion{ii}& 1260.422& 1.007E+00& -80& 120& $107\pm34$& \textbf{1}\\
Fe\ion{ii}& 1144.938& 1.060E-01& -50& 50& $<65$& \textbf{5}\\
\hline
\end{tabular}
\\$^{\star}$The data quality flags represent a sum of whether or not the line is:\\ adopted ($+1$), blended ($+2$), undetected ($+4$), or saturated ($+8$).\\\end{center}
\end{table}

\begin{figure*}
\begin{center}
\includegraphics[width=\textwidth]{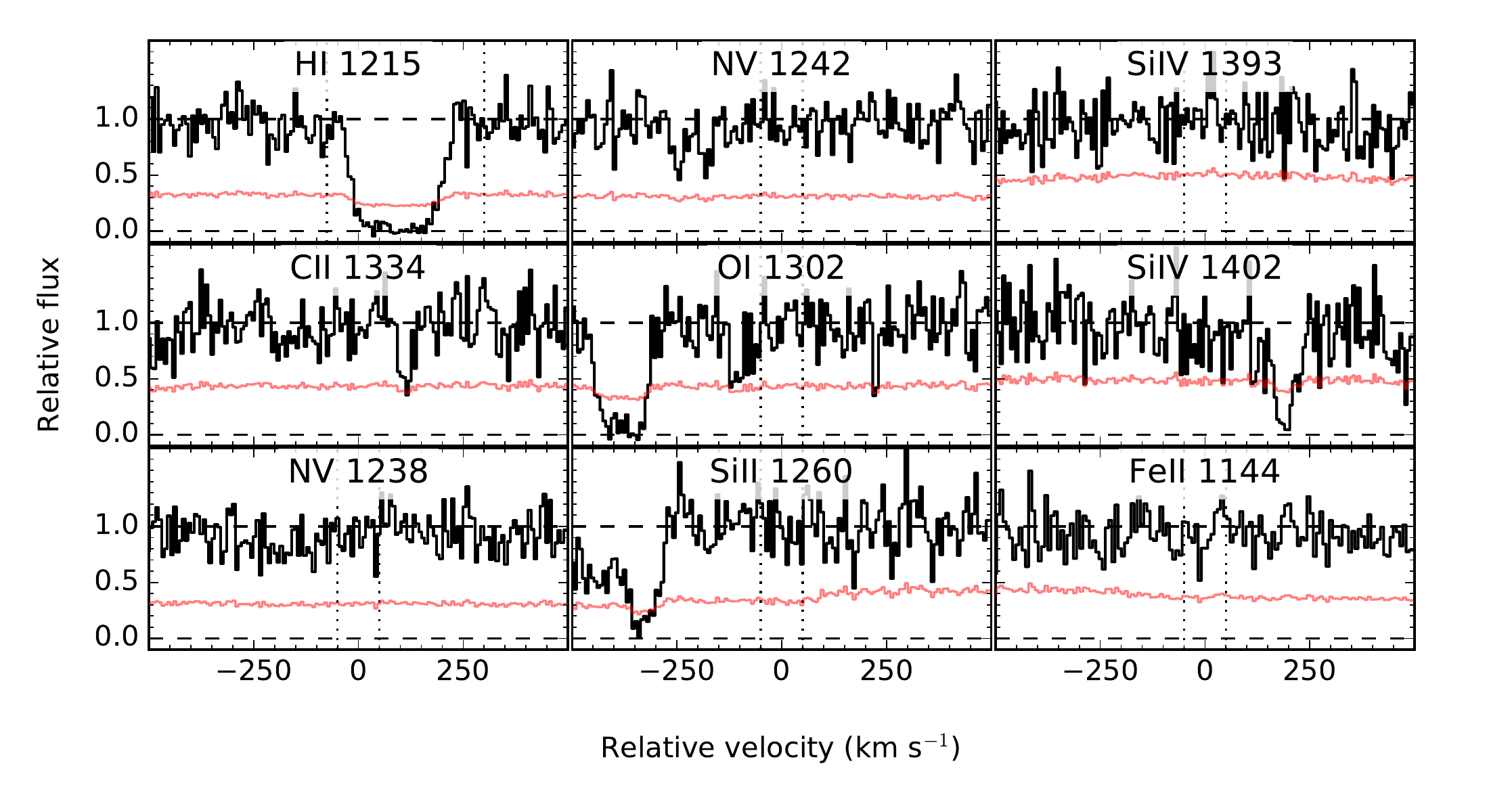}
\caption{Velocity profiles for the sightline towards J1404+3353 (z$_{\rm gal}$=0.026).}
\label{fig:J1404}
\end{center}
\end{figure*}

\begin{table}
\scriptsize
\begin{center}
\caption{Measured EWs for J1404+3353 (z=0.026)}
\label{tab:J1404+3353,3_210}
\begin{tabular}{lcccccc}
\hline
Ion& $\lambda$& $f$& $v_{\rm min}$& $v_{\rm max}$& EW& flag$^{\star}$\\
& [\AA{}]& & [\kms{}]& [\kms{}]& [m\AA{}]& \\
\hline
H\ion{i}& 1215.670& 4.164E-01& -75& 300& $736\pm58$& \textbf{9}\\
C\ion{ii}& 1334.532& 1.278E-01& 0& 200& \nodata{}& 2\\
N\ion{v}& 1238.821& 1.570E-01& -50& 50& $<101$& \textbf{5}\\
N\ion{v}& 1242.804& 7.823E-02& -50& 50& $<103$& \textbf{5}\\
O\ion{i}& 1302.168& 4.887E-02& -50& 50& $<147$& \textbf{5}\\
Si\ion{ii}& 1260.422& 1.007E+00& -50& 50& $<108$& \textbf{5}\\
Si\ion{iv}& 1393.755& 5.280E-01& -50& 50& $<179$& \textbf{5}\\
Si\ion{iv}& 1402.770& 2.620E-01& 70& 300& \nodata{}& 10\\
Fe\ion{ii}& 1144.938& 1.060E-01& -50& 50& $<117$& \textbf{5}\\
\hline
\end{tabular}
\\$^{\star}$The data quality flags represent a sum of whether or not the line is:\\ adopted ($+1$), blended ($+2$), undetected ($+4$), or saturated ($+8$).\\\end{center}
\end{table}

\begin{figure*}
\begin{center}
\includegraphics[width=\textwidth]{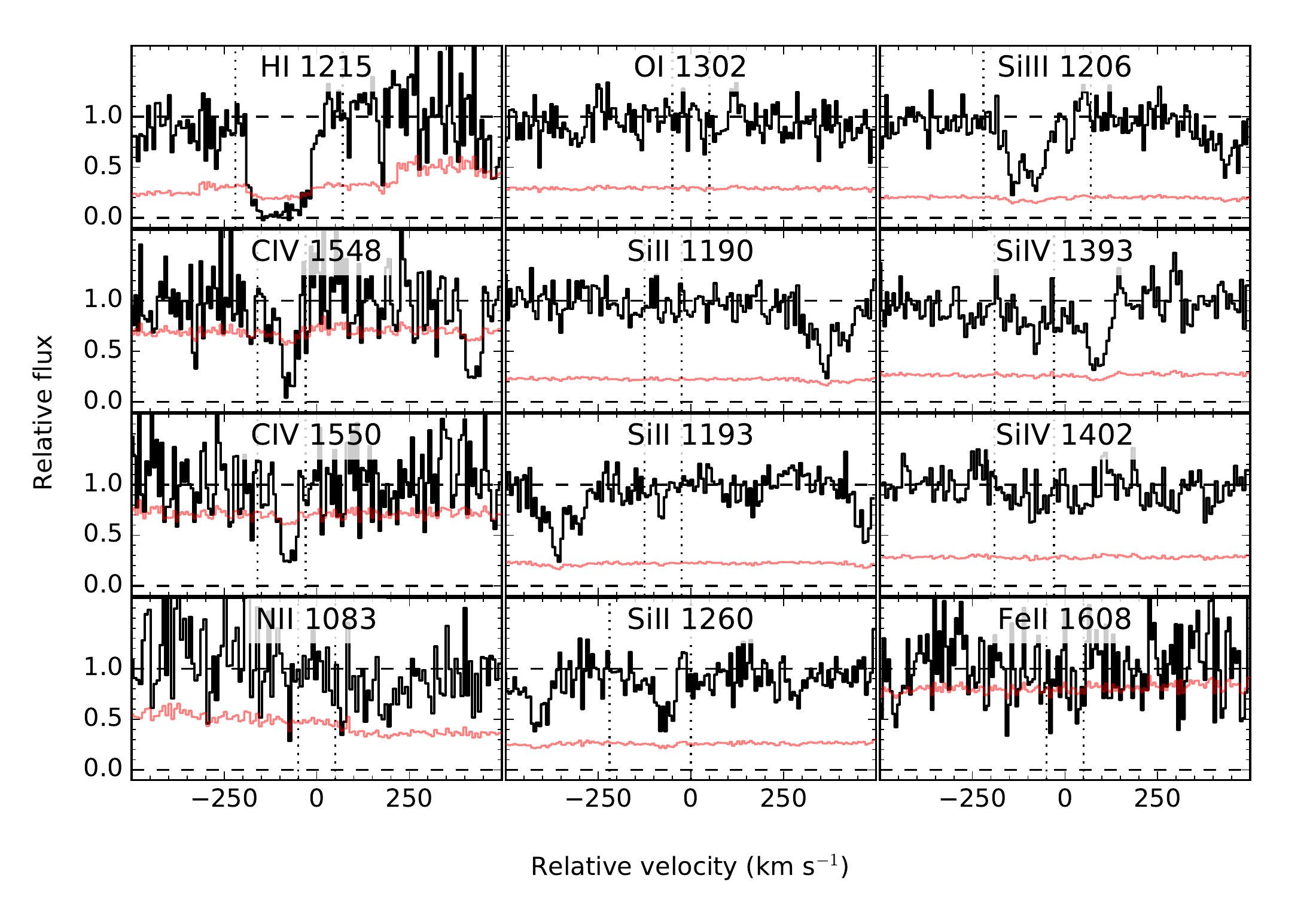}
\caption{Velocity profiles for the sightline towards J1419+0606 (z$_{\rm gal}$=0.049).}
\label{fig:J1419}
\end{center}
\end{figure*}

\begin{table}
\scriptsize
\begin{center}
\caption{Measured EWs for J1419+0606 (z=0.049)}
\label{tab:J1419+0606,293_184}
\begin{tabular}{lcccccc}
\hline
Ion& $\lambda$& $f$& $v_{\rm min}$& $v_{\rm max}$& EW& flag$^{\star}$\\
& [\AA{}]& & [\kms{}]& [\kms{}]& [m\AA{}]& \\
\hline
H\ion{i}& 1215.670& 4.164E-01& -220& 70& $575\pm48$& \textbf{9}\\
C\ion{iv}& 1548.195& 1.908E-01& -160& -30& $135\pm100$& \textbf{9}\\
C\ion{iv}& 1550.770& 9.522E-02& -160& -30& $107\pm103$& \textbf{9}\\
N\ion{ii}& 1083.990& 1.031E-01& -50& 50& $<146$& \textbf{5}\\
O\ion{i}& 1302.168& 4.887E-02& -50& 50& $<97$& \textbf{5}\\
Si\ion{ii}& 1190.416& 2.502E-01& -125& -25& $<71$& \textbf{5}\\
Si\ion{ii}& 1193.290& 4.991E-01& -125& -25& $<70$& \textbf{5}\\
Si\ion{ii}& 1260.422& 1.007E+00& -220& 0& $158\pm41$& \textbf{1}\\
Si\ion{iii}& 1206.500& 1.660E+00& -220& 70& $294\pm34$& \textbf{1}\\
Si\ion{iv}& 1393.755& 5.280E-01& -190& -30& $119\pm43$& \textbf{1}\\
Si\ion{iv}& 1402.770& 2.620E-01& -190& -30& $93\pm44$& \textbf{1}\\
Fe\ion{ii}& 1608.451& 5.800E-02& -50& 50& $<331$& \textbf{5}\\
\hline
\end{tabular}
\\$^{\star}$The data quality flags represent a sum of whether or not the line is:\\ adopted ($+1$), blended ($+2$), undetected ($+4$), or saturated ($+8$).\\\end{center}
\end{table}

\clearpage

\begin{figure*}
\begin{center}
\includegraphics[width=\textwidth]{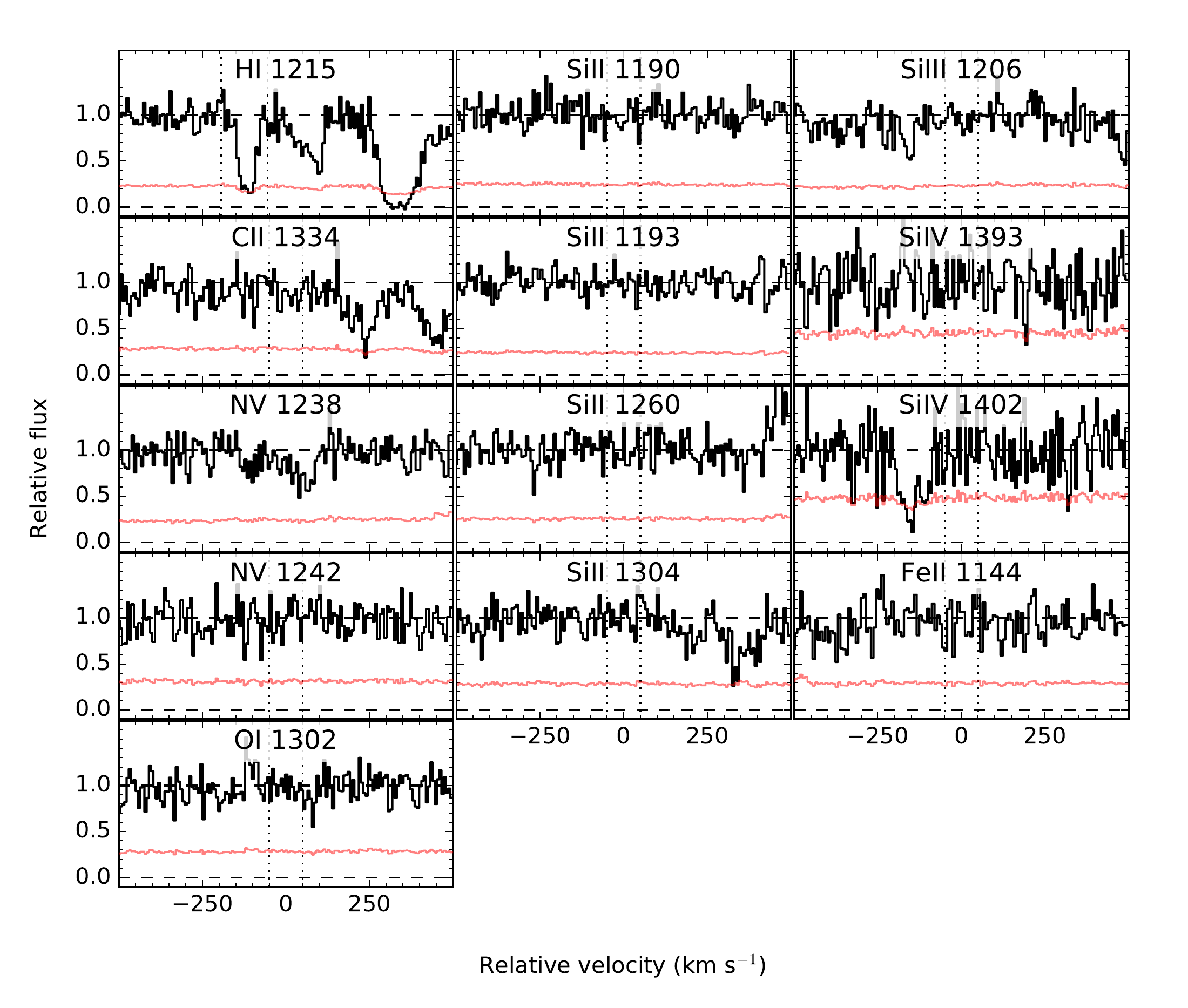}
\caption{Velocity profiles for the sightline towards J1454+3046 (z$_{\rm gal}$=0.031).}
\label{fig:J1454}
\end{center}
\end{figure*}

\begin{table}
\scriptsize
\begin{center}
\caption{Measured EWs for J1454+3046 (z=0.031)}
\label{tab:J1454+3046,4_453}
\begin{tabular}{lcccccc}
\hline
Ion& $\lambda$& $f$& $v_{\rm min}$& $v_{\rm max}$& EW& flag$^{\star}$\\
& [\AA{}]& & [\kms{}]& [\kms{}]& [m\AA{}]& \\
\hline
H\ion{i}& 1215.670& 4.164E-01& -195& -55& $194\pm27$& \textbf{9}\\
C\ion{ii}& 1334.532& 1.278E-01& -50& 50& $<95$& \textbf{5}\\
N\ion{v}& 1238.821& 1.570E-01& -50& 50& \nodata{}& 2\\
N\ion{v}& 1242.804& 7.823E-02& -50& 50& $<105$& \textbf{5}\\
O\ion{i}& 1302.168& 4.887E-02& -50& 50& $<97$& \textbf{5}\\
Si\ion{ii}& 1190.416& 2.502E-01& -50& 50& $<80$& \textbf{5}\\
Si\ion{ii}& 1193.290& 4.991E-01& -50& 50& $<78$& \textbf{5}\\
Si\ion{ii}& 1260.422& 1.007E+00& -50& 50& $<84$& \textbf{5}\\
Si\ion{ii}& 1304.370& 9.400E-02& -50& 50& $<97$& \textbf{5}\\
Si\ion{iii}& 1206.500& 1.660E+00& -50& 50& $<76$& \textbf{5}\\
Si\ion{iv}& 1393.755& 5.280E-01& -50& 50& $<160$& \textbf{5}\\
Si\ion{iv}& 1402.770& 2.620E-01& -50& 50& $<170$& \textbf{5}\\
Fe\ion{ii}& 1144.938& 1.060E-01& -50& 50& $<92$& \textbf{5}\\
\hline
\end{tabular}
\\$^{\star}$The data quality flags represent a sum of whether or not the line is:\\ adopted ($+1$), blended ($+2$), undetected ($+4$), or saturated ($+8$).\\\end{center}
\end{table}

\begin{figure*}
\begin{center}
\includegraphics[width=\textwidth]{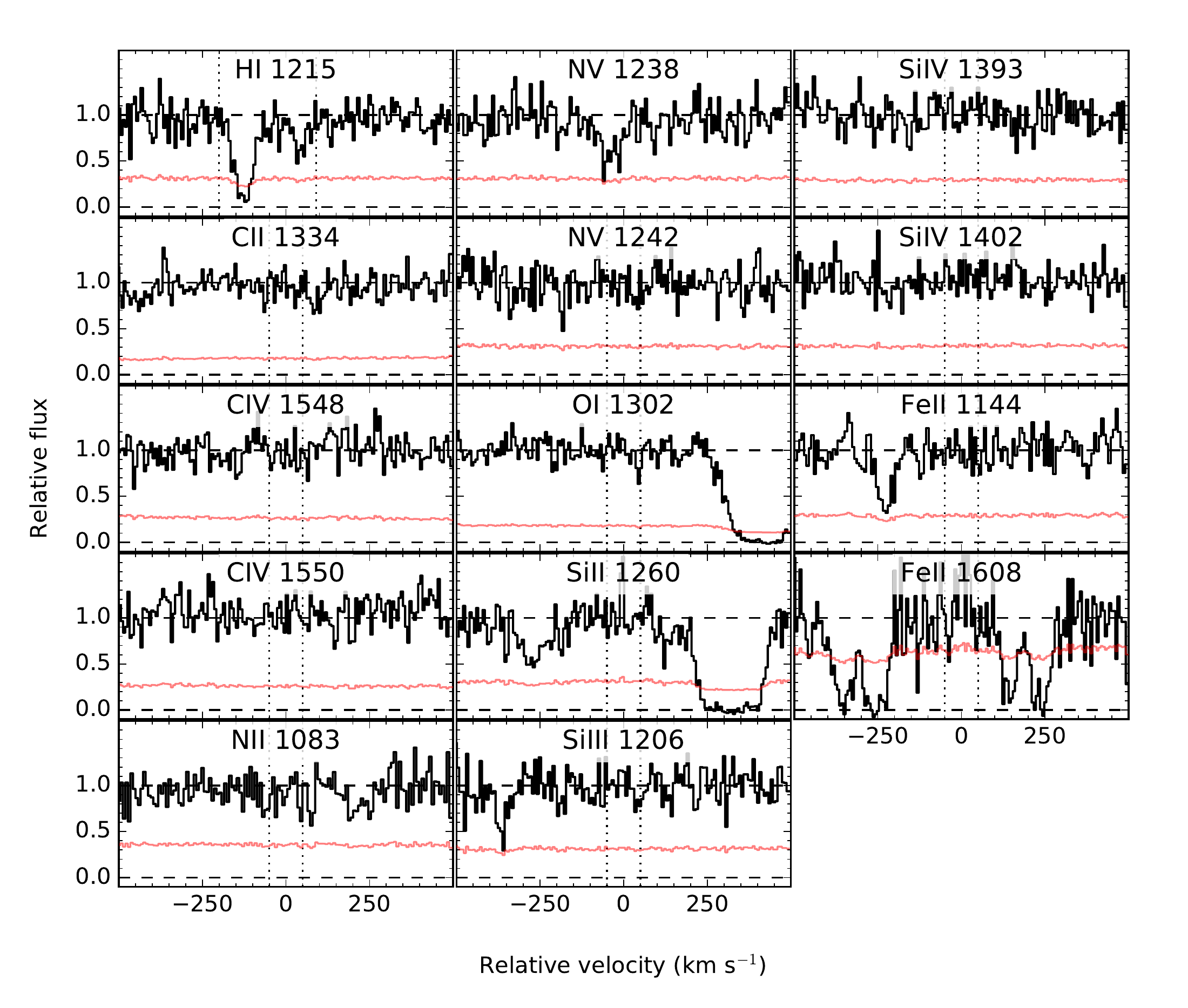}
\caption{Velocity profiles for the sightline towards J1536+1412 (z$_{\rm gal}$=0.093).}
\label{fig:J1536}
\end{center}
\end{figure*}

\begin{table}
\scriptsize
\begin{center}
\caption{Measured EWs for J1536+1412 (z=0.093)}
\label{tab:J1536+1412,321_86}
\begin{tabular}{lcccccc}
\hline
Ion& $\lambda$& $f$& $v_{\rm min}$& $v_{\rm max}$& EW& flag$^{\star}$\\
& [\AA{}]& & [\kms{}]& [\kms{}]& [m\AA{}]& \\
\hline
H\ion{i}& 1215.670& 4.164E-01& -200& 90& $367\pm52$& \textbf{9}\\
C\ion{ii}& 1334.532& 1.278E-01& -50& 50& $<61$& \textbf{5}\\
C\ion{iv}& 1548.195& 1.908E-01& -50& 50& $<105$& \textbf{5}\\
C\ion{iv}& 1550.770& 9.522E-02& -50& 50& $<101$& \textbf{5}\\
N\ion{ii}& 1083.990& 1.031E-01& -50& 50& $<108$& \textbf{5}\\
N\ion{v}& 1238.821& 1.570E-01& -175& -75& \nodata{}& 6\\
N\ion{v}& 1242.804& 7.823E-02& -50& 50& $<97$& \textbf{5}\\
O\ion{i}& 1302.168& 4.887E-02& -50& 50& $<59$& \textbf{5}\\
Si\ion{ii}& 1260.422& 1.007E+00& -50& 50& $<100$& \textbf{5}\\
Si\ion{iii}& 1206.500& 1.660E+00& -50& 50& $<96$& \textbf{5}\\
Si\ion{iv}& 1393.755& 5.280E-01& -50& 50& $<111$& \textbf{5}\\
Si\ion{iv}& 1402.770& 2.620E-01& -50& 50& $<119$& \textbf{5}\\
Fe\ion{ii}& 1144.938& 1.060E-01& -50& 50& $<89$& \textbf{5}\\
Fe\ion{ii}& 1608.451& 5.800E-02& -50& 50& $<267$& \textbf{5}\\
\hline
\end{tabular}
\\$^{\star}$The data quality flags represent a sum of whether or not the line is:\\ adopted ($+1$), blended ($+2$), undetected ($+4$), or saturated ($+8$).\\\end{center}
\end{table}

\begin{figure*}
\begin{center}
\includegraphics[width=\textwidth]{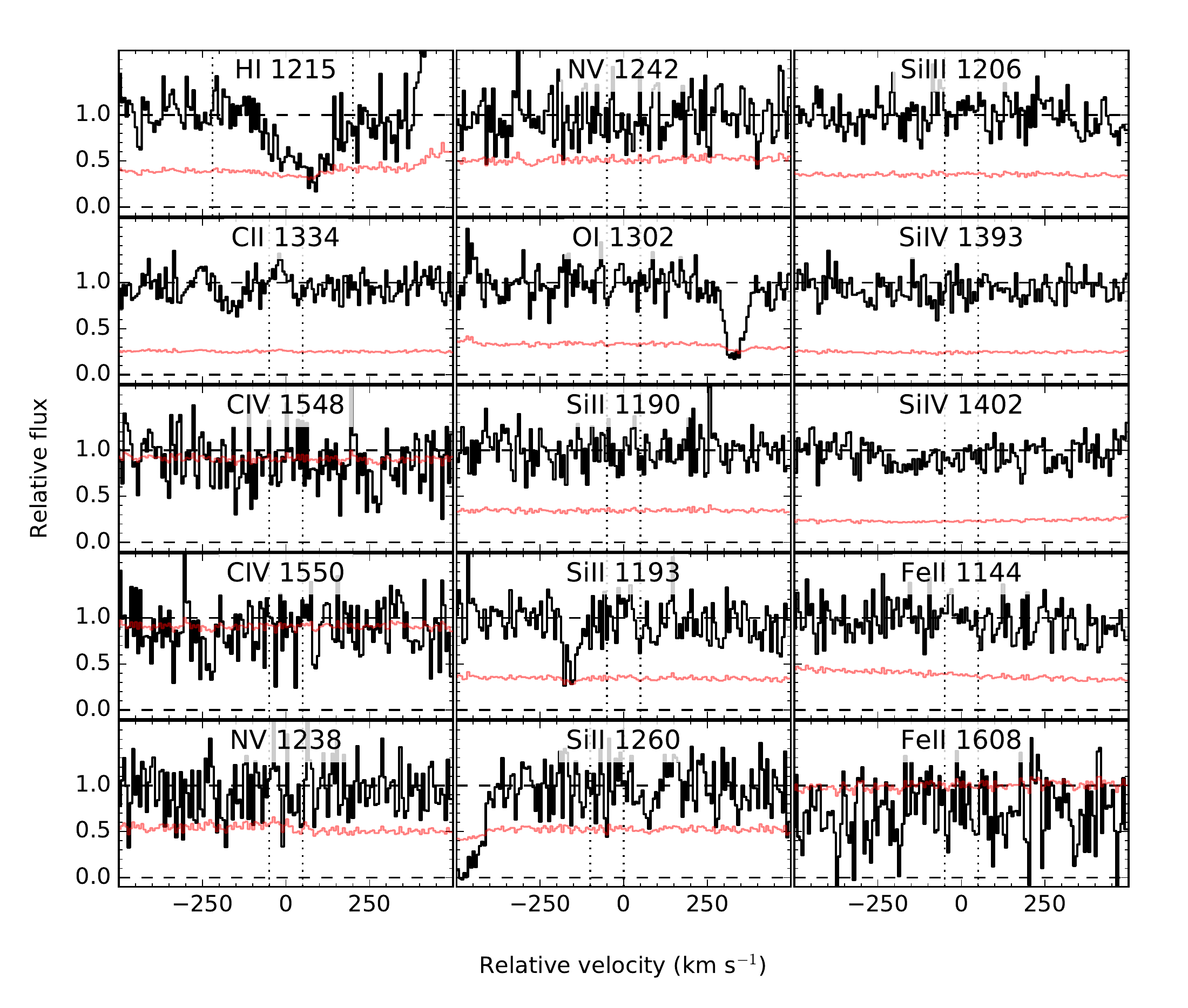}
\caption{Velocity profiles for the sightline towards J1607+1334 (z$_{\rm gal}$=0.069).}
\label{fig:J1607}
\end{center}
\end{figure*}

\begin{table}
\scriptsize
\begin{center}
\caption{Measured EWs for J1607+1334 (z=0.069)}
\label{tab:J1607+1334,102_118}
\begin{tabular}{lcccccc}
\hline
Ion& $\lambda$& $f$& $v_{\rm min}$& $v_{\rm max}$& EW& flag$^{\star}$\\
& [\AA{}]& & [\kms{}]& [\kms{}]& [m\AA{}]& \\
\hline
H\ion{i}& 1215.670& 4.164E-01& -220& 200& $400\pm82$& \textbf{9}\\
C\ion{ii}& 1334.532& 1.278E-01& -50& 50& $<87$& \textbf{5}\\
C\ion{iv}& 1548.195& 1.908E-01& -50& 50& $<363$& \textbf{5}\\
C\ion{iv}& 1550.770& 9.522E-02& -50& 50& $<361$& \textbf{5}\\
N\ion{v}& 1238.821& 1.570E-01& -50& 50& $<182$& \textbf{5}\\
N\ion{v}& 1242.804& 7.823E-02& -50& 50& $<165$& \textbf{5}\\
O\ion{i}& 1302.168& 4.887E-02& -50& 50& $<108$& \textbf{5}\\
Si\ion{ii}& 1190.416& 2.502E-01& -50& 50& $<112$& \textbf{5}\\
Si\ion{ii}& 1193.290& 4.991E-01& -50& 50& $<111$& \textbf{5}\\
Si\ion{ii}& 1260.422& 1.007E+00& -100& 0& $<172$& \textbf{5}\\
Si\ion{iii}& 1206.500& 1.660E+00& -50& 50& $<113$& \textbf{5}\\
Si\ion{iv}& 1393.755& 5.280E-01& -50& 50& $<93$& \textbf{5}\\
Si\ion{iv}& 1402.770& 2.620E-01& -50& 50& $<85$& \textbf{5}\\
Fe\ion{ii}& 1144.938& 1.060E-01& -50& 50& $<121$& \textbf{5}\\
Fe\ion{ii}& 1608.451& 5.800E-02& -50& 50& $<412$& \textbf{5}\\
\hline
\end{tabular}
\\$^{\star}$The data quality flags represent a sum of whether or not the line is:\\ adopted ($+1$), blended ($+2$), undetected ($+4$), or saturated ($+8$).\\\end{center}
\end{table}

\begin{figure*}
\begin{center}
\includegraphics[width=\textwidth]{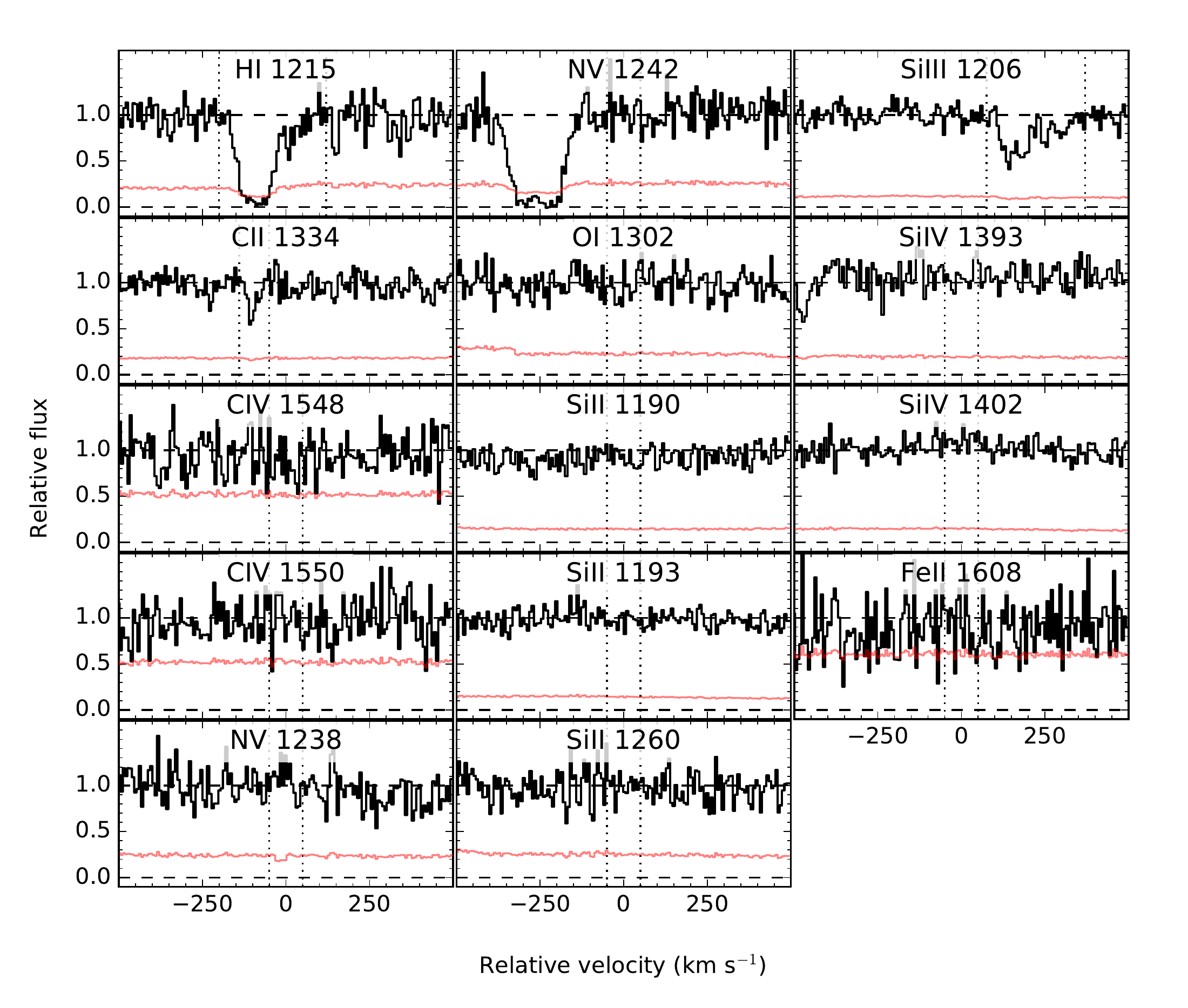}
\caption{Velocity profiles for the sightline towards J2133$-$0712 (z$_{\rm gal}$=0.064).}
\label{fig:J2133}
\end{center}
\end{figure*}

\begin{table}
\scriptsize
\begin{center}
\caption{Measured EWs for J2133-0712 (z=0.064)}
\label{tab:J2133-0712,325_111}
\begin{tabular}{lcccccc}
\hline
Ion& $\lambda$& $f$& $v_{\rm min}$& $v_{\rm max}$& EW& flag$^{\star}$\\
& [\AA{}]& & [\kms{}]& [\kms{}]& [m\AA{}]& \\
\hline
H\ion{i}& 1215.670& 4.164E-01& -200& 120& $501\pm37$& \textbf{9}\\
C\ion{ii}& 1334.532& 1.278E-01& -140& -50& $53\pm18$& \textbf{1}\\
C\ion{iv}& 1548.195& 1.908E-01& -50& 50& $<205$& \textbf{5}\\
C\ion{iv}& 1550.770& 9.522E-02& -50& 50& $<212$& \textbf{5}\\
N\ion{v}& 1238.821& 1.570E-01& -50& 50& $<74$& \textbf{5}\\
N\ion{v}& 1242.804& 7.823E-02& -50& 50& $<83$& \textbf{5}\\
O\ion{i}& 1302.168& 4.887E-02& -50& 50& $<73$& \textbf{5}\\
Si\ion{ii}& 1190.416& 2.502E-01& -50& 50& $<46$& \textbf{5}\\
Si\ion{ii}& 1193.290& 4.991E-01& -50& 50& $<47$& \textbf{5}\\
Si\ion{ii}& 1260.422& 1.007E+00& -50& 50& $<84$& \textbf{5}\\
Si\ion{iii}& 1206.500& 1.660E+00& 75& 370& $243\pm18$& \textbf{1}\\
Si\ion{iv}& 1393.755& 5.280E-01& -50& 50& $<75$& \textbf{5}\\
Si\ion{iv}& 1402.770& 2.620E-01& -50& 50& $<59$& \textbf{5}\\
Fe\ion{ii}& 1608.451& 5.800E-02& -50& 50& $<252$& \textbf{5}\\
\hline
\end{tabular}
\\$^{\star}$The data quality flags represent a sum of whether or not the line is:\\ adopted ($+1$), blended ($+2$), undetected ($+4$), or saturated ($+8$).\\\end{center}
\end{table}

\begin{figure*}
\begin{center}
\includegraphics[width=\textwidth]{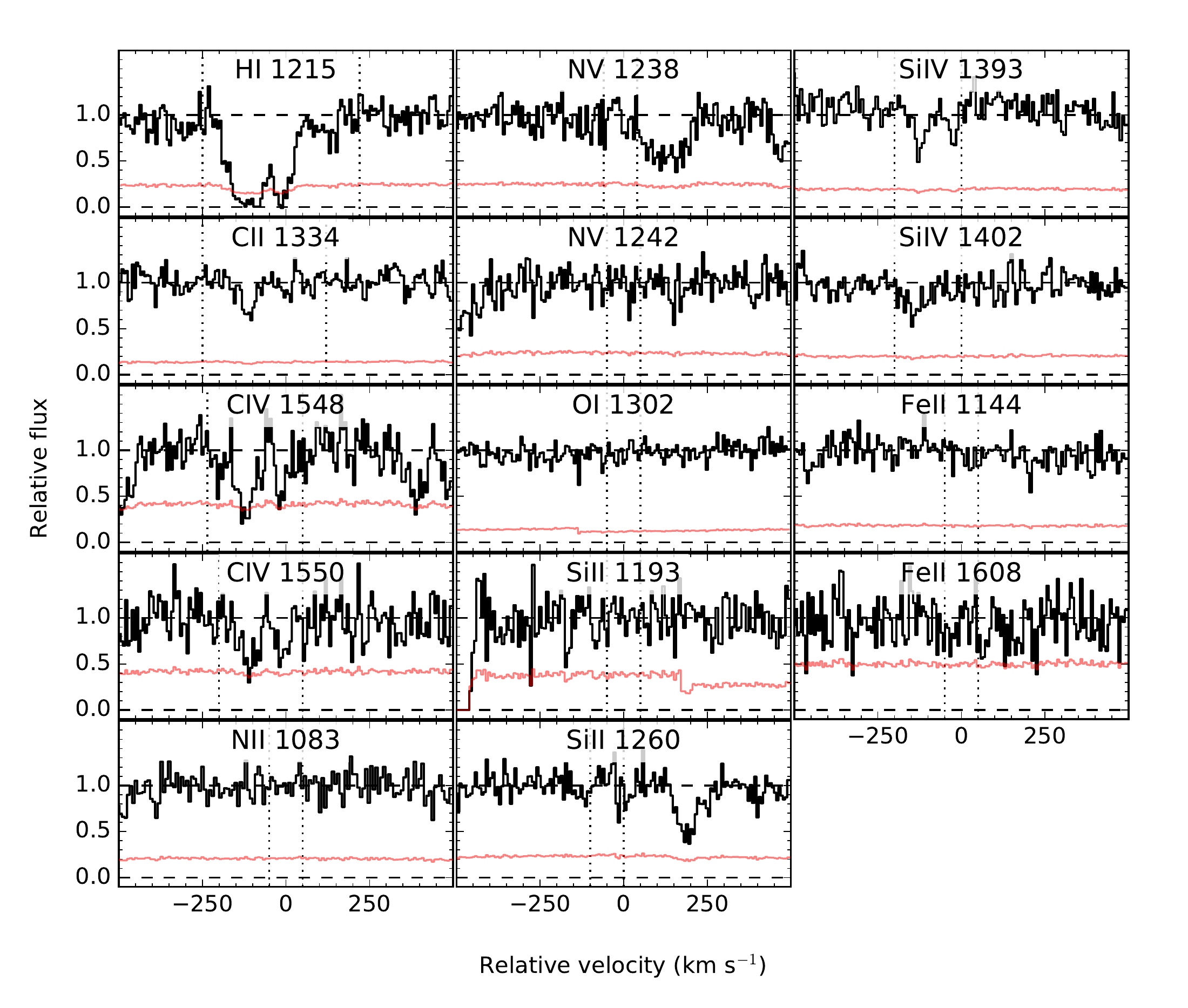}
\caption{Velocity profiles for the sightline towards J2322$-$0053 (z$_{\rm gal}$=0.081).}
\label{fig:J2322}
\end{center}
\end{figure*}

\begin{table}
\scriptsize
\begin{center}
\caption{Measured EWs for J2322-0053 (z=0.081)}
\label{tab:J2322-0053,289_77}
\begin{tabular}{lcccccc}
\hline
Ion& $\lambda$& $f$& $v_{\rm min}$& $v_{\rm max}$& EW& flag$^{\star}$\\
& [\AA{}]& & [\kms{}]& [\kms{}]& [m\AA{}]& \\
\hline
H\ion{i}& 1215.670& 4.164E-01& -250& 220& $785\pm48$& \textbf{9}\\
C\ion{ii}& 1334.532& 1.278E-01& -250& 120& $63\pm29$& \textbf{1}\\
C\ion{iv}& 1548.195& 1.908E-01& -235& 50& $350\pm90$& \textbf{1}\\
C\ion{iv}& 1550.770& 9.522E-02& -200& 50& $244\pm85$& \textbf{1}\\
N\ion{ii}& 1083.990& 1.031E-01& -50& 50& $<63$& \textbf{5}\\
N\ion{v}& 1238.821& 1.570E-01& -60& 40& $<78$& \textbf{5}\\
N\ion{v}& 1242.804& 7.823E-02& -50& 50& $<77$& \textbf{5}\\
O\ion{i}& 1302.168& 4.887E-02& -50& 50& $<39$& \textbf{5}\\
Si\ion{ii}& 1193.290& 4.991E-01& -50& 50& $<119$& \textbf{5}\\
Si\ion{ii}& 1260.422& 1.007E+00& -100& 0& $<78$& \textbf{5}\\
Si\ion{iv}& 1393.755& 5.280E-01& -200& 0& $79\pm33$& \textbf{1}\\
Si\ion{iv}& 1402.770& 2.620E-01& -200& 0& $137\pm34$& \textbf{1}\\
Fe\ion{ii}& 1144.938& 1.060E-01& -50& 50& $<56$& \textbf{5}\\
Fe\ion{ii}& 1608.451& 5.800E-02& -50& 50& $<202$& \textbf{5}\\
\hline
\end{tabular}
\\$^{\star}$The data quality flags represent a sum of whether or not the line is:\\ adopted ($+1$), blended ($+2$), undetected ($+4$), or saturated ($+8$).\\\end{center}
\end{table}

\onecolumn
\begin{longtable}{lccccccc}
\caption{Adopted EWs for COS-AGN sightlines}
\label{tab:all_ews}\\
\hline
QSO& z$_{\rm gal}$& Ion& $\lambda$& $v_{\rm min}$& $v_{\rm max}$& EW& flag$^{\star}$\\
& & & [\AA{}]& [\kms{}]& [\kms{}]& [m\AA{}]& \\
\hline
\endfirsthead
\multicolumn{8}{l}{{\bfseries \tablename\ \thetable{} -- continued from previous page}}\\
\hline
QSO& z$_{\rm gal}$& Ion& $\lambda$& $v_{\rm min}$& $v_{\rm max}$& EW& flag$^{\star}$\\
& & & [\AA{}]& [\kms{}]& [\kms{}]& [m\AA{}]& \\
\hline
\endhead
\hline \multicolumn{8}{l}{\bfseries Continued on next page}\\
\hline
\multicolumn{8}{l}{$^{\star}$The data quality flags represent a sum of whether or not the line is:}\\
\multicolumn{8}{l}{adopted ($+1$), blended ($+2$), undetected ($+4$), or saturated ($+8$).}\\
\endfoot
\hline
\multicolumn{8}{l}{$^{\star}$The data quality flags represent a sum of whether or not the line is:}\\
\multicolumn{8}{l}{adopted ($+1$), blended ($+2$), undetected ($+4$), or saturated ($+8$).}\\
\endlastfoot
\\
J0116+1429& 0.060& H\ion{i}& 1215& -200& 50& $306\pm38$& 9\\
J0116+1429& 0.060& C\ion{ii}& 1334& -50& 50& $<45$& 5\\
J0116+1429& 0.060& C\ion{iv}& 1548& -50& 50& $<138$& 5\\
J0116+1429& 0.060& C\ion{iv}& 1550& -50& 50& $<141$& 5\\
J0116+1429& 0.060& N\ion{v}& 1238& -50& 50& $<103$& 5\\
J0116+1429& 0.060& N\ion{v}& 1242& -50& 50& $<89$& 5\\
J0116+1429& 0.060& O\ion{i}& 1302& -50& 50& $<106$& 5\\
J0116+1429& 0.060& Si\ion{ii}& 1190& -50& 50& $<84$& 5\\
J0116+1429& 0.060& Si\ion{ii}& 1193& -50& 50& $<83$& 5\\
J0116+1429& 0.060& Si\ion{ii}& 1260& -200& 10& $<190$& 3\\
J0116+1429& 0.060& Si\ion{iii}& 1206& -50& 50& $<85$& 5\\
J0116+1429& 0.060& Si\ion{iv}& 1393& -50& 50& $<72$& 5\\
J0116+1429& 0.060& Si\ion{iv}& 1402& -50& 50& $<75$& 5\\
J0116+1429& 0.060& Fe\ion{ii}& 1608& -50& 50& $<100$& 5\\
J0843+4117& 0.068& H\ion{i}& 1215& -50& 50& $<131$& 5\\
J0843+4117& 0.068& C\ion{iv}& 1548& -50& 50& $<188$& 5\\
J0843+4117& 0.068& C\ion{iv}& 1550& -50& 50& $<188$& 5\\
J0843+4117& 0.068& N\ion{ii}& 1083& -50& 50& $<123$& 5\\
J0843+4117& 0.068& N\ion{v}& 1238& -50& 50& $<124$& 5\\
J0843+4117& 0.068& N\ion{v}& 1242& -50& 50& $<127$& 5\\
J0843+4117& 0.068& Si\ion{ii}& 1190& -50& 50& $<104$& 5\\
J0843+4117& 0.068& Si\ion{ii}& 1193& -50& 50& $<156$& 5\\
J0843+4117& 0.068& Si\ion{ii}& 1260& -50& 50& $<128$& 5\\
J0843+4117& 0.068& Si\ion{iii}& 1206& 0& 100& $<172$& 5\\
J0843+4117& 0.068& Fe\ion{ii}& 1144& -50& 50& $<129$& 5\\
J0851+4243& 0.024& H\ion{i}& 1215& -275& 250& $1066\pm53$& 9\\
J0851+4243& 0.024& C\ion{ii}& 1334& -50& 50& $<64$& 7\\
J0851+4243& 0.024& C\ion{iv}& 1548& -50& 50& $<84$& 5\\
J0851+4243& 0.024& C\ion{iv}& 1550& -50& 50& $<85$& 5\\
J0851+4243& 0.024& N\ion{v}& 1238& -50& 50& $<92$& 5\\
J0851+4243& 0.024& N\ion{v}& 1242& -50& 50& $<87$& 5\\
J0851+4243& 0.024& Si\ion{ii}& 1193& -50& 50& $<98$& 5\\
J0851+4243& 0.024& Si\ion{ii}& 1260& -50& 50& $<92$& 5\\
J0851+4243& 0.024& Si\ion{iii}& 1206& -200& 75& $165\pm44$& 1\\
J0851+4243& 0.024& Si\ion{iv}& 1393& -50& 50& $<56$& 5\\
J0851+4243& 0.024& Fe\ion{ii}& 1144& -50& 50& $<91$& 5\\
J0851+4243& 0.024& Fe\ion{ii}& 1608& -50& 50& $<128$& 5\\
J0852+0313& 0.129& H\ion{i}& 1025& -250& 300& $752\pm91$& 9\\
J0852+0313& 0.129& H\ion{i}& 1215& -275& 300& $1356\pm56$& 9\\
J0852+0313& 0.129& C\ion{ii}& 1036& -155& 230& $<366$& 11\\
J0852+0313& 0.129& C\ion{iv}& 1548& -200& 190& $657\pm66$& 1\\
J0852+0313& 0.129& C\ion{iv}& 1550& -210& 190& $474\pm71$& 1\\
J0852+0313& 0.129& N\ion{ii}& 1083& -30& 215& $200\pm41$& 1\\
J0852+0313& 0.129& N\ion{v}& 1238& -130& 195& $260\pm56$& 1\\
J0852+0313& 0.129& O\ion{i}& 1302& -250& 150& $<264$& 3\\
J0852+0313& 0.129& O\ion{vi}& 1031& -200& 240& $<451$& 11\\
J0852+0313& 0.129& O\ion{vi}& 1037& -135& 235& $273\pm80$& 1\\
J0852+0313& 0.129& Si\ion{ii}& 1190& -50& 200& $234\pm30$& 1\\
J0852+0313& 0.129& Si\ion{ii}& 1193& -145& 200& $378\pm40$& 9\\
J0852+0313& 0.129& Si\ion{ii}& 1260& -130& 200& $458\pm18$& 1\\
J0852+0313& 0.129& Si\ion{iii}& 1206& -225& 215& $804\pm52$& 9\\
J0852+0313& 0.129& Si\ion{iv}& 1393& -155& 230& $235\pm19$& 1\\
J0852+0313& 0.129& Si\ion{iv}& 1402& -125& 150& $115\pm28$& 1\\
J0852+0313& 0.129& Fe\ion{ii}& 1144& 50& 200& $87\pm39$& 1\\
J0853+4349& 0.090& H\ion{i}& 1025& 130& 340& $185\pm19$& 9\\
J0853+4349& 0.090& H\ion{i}& 1215& 85& 350& $512\pm9$& 9\\
J0853+4349& 0.090& C\ion{ii}& 1036& 50& 150& $<30$& 5\\
J0853+4349& 0.090& N\ion{ii}& 1083& 150& 250& $<20$& 5\\
J0853+4349& 0.090& N\ion{v}& 1238& 150& 250& $<21$& 5\\
J0853+4349& 0.090& N\ion{v}& 1242& -50& 50& $<22$& 5\\
J0853+4349& 0.090& Si\ion{ii}& 1190& 150& 250& $<22$& 5\\
J0853+4349& 0.090& Si\ion{iii}& 1206& 130& 295& $71\pm9$& 1\\
J0853+4349& 0.090& Fe\ion{ii}& 1144& 150& 250& $<20$& 5\\
J0948+5800& 0.084& H\ion{i}& 1215& -220& 0& $398\pm39$& 9\\
J0948+5800& 0.084& C\ion{ii}& 1334& -50& 50& $<64$& 5\\
J0948+5800& 0.084& C\ion{iv}& 1548& -170& -70& $107\pm67$& 1\\
J0948+5800& 0.084& C\ion{iv}& 1550& -170& -70& $29\pm70$& 1\\
J0948+5800& 0.084& N\ion{ii}& 1083& -50& 50& $<77$& 5\\
J0948+5800& 0.084& N\ion{v}& 1238& -50& 50& $<100$& 5\\
J0948+5800& 0.084& O\ion{i}& 1302& -50& 50& $<49$& 5\\
J0948+5800& 0.084& Si\ion{ii}& 1190& -50& 50& $<221$& 5\\
J0948+5800& 0.084& Si\ion{ii}& 1193& -50& 50& $<106$& 5\\
J0948+5800& 0.084& Si\ion{ii}& 1260& -50& 50& $<98$& 5\\
J0948+5800& 0.084& Si\ion{iv}& 1402& -50& 50& $<74$& 5\\
J0948+5800& 0.084& Fe\ion{ii}& 1608& -50& 50& $<335$& 5\\
J1117+2634& 0.065& H\ion{i}& 1215& -75& 200& $453\pm38$& 9\\
J1117+2634& 0.065& C\ion{ii}& 1334& -50& 50& $<38$& 5\\
J1117+2634& 0.065& C\ion{iv}& 1548& -50& 50& $<113$& 5\\
J1117+2634& 0.065& C\ion{iv}& 1550& -50& 50& $<118$& 5\\
J1117+2634& 0.065& N\ion{ii}& 1083& -50& 50& $<96$& 5\\
J1117+2634& 0.065& N\ion{v}& 1242& -50& 50& $<61$& 5\\
J1117+2634& 0.065& Si\ion{ii}& 1193& -50& 50& $<54$& 5\\
J1117+2634& 0.065& Si\ion{ii}& 1260& -50& 50& $<63$& 5\\
J1117+2634& 0.065& Si\ion{iii}& 1206& -50& 50& $<87$& 5\\
J1117+2634& 0.065& Si\ion{iv}& 1393& -50& 50& $<68$& 5\\
J1117+2634& 0.065& Si\ion{iv}& 1402& -50& 50& $<68$& 5\\
J1117+2634& 0.065& Fe\ion{ii}& 1608& -50& 50& $<68$& 5\\
J1117+2634& 0.029& H\ion{i}& 1215& 230& 435& $256\pm25$& 9\\
J1117+2634& 0.029& C\ion{ii}& 1334& -50& 50& $<73$& 5\\
J1117+2634& 0.029& C\ion{iv}& 1548& -50& 50& $<138$& 5\\
J1117+2634& 0.029& C\ion{iv}& 1550& -50& 50& $<148$& 5\\
J1117+2634& 0.029& N\ion{v}& 1242& -50& 50& $<86$& 5\\
J1117+2634& 0.029& Si\ion{ii}& 1190& -50& 50& $<77$& 5\\
J1117+2634& 0.029& Si\ion{ii}& 1193& -50& 50& $<74$& 5\\
J1117+2634& 0.029& Si\ion{ii}& 1260& -50& 50& $<82$& 5\\
J1117+2634& 0.029& Si\ion{iv}& 1393& -50& 50& $<41$& 5\\
J1117+2634& 0.029& Si\ion{iv}& 1402& -50& 50& $<42$& 5\\
J1117+2634& 0.029& Fe\ion{ii}& 1144& -50& 50& $<68$& 5\\
J1117+2634& 0.029& Fe\ion{ii}& 1608& -50& 50& $<124$& 5\\
J1127+2654& 0.033& H\ion{i}& 1215& -200& 200& $723\pm36$& 9\\
J1127+2654& 0.033& C\ion{ii}& 1334& -50& 50& $<83$& 5\\
J1127+2654& 0.033& C\ion{iv}& 1550& -50& 50& $<79$& 5\\
J1127+2654& 0.033& N\ion{v}& 1238& -50& 50& $<88$& 5\\
J1127+2654& 0.033& O\ion{i}& 1302& -50& 50& $<80$& 5\\
J1127+2654& 0.033& Si\ion{ii}& 1190& -50& 50& $<66$& 5\\
J1127+2654& 0.033& Si\ion{ii}& 1193& -50& 50& $<90$& 5\\
J1127+2654& 0.033& Si\ion{iii}& 1206& -50& 50& $<66$& 5\\
J1127+2654& 0.033& Si\ion{iv}& 1393& 0& 100& $<34$& 5\\
J1127+2654& 0.033& Si\ion{iv}& 1402& -50& 50& $<39$& 5\\
J1127+2654& 0.033& Fe\ion{ii}& 1144& -25& 75& $<83$& 5\\
J1127+2654& 0.033& Fe\ion{ii}& 1608& -50& 50& $<80$& 5\\
J1142+3016& 0.032& H\ion{i}& 1215& -130& 230& $927\pm36$& 9\\
J1142+3016& 0.032& C\ion{iv}& 1548& -130& 230& $167\pm53$& 1\\
J1142+3016& 0.032& C\ion{iv}& 1550& -130& 230& $129\pm54$& 1\\
J1142+3016& 0.032& N\ion{v}& 1238& -50& 50& $<123$& 5\\
J1142+3016& 0.032& N\ion{v}& 1242& -50& 50& $<114$& 5\\
J1142+3016& 0.032& O\ion{i}& 1302& -50& 50& $<100$& 5\\
J1142+3016& 0.032& Si\ion{ii}& 1190& -50& 50& $<82$& 5\\
J1142+3016& 0.032& Si\ion{ii}& 1193& -50& 50& $<76$& 5\\
J1142+3016& 0.032& Si\ion{ii}& 1260& -50& 50& $<86$& 5\\
J1142+3016& 0.032& Si\ion{iii}& 1206& -130& 170& $268\pm41$& 1\\
J1142+3016& 0.032& Si\ion{iv}& 1393& -50& 50& $<36$& 5\\
J1142+3016& 0.032& Si\ion{iv}& 1402& -50& 50& $<38$& 5\\
J1142+3016& 0.032& Fe\ion{ii}& 1144& -50& 50& $<98$& 5\\
J1155+2922& 0.046& H\ion{i}& 1215& -300& 175& $721\pm43$& 9\\
J1155+2922& 0.046& C\ion{ii}& 1334& -50& 50& $<109$& 5\\
J1155+2922& 0.046& N\ion{v}& 1238& -50& 50& $<100$& 5\\
J1155+2922& 0.046& O\ion{i}& 1302& -50& 50& $<97$& 5\\
J1155+2922& 0.046& Si\ion{ii}& 1190& -50& 50& $<71$& 5\\
J1155+2922& 0.046& Si\ion{ii}& 1193& -50& 50& $<69$& 5\\
J1155+2922& 0.046& Si\ion{ii}& 1260& -50& 50& $<124$& 5\\
J1155+2922& 0.046& Si\ion{iii}& 1206& -70& 120& $73\pm33$& 1\\
J1155+2922& 0.046& Si\ion{iv}& 1393& -50& 50& $<177$& 5\\
J1155+2922& 0.046& Fe\ion{ii}& 1144& -300& 175& $85\pm52$& 1\\
J1214+0825& 0.074& H\ion{i}& 1215& -120& 120& $229\pm34$& 9\\
J1214+0825& 0.074& C\ion{ii}& 1334& -70& 120& $154\pm41$& 1\\
J1214+0825& 0.074& N\ion{ii}& 1083& -50& 50& $<92$& 5\\
J1214+0825& 0.074& N\ion{v}& 1238& -50& 50& $<74$& 5\\
J1214+0825& 0.074& O\ion{i}& 1302& -50& 50& $<83$& 5\\
J1214+0825& 0.074& Si\ion{ii}& 1190& -50& 50& $<66$& 5\\
J1214+0825& 0.074& Si\ion{ii}& 1260& -80& 120& $107\pm34$& 1\\
J1214+0825& 0.074& Fe\ion{ii}& 1144& -50& 50& $<65$& 5\\
J1404+3353& 0.026& H\ion{i}& 1215& -75& 300& $736\pm58$& 9\\
J1404+3353& 0.026& N\ion{v}& 1238& -50& 50& $<101$& 5\\
J1404+3353& 0.026& N\ion{v}& 1242& -50& 50& $<103$& 5\\
J1404+3353& 0.026& O\ion{i}& 1302& -50& 50& $<147$& 5\\
J1404+3353& 0.026& Si\ion{ii}& 1260& -50& 50& $<108$& 5\\
J1404+3353& 0.026& Si\ion{iv}& 1393& -50& 50& $<179$& 5\\
J1404+3353& 0.026& Fe\ion{ii}& 1144& -50& 50& $<117$& 5\\
J1419+0606& 0.049& H\ion{i}& 1215& -220& 70& $575\pm48$& 9\\
J1419+0606& 0.049& C\ion{iv}& 1548& -160& -30& $135\pm100$& 9\\
J1419+0606& 0.049& C\ion{iv}& 1550& -160& -30& $107\pm103$& 9\\
J1419+0606& 0.049& N\ion{ii}& 1083& -50& 50& $<146$& 5\\
J1419+0606& 0.049& O\ion{i}& 1302& -50& 50& $<97$& 5\\
J1419+0606& 0.049& Si\ion{ii}& 1190& -125& -25& $<71$& 5\\
J1419+0606& 0.049& Si\ion{ii}& 1193& -125& -25& $<70$& 5\\
J1419+0606& 0.049& Si\ion{ii}& 1260& -220& 0& $158\pm41$& 1\\
J1419+0606& 0.049& Si\ion{iii}& 1206& -220& 70& $294\pm34$& 1\\
J1419+0606& 0.049& Si\ion{iv}& 1393& -190& -30& $119\pm43$& 1\\
J1419+0606& 0.049& Si\ion{iv}& 1402& -190& -30& $93\pm44$& 1\\
J1419+0606& 0.049& Fe\ion{ii}& 1608& -50& 50& $<331$& 5\\
J1454+3046& 0.031& H\ion{i}& 1215& -195& -55& $194\pm27$& 9\\
J1454+3046& 0.031& C\ion{ii}& 1334& -50& 50& $<95$& 5\\
J1454+3046& 0.031& N\ion{v}& 1242& -50& 50& $<105$& 5\\
J1454+3046& 0.031& O\ion{i}& 1302& -50& 50& $<97$& 5\\
J1454+3046& 0.031& Si\ion{ii}& 1190& -50& 50& $<80$& 5\\
J1454+3046& 0.031& Si\ion{ii}& 1193& -50& 50& $<78$& 5\\
J1454+3046& 0.031& Si\ion{ii}& 1260& -50& 50& $<84$& 5\\
J1454+3046& 0.031& Si\ion{ii}& 1304& -50& 50& $<97$& 5\\
J1454+3046& 0.031& Si\ion{iii}& 1206& -50& 50& $<76$& 5\\
J1454+3046& 0.031& Si\ion{iv}& 1393& -50& 50& $<160$& 5\\
J1454+3046& 0.031& Si\ion{iv}& 1402& -50& 50& $<170$& 5\\
J1454+3046& 0.031& Fe\ion{ii}& 1144& -50& 50& $<92$& 5\\
J1536+1412& 0.093& H\ion{i}& 1215& -200& 90& $367\pm52$& 9\\
J1536+1412& 0.093& C\ion{ii}& 1334& -50& 50& $<61$& 5\\
J1536+1412& 0.093& C\ion{iv}& 1548& -50& 50& $<105$& 5\\
J1536+1412& 0.093& C\ion{iv}& 1550& -50& 50& $<101$& 5\\
J1536+1412& 0.093& N\ion{ii}& 1083& -50& 50& $<108$& 5\\
J1536+1412& 0.093& N\ion{v}& 1242& -50& 50& $<97$& 5\\
J1536+1412& 0.093& O\ion{i}& 1302& -50& 50& $<59$& 5\\
J1536+1412& 0.093& Si\ion{ii}& 1260& -50& 50& $<100$& 5\\
J1536+1412& 0.093& Si\ion{iii}& 1206& -50& 50& $<96$& 5\\
J1536+1412& 0.093& Si\ion{iv}& 1393& -50& 50& $<111$& 5\\
J1536+1412& 0.093& Si\ion{iv}& 1402& -50& 50& $<119$& 5\\
J1536+1412& 0.093& Fe\ion{ii}& 1144& -50& 50& $<89$& 5\\
J1536+1412& 0.093& Fe\ion{ii}& 1608& -50& 50& $<267$& 5\\
J1607+1334& 0.069& H\ion{i}& 1215& -220& 200& $400\pm82$& 9\\
J1607+1334& 0.069& C\ion{ii}& 1334& -50& 50& $<87$& 5\\
J1607+1334& 0.069& C\ion{iv}& 1548& -50& 50& $<363$& 5\\
J1607+1334& 0.069& C\ion{iv}& 1550& -50& 50& $<361$& 5\\
J1607+1334& 0.069& N\ion{v}& 1238& -50& 50& $<182$& 5\\
J1607+1334& 0.069& N\ion{v}& 1242& -50& 50& $<165$& 5\\
J1607+1334& 0.069& O\ion{i}& 1302& -50& 50& $<108$& 5\\
J1607+1334& 0.069& Si\ion{ii}& 1190& -50& 50& $<112$& 5\\
J1607+1334& 0.069& Si\ion{ii}& 1193& -50& 50& $<111$& 5\\
J1607+1334& 0.069& Si\ion{ii}& 1260& -100& 0& $<172$& 5\\
J1607+1334& 0.069& Si\ion{iii}& 1206& -50& 50& $<113$& 5\\
J1607+1334& 0.069& Si\ion{iv}& 1393& -50& 50& $<93$& 5\\
J1607+1334& 0.069& Si\ion{iv}& 1402& -50& 50& $<85$& 5\\
J1607+1334& 0.069& Fe\ion{ii}& 1144& -50& 50& $<121$& 5\\
J1607+1334& 0.069& Fe\ion{ii}& 1608& -50& 50& $<412$& 5\\
J2133-0712& 0.064& H\ion{i}& 1215& -200& 120& $501\pm37$& 9\\
J2133-0712& 0.064& C\ion{ii}& 1334& -140& -50& $53\pm18$& 1\\
J2133-0712& 0.064& C\ion{iv}& 1548& -50& 50& $<205$& 5\\
J2133-0712& 0.064& C\ion{iv}& 1550& -50& 50& $<212$& 5\\
J2133-0712& 0.064& N\ion{v}& 1238& -50& 50& $<74$& 5\\
J2133-0712& 0.064& N\ion{v}& 1242& -50& 50& $<83$& 5\\
J2133-0712& 0.064& O\ion{i}& 1302& -50& 50& $<73$& 5\\
J2133-0712& 0.064& Si\ion{ii}& 1190& -50& 50& $<46$& 5\\
J2133-0712& 0.064& Si\ion{ii}& 1193& -50& 50& $<47$& 5\\
J2133-0712& 0.064& Si\ion{ii}& 1260& -50& 50& $<84$& 5\\
J2133-0712& 0.064& Si\ion{iii}& 1206& 75& 370& $243\pm18$& 1\\
J2133-0712& 0.064& Si\ion{iv}& 1393& -50& 50& $<75$& 5\\
J2133-0712& 0.064& Si\ion{iv}& 1402& -50& 50& $<59$& 5\\
J2133-0712& 0.064& Fe\ion{ii}& 1608& -50& 50& $<252$& 5\\
J2322-0053& 0.081& H\ion{i}& 1215& -250& 220& $785\pm48$& 9\\
J2322-0053& 0.081& C\ion{ii}& 1334& -250& 120& $63\pm29$& 1\\
J2322-0053& 0.081& C\ion{iv}& 1548& -235& 50& $350\pm90$& 1\\
J2322-0053& 0.081& C\ion{iv}& 1550& -200& 50& $244\pm85$& 1\\
J2322-0053& 0.081& N\ion{ii}& 1083& -50& 50& $<63$& 5\\
J2322-0053& 0.081& N\ion{v}& 1238& -60& 40& $<78$& 5\\
J2322-0053& 0.081& N\ion{v}& 1242& -50& 50& $<77$& 5\\
J2322-0053& 0.081& O\ion{i}& 1302& -50& 50& $<39$& 5\\
J2322-0053& 0.081& Si\ion{ii}& 1193& -50& 50& $<119$& 5\\
J2322-0053& 0.081& Si\ion{ii}& 1260& -100& 0& $<78$& 5\\
J2322-0053& 0.081& Si\ion{iv}& 1393& -200& 0& $79\pm33$& 1\\
J2322-0053& 0.081& Si\ion{iv}& 1402& -200& 0& $137\pm34$& 1\\
J2322-0053& 0.081& Fe\ion{ii}& 1144& -50& 50& $<56$& 5\\
J2322-0053& 0.081& Fe\ion{ii}& 1608& -50& 50& $<202$& 5\\
\end{longtable}
\twocolumn

\end{document}